\documentclass[aps,
pra,
twocolumn,
amsmath,
amssymb,
graphicx,
showpacs,
floatfix
]
{revtex4-1} 

\usepackage{hyperref}
\usepackage{mathtools}
\usepackage{amsmath}
\usepackage{epsfig}
\usepackage{wrapfig} 
\usepackage{mathrsfs}
\usepackage{amssymb} 
\usepackage[x11names]{xcolor}
\usepackage{amsbsy}
\usepackage{subfigure}
\usepackage{marginnote}
\usepackage{slashed}
\usepackage{marvosym}
\usepackage{pifont}
\usepackage{mathbbol}
\usepackage{txfonts}
\usepackage{wrapfig}
\usepackage{upgreek}
\usepackage{bbm}
\usepackage{yfonts}
\usepackage{xspace}

\newcommand{\bcen}{\begin{center}}
\newcommand{\ecen}{\end{center}}
\newcommand{\btab}{\begin{tabular}}
\newcommand{\etab}{\end{tabular}}
\newcommand{\bdes}{\begin{description}}
\newcommand{\edes}{\end{description}}

\newcommand{\beq}{\begin{equation}}
\newcommand{\eeq}{\end{equation}}
\newcommand{\bea}{\begin{eqnarray}}
\newcommand{\eea}{\end{eqnarray}}

\newcommand{\half}{\frac{1}{2}}
\newcommand{\bary}{\begin{array}}
\newcommand{\eary}{\end{array}}
\newcommand{\benum}{\begin{enumerate}}
\newcommand{\eenum}{\end{enumerate}}
\newcommand{\bitem}{\begin{itemize}}
\newcommand{\eitem}{\end{itemize}}

\newcommand{\bOne}{{\boldsymbol{1}}}
\newcommand{\bR} { {\mathbf R} }

\newcommand{\bzero} { {\boldsymbol{0}}}

\newcommand{\ket}[1]{{| #1 \rangle}}
\newcommand{\braket}[2]{\langle #1 | #2 \rangle}

\newcommand{\eqn}[1] {eqn.~(\ref{#1})}
\newcommand{\prn}[1] {(\ref{#1})}
\newcommand{\sect}[1] {section~\ref{#1}}
\newcommand{\Sect}[1] {Section~\ref{#1}}
\newcommand{\fig}[1]{fig.~\ref{#1}}
\newcommand{\Fig}[1]{Fig.~\ref{#1}}

\newcommand{\vPsi}{{{\Psi}}}
\newcommand{\vtPsi}{{\widetilde{\Psi}}}
\newcommand{\vH}{{\mathscr{H}}}
\newcommand{\vU}{{\mathscr{U}}}
\newcommand{\vI}{{\mathscr{I}}}
\newcommand{\vN}{{\mathscr{N}}}
\newcommand{\TR}{{\mathscr{T}}}
\newcommand{\CC}{{\mathscr{C}}}
\newcommand{\SL}{{\mathscr{S}}}

\newcommand{\mA}{{\mathbf{A}}}
\newcommand{\mR}{\mathbf{R}}
\newcommand{\mH}{\mathbf{H}}

\newcommand{\mHc}{{\breve{\mH}}}
\newcommand{\mHr}{{\mathring{\mH}}}

\newcommand{\mF}{\mathbf{F}}
\newcommand{\mOne}{\mathbf{1}}
\newcommand{\mJ}{\mathbf{J}}
\newcommand{\mU}{\mathbf{U}}
\newcommand{\mZero}{\mathbf{0}}
\newcommand{\calH}{\mathcal{H}}
\newcommand{\calV}{\mathcal{V}}

\newcommand{\vUUsl}{{{\vU}_{\textup{ {\sc USL}}}}}
\newcommand{\vUTrn}{{{\vU}_{\textup{ {\sc TRN}}}}}
\newcommand{\vUUsli}{{{\vU}_{\textup{ {\sc USL}}}^{-1}}}
\newcommand{\vUTrni}{{{\vU}_{\textup{ {\sc TRN}}}^{-1}}}

\newcommand{\mUUsl}{{{\mU}_{\textup{ {\sc USL}}}}}
\newcommand{\mUTrn}{{{\mU}_{\textup{ {\sc TRN}}}}}

\newcommand{\mUTrns}{{{\mU}_{\textup{ {\sc TRN}}}^*}}

\newcommand{\mUT}{{{\mU}_{\textup{T}}}}
\newcommand{\mUC}{{{\mU}_{\textup{C}}}}
\newcommand{\mUS}{{{\mU}_{\textup{S}}}}
\newcommand{\mUTd}{{{\mU}_{\textup{T}}^\dagger}}
\newcommand{\mUCd}{{{\mU}_{\textup{C}}^\dagger}}
\newcommand{\mUSd}{{{\mU}_{\textup{S}}^\dagger}}
\newcommand{\mUTs}{{{\mU}_{\textup{T}}^*}}
\newcommand{\mUCs}{{{\mU}_{\textup{C}}^*}}
\newcommand{\mUSs}{{{\mU}_{\textup{S}}^*}}

\newcommand{\mtUT}{{{\widetilde{\mU}}_{\textup{T}}}}

\newcommand{\mtUS}{{{\widetilde{\mU}}_{\textup{S}}}}

\newcommand{\mT}{{\mathbf{T}}}
\newcommand{\mC}{{\mathbf{C}}}
\newcommand{\mS}{{\mathbf{S}}}
\newcommand{\mK}{{\mathbf{K}}}

\newcommand{\mV}{{\mathbf{V}}}

\newcommand{\tT}{{\textup{T}}}
\newcommand{\tC}{{\textup{C}}}
\newcommand{\tS}{{\textup{S}}}

\newcommand{\UA}{{UA}}
\newcommand{\UL}{{UL}}
\newcommand{\TL}{{TL}}
\newcommand{\TA}{{TA}}

\newcommand{\UAm}{\textup{UA}}
\newcommand{\ULm}{\textup{UL}}
\newcommand{\TLm}{\textup{TL}}
\newcommand{\TAm}{\textup{TA}}

\newcommand{\GUL}{G_\calV^{\ULm}}
\newcommand{\GUA}{G_\calV^{\UAm}}
\newcommand{\GTL}{G_\calV^{\TLm}}
\newcommand{\GTA}{G_\calV^{\TAm}}

\newcommand{\Ac}{{\sf{A}}}
\newcommand{\AI}{{\sf{AI}}}
\newcommand{\AII}{{\sf{AII}}}
\newcommand{\AIII}{{\sf{AIII}}}
\newcommand{\Dc}{{\sf{D}}}
\newcommand{\C}{{\sf{C}}}
\newcommand{\BDI}{{\sf{BDI}}}
\newcommand{\CI}{{\sf{CI}}}
\newcommand{\DIII}{{\sf{DIII}}}
\newcommand{\CII}{{\sf{CII}}}

\newcommand{\absI}{\textup{$I$}}
\newcommand{\absT}{\textup{$\Theta$}}
\newcommand{\absC}{\textup{$\Xi$}}
\newcommand{\absS}{\textup{$\Sigma$}}
\newcommand{\absg}{\textup{$g$}}

\newcommand{\symS}{{\mathcal{S}}}
\newcommand{\symI}{{\mathcal{I}}}
\newcommand{\symZ}{{\mathcal{Z}}}
\newcommand{\calW}{{\mathcal{W}}}
\newcommand{\symK}{{\mathcal{K}}}
\newcommand{\Klein}{{\symK_4}}

\newcommand{\ci}{\mathbbm{i}}

\newcommand{\lieu}{\boldsymbol{\mathfrak{u}}}

\newcommand{\NP}{N_P}

\newcommand{\nbdy}[3]{{#1}^{(#2)}_{#3}}
\newcommand{\mnbdy}[2]{{#1}^{(#2)}}
\newcommand{\combi}[2]{\small{{#1}\choose{#2}}}

\DeclareMathOperator{\tr}{tr}

\newcommand{\mylabel}[1]{\label{#1}}

\newcommand{\mycite}[1]{\cite{#1}}

\newcommand{\tbtmat}[4]{{\left(\begin{array}{cc}#1 & #2 \\#3 & #4 \end{array} \right)}}

\newcommand{\Hmat}{{\mathbb{H}}}
\newcommand{\Schrod}{{\mathbb{U}_{\textup{Schr\"od}}}}
\newcommand{\Hone}{\mnbdy{\mH}{1}}
\newcommand{\class}[1]{{\medskip\noindent{\bf Class #1:}}}
\newcommand{\ciH}{{\ci \calH}}
\newcommand{\ciHone}{\mnbdy{\ciH}{1}}
\newcommand{\mh}{\mathbf{h}}
\newcommand{\lieo}{\boldsymbol{\mathfrak{o}}}
\newcommand{\lieusp}{\boldsymbol{\mathfrak{usp}}}

\newcommand{\mytitle}{{The tenfold way redux: Fermionic systems with $N$-body interactions}}

\newcommand{\myauthors}{{Adhip Agarwala, Arijit Haldar and Vijay B.~Shenoy}}
\newcommand{\myaffl}{{Department of Physics, Indian Institute of Science, Bangalore 560012, India.}}

\usepackage{hyperref}
\usepackage{amsmath}
\usepackage{epsfig}
\usepackage{wrapfig} 
\usepackage{mathrsfs}
\usepackage{amssymb} 
\usepackage{color}
\usepackage{amsbsy}
\usepackage{subfigure}
\usepackage{marginnote}
\usepackage{slashed}
\usepackage{marvosym}
\usepackage{pifont}
\usepackage{mathbbol}
\usepackage{txfonts}
\usepackage{wrapfig}
\usepackage{upgreek}
\usepackage{bbm}

\newcommand{\usenomenclature}{}
\ifdefined\usenomenclature
\usepackage[noprefix]{nomencl}
\fi
\newcommand{\makenomcl}{\ifdefined\usenomenclature\makenomenclature\fi}

\newcommand{\nomcls}[3]{\ifdefined\usenomenclature \nomenclature[#1]{#2}{#3}\fi}

\newcommand{\printnomcl}{\ifdefined\usenomenclature\printnomenclature\fi}
\newcommand{\symref}[1]{First appearance in sec\ref{#1}, page \pageref{#1}.}
\newcommand{\symeqref}[1]{First appearance in \eqn{#1}, page \pageref{#1}.}
\newcommand{\symdefref}[1]{Defined in \eqn{#1}, page \pageref{#1}.}

\newcommand{\No}{L} 
\newcommand{\Mo}{M} 
\newcommand{\mb}{N} 
\newcommand{\tbtarray}[4]{{\begin{array}{cc}#1 & #2 \\#3 & #4 \end{array}}} 
\newcommand{\tbtpart}[4]{{\left(\begin{array}{c|c}#1 & #2 \\ \hline #3 & #4 \end{array} \right)}} 

\newcommand{\mUTmb}[1]{{{{\mU}}_{\textup{T}}^{(#1)}}}
\newcommand{\mUCmb}[1]{{{{\mU}}_{\textup{C}}^{(#1)}}}
\newcommand{\mUSmb}[1]{{{{\mU}}_{\textup{S}}^{(#1)}}}
\newcommand{\mUTdmb}[1]{{{{\mU}}_{\textup{T}}^{{(#1)}\dagger}}}
\newcommand{\mUCdmb}[1]{{{{\mU}}_{\textup{C}}^{{(#1)}\dagger}}}
\newcommand{\mUSdmb}[1]{{{{\mU}}_{\textup{S}}^{{(#1)}\dagger}}}

\newcommand{\mHmb}[1]{\mathbf{H}^{(#1)}}

\newcommand{\ta}{\alpha} 
\newcommand{\tb}{\beta}
\newcommand{\pa}{{\alpha_p}} 
\newcommand{\pb}{{\beta_p}}
\newcommand{\qa}{{\alpha_q}} 
\newcommand{\qb}{{\beta_q}}

\newcommand{\sysign}[4]{$\begin{array}{c} $#1$ \\ $(#2,#3,#4)$ \end{array}$} 
\newcommand{\Sys}{\sigma} 
\newcommand{\Ans}{\pi} 

\newcommand{\mHt}{\mathbf{H}^{(2)}}
\newcommand{\mht}{\mathbf{h}^{(2)}}
\newcommand{\mHm}{\mathbf{H}^{(\mb)}}
\newcommand{\mhm}{\mathbf{h}^{(\mb)}}

\renewcommand{\symS}{{\mathcal{G}}} 
\renewcommand{\mV}{\mathbf{u}} 
\newcommand{\mr}{\mathbf{r}} 

\newcommand{\USp}{\text{\it USp}}

\newcommand{\transp}[1]{\left[{#1}\right]^T} 
\newcommand{\conj}[1]{\left[{#1}\right]^*} 
\newcommand{\dagg}[1]{\left[{#1}\right]^\dagger} 

\newcommand{\mHPLe}[1]{\mathbf{H}^{(#1)}_{+}}
\newcommand{\mHMIe}[1]{\mathbf{H}^{(#1)}_{-}}

\newcommand{\mUTt}{{{{\mU}}_{\textup{T}}^T}}

\makenomcl

\let\chapter\section
\let\section\subsection
\let\subsection\subsubsection

\newcommand{\oibook}[1]{}
\newcommand{\Grot}{Grotesque}
\newcommand{\grot}{grotesque}
\newcommand{\GFS}{GFS}

\renewcommand{\le}{\leqslant}

\begin{document}

\title{\mytitle}
\author{\myauthors}\email{shenoy@physics.iisc.ernet.in}
\affiliation{\myaffl}

\date{\today{}}
\begin{abstract} 
We provide a systematic treatment of the tenfold way of classifying fermionic systems that naturally allows for the study of those with arbitrary $\mb$-body interactions. We identify four types of symmetries that such systems can possess, which consist of one ordinary type (usual unitary symmetries), and three {\em non}-ordinary symmetries (such as time reversal, charge conjugation and sublattice). Focusing on systems  that possess no non-trivial ordinary symmetries, we demonstrate that the non-ordinary symmetries are strongly constrained. This approach not only leads very naturally to the tenfold classes, but also obtains the canonical representations of these symmetries in each of the ten classes. We also provide a group cohomological perspective of our results in terms of projective representations. We then use the canonical representations of the symmetries to obtain the structure of Hamiltonians with arbitrary $\mb$-body interactions  in each of the ten classes. We show that the space of $\mb$-body Hamiltonians has an affine subspace (of a vector space) structure in classes which have either or both charge conjugation and sublattice symmetries. Our results can help address open questions on the topological classification of interacting fermionic systems. 

\end{abstract}

\pacs{71.10.-w, 71.27.+a, 71.10.Fd}

\maketitle 




\section{Introduction}\mylabel{sec:Intro}

Until very recently, our understanding of fermionic many body systems, for most part, could be traced to a handful of ground states and their excitations, e.~g., the Fermi sea, the band insulator, the filled Landau level, and the BCS superconducting state. This picture has been drastically overhauled in the last decade initiated by the discovery of the two dimensional spin Hall insulators \mycite{Murakami2004,Kane2005a,Kane2005b,Bernevig2006,Bernevig2006a}, and bolstered by its experimental realization \mycite{Konig2007a}. It soon became clear \mycite{FuKaneMele2007,Moore2007,Roy2009b,Hsieh2008}
 that systems with time reversal symmetry have new types of ground states -- the topological insulator -- in three dimensions as well (see  \mycite{Moore2010,Qi2010,Hasan2010,Qi2011,Bernevig2013} for a review). These developments naturally motivated the question of the classification of gapped states of fermionic many body systems which has now firmly established itself as a key research direction of condensed matter physics.

For systems of noninteracting fermions, a comprehensive classification has been achieved (see \citep{Beenakker2015,Chiu2015,Ludwig2015} for an overview), and indeed marks a milestone in condensed matter research. Central to this success is the symmetry classification of fermionic systems based on a set of ``intrinsic'' symmetries -- the tenfold way of \citeauthor{Altland1997}\mycite{Altland1997} (see \cite{Heinzner2005,Zirnbauer2010} for a more formal treatment) -- which places any fermionic system into one of ten symmetry classes. From the point of view of fermionic physics, this work represents the culmination of a program of classification initiated by Dyson \cite{Dyson1962} via the threefold way. In each symmetry class, a gapped fermionic system may possess ground states which are topologically distinct.  Early work in this direction \mycite{Qi2008,Schnyder2008} was developed into a complete picture \mycite{Ryu2010}, culminating in the ``periodic table'' of Kitaev\mycite{Kitaev2009}. The key result is that in any spatial dimension, there are only five symmetry classes that support nontrivial topological phases of noninteracting fermionic systems. The presence or absence of a topological phase is characterized by a nontrivial abelian group such as $\mathbb{Z}, \mathbb{Z}_2$ or $2\mathbb{Z}$.  Even more remarkably, the pattern of nontrivial groups has a periodicity (in spatial dimensions)  of 2 for the so called complex classes, and 8 in the real classes with a very specific relationship between nontrivial classes in a given dimension and ones just above and just below. These ideas have also been generalized to include defects\mycite{Freedman2011}, and have also been visited again from more formal perspectives\mycite{Freed2013,Kennedy2016} (see also \mycite{Prodan2016}). Further developments in the physics of noninteracting systems came up with the study of the interplay of intrinsic symmetries with those of the environment (such as crystalline space groups) that have resulted in a more intricate classification\cite{Fu2010,Slager2013,Alexandradinata2016}.

A most intriguing story emerges up on the inclusion of interactions, i.e., for a system of fermions  with ``nonquadratic'' terms in their Hamiltonian. \citeauthor{Fidkowski2010}\cite{Fidkowski2010} showed that the presence of two-body interactions results in a ``collapse'' of the noninteracting topological classification -- for example,  Kitaev's Majorana chain\cite{Kitaev2001}, described by the group $\mathbb{Z}$ collapses to $\mathbb{Z}_8$ upon the inclusion of interactions. Following this, the natural question that arises is regarding the principles of classification of topological phases in the presence of interaction. Ideas based on group cohomology\cite{Chen2013}, and supercohomology for fermions\cite{Gu2014} have been put forth. While these are important steps towards the final goal of topological classification of interacting systems, the problem remains at the very frontier of condensed matter research\cite{Ryu2015,Senthil2015}.

A crucial aspect of the tenfold symmetry classification of noninteracting systems is the determination of the structure of the Hamiltonians allowed in each class. The knowledge of this structure then provides ways for viewing these systems from various perspectives e.~g., classifying spaces\mycite{Kennedy2015}, structure of the target manifolds of nonlinear $\sigma$-model descriptions\mycite{Ryu2010} etc. To the best of our knowledge, a study of the structure of fermionic Hamiltonians in each of the ten symmetry classes with arbitrary $\mb$\mylabel{sym:mb}-body interactions is not available in the literature. An understanding of the structure of the interacting Hamiltonians will not only aid the analysis of these systems, but also help motivate models that could be crucial to build phenomenology and intuition (possibly through numerics)  much like what the Kane-Mele and BHZ models\cite{Kane2005a,Bernevig2006a} did in the context of noninteracting systems. This paper aims to fill this lacuna. Enroute this pursuit, we build a transparent framework that allows for a systematic approach to these problems, which even provides further insights for the noninteracting case.
\nomcls{10}{$\mb$}{Hamiltonian contains upto $\mb$-body interaction terms.  \symref{sym:mb}}
Our framework is developed for a system comprising of $\No$\mylabel{sym:No} one-particle states (dubbed ``orbitals'') which can be populated by fermions. Symmetries of such a system can be either {\em linear} or {\em antilinear}, and {\em non-transposing} (occupied orbitals mapped to occupied orbitals) or {\em transposing} (occupied orbitals mapped to unoccupied ones), i.~e., four distinct types of symmetries. We call the linear non-transposing type of symmetries as {\em ordinary}, and the remaining three types (antilinear non-transposing, linear transposing, antilinear transposing) as {\em non-ordinary} symmetries. The focus of our work are  those fermionic systems -- which we dub as {\em ``\grot''} -- which do not possess any nontrivial ordinary symmetries. We show that non-ordinary symmetries of a \grot~fermionic system (\GFS) are {\em solitary}, i.~e., a \GFS~ can possess at most one from each type of non-ordinary symmetries. Identifying these solitary non-ordinary symmetries with time reversal (T), charge conjugation (C), and sublattice (S), leads us to the familiar \citeauthor{Altland1997} tenfold symmetry classes. We then address the key question of determining the structure of the Hamiltonian in these classes, by constructing ``canonical'' representations (by unitary matrices) for the symmetry operations in each class. This construction allows for a systematic and efficient determination of the structure of any arbitrary $N$-body interaction term in the Hamiltonian of any class. Not surprisingly we recover all known results of noninteracting systems and, more importantly, we provide physical insights that can, inter alia, aid model building thereby helping address the outstanding open problems vis a vis topological phases of interacting fermionic systems.
\nomcls{20}{$\No$}{No. of orbitals. \symref{sym:No}}

We begin the discussion with the setting of the problem in \sect{sec:Setting}. This is followed in \sect{sec:Symmetries} by a discussion of the four types of symmetry operations that a fermionic system can possess. \Sect{sec:Grot} introduces and studies those fermionic systems that possess no nontrivial ordinary symmetries and obtains the constraints that the symmetries have to satisfy. The tenfold way is elucidated in \sect{sec:TenFold} which includes a treatment of the canonical representation of the symmetry in each of the symmetry classes, the result of which is displayed in table ~\ref{tab:tenfold}. This is followed by \sect{sec:GrpCohom} that provides an understanding of the results shown in table ~\ref{tab:tenfold} from the point of view of group cohomology. It is shown that every entry of table ~\ref{tab:tenfold} is made up of copies of irreducible projective representations of the Klein group or its subgroups. Known results of noninteracting systems are reproduced in \sect{sec:NonInt} (see table~\ref{tab:NonInt}). \Sect{sec:IntGen} lays down the framework for obtaining the structure of a Hamiltonian with $N$-body interactions by reducing the problem to the determination of certain vector subspaces and arbitrary $K$-body interaction term is shown to be an element of an affine subspace. Techniques needed for performing the analysis for determining the subspace structure are developed first for the two-body case in \sect{sec:Twobody} and later generalized to the arbitrary $N$-body case in \sect{sec:Nbody}. Tables~\ref{tab:NevenH} and \ref{tab:NoddH} contain the structures of the $N$-body interaction term in each class. The paper is concluded in \sect{sec:Conc} where the significance and scope of our results are highlighted including their use in other problems of interest. For the convenience of the reader, all important symbols used in the paper are listed in appendix \ref{sec:ListOfSymbols}.

The  paper is structured to be self contained  and easy to use in that we develop the discussion from the very basic building blocks. This desideratum naturally results in some overlap with the results from the previous works.  We shall specifically mention only those results that we have used explicitly in our discussion. 

\section{The Setting}\mylabel{sec:Setting}

Our system consists of $\No$ one particle states $\ket{i}$\mylabel{sym:keti}, $i=1,\ldots,\No$-- which we call ``orbitals'' --  that are orthonormal $\braket{i}{j} = \delta_{ij}$  ($\delta_{ij}$ is the Kronecker delta symbol). Note that these states could denote a variety of situations; $i$  could be orbitals at different sites of a lattice, or different atomic orbitals, or even flavor states of an elementary particle -- our use of the term orbital denotes one particle states in any context including those just stated.\nomcls{30}{$\ket{i}$}{One-particle state. \symref{sym:keti}} 
 Starting from the vacuum state $\ket{0}$\mylabel{sym:vacuum}, we can create the one particle state\nomcls{40}{$\ket{0}$}{Vacuum state. \symref{sym:vacuum}} 
\beq\mylabel{eqn:PsiDef}
\ket{i} \, \equiv \, \psi^\dagger_i \ket{0},
\eeq
where $\psi^\dagger_i (\psi_i) $\mylabel{sym:psipsid}  is a fermion creation(annihilation) operator that creates(destroys) a particle in the one particle state $\ket{i}$. These operators satisfy the well known fermion anticommutation relations\nomcls{50}{$\psi_i,\psi^\dagger_i$}{Annihilation and creation operators for the $i$-th orbital. \symref{sym:psipsid}} 
\beq\mylabel{eqn:AntiComm}
\psi_i \psi_j^\dagger + \psi_j^\dagger \psi_i = \{\psi_i,\psi_j^\dagger\} = \delta_{ij},
\eeq
 and
\beq\mylabel{eqn:AntiCommAnn}
\{\psi_i,\psi_j\} = 0.
\eeq
We collect these fermionic operators into convenient arrays
\beq \mylabel{eqn:BigPsiDefn}
\vPsi = \left[ \begin{array}{c}
\psi_1 \\
\vdots \\
\psi_\No 
\end{array}
\right], \;\;\;\;
\vPsi^\dagger = \left[ \begin{array}{ccc}
\psi^\dagger_1 & \ldots & \psi^\dagger_\No
\end{array}
\right].
\eeq
\nomcls{60}{$\vPsi$}{Row matrix of $\psi_i$. \symeqref{eqn:BigPsiDefn}} 
\nomcls{70}{$\vPsi^\dagger$}{Column matrix of $\psi_i^\dagger$. \symeqref{eqn:BigPsiDefn}} 

A different set of orthonormal one particle states $\ket{\tilde{i}}$ ($i = 1,\ldots,\No$) can, of course, be used as effectively. The states $\ket{\tilde{i}}$ are related to $\ket{i}$ via $\ket{\tilde{i}} = \sum_{j=1}^\No R_{ji} \ket{j}$, where $R_{ij}$ are the components of an $\No \times \No$ unitary matrix   $\mR$\mylabel{sym:mR}.\footnote{Throughout the paper a bold roman symbol (e.g.~$\mR$) is used to denote a matrix, and the light symbol with indices shown (e.g. $R_{ij}$) will denote its components.}  In terms of the operators $\vPsi$, we have \nomcls{80}{$\mR$}{Matrix for basis transformation of $\vPsi$. \symref{sym:mR}} 
\beq\mylabel{eqn:BasisChange}
\vtPsi^\dagger = \vPsi^\dagger \mR,
\eeq
along with other useful relations
\beq\mylabel{eqn:BasisChangeUseful}
\vtPsi = \mR^\dagger \vPsi,  \;\;\;\; \vtPsi^{\dagger T} = \mR^T \vPsi^{\dagger T},  \;\;\;\; \vtPsi^T = \vPsi^T \mR^*
\eeq
where $(~)^T$ denotes the transpose.
\nomcls{90}{$\vtPsi$}{Basis transformed $\vPsi$. \symeqref{eqn:BasisChange}} 

The system can have any number of particles $\NP$\mylabel{sym:NP} ranging from $0$ to $\No$. For each particle number $\NP$, the set of allowed states is spanned by $\combi{\No}{\NP}$ states obtained by creating $\NP$-particle states from the vacuum using distinct combinations of the operators $\psi_i^\dagger$. The vector space of $\NP$ particle states is denoted by $\calV_{\NP}$\mylabel{sym:calVNP}. The full Hilbert-Fock space of the system is given by\nomcls{a10}{$\NP$}{Number of particles. \symref{sym:NP}}\nomcls{a20}{$\calV_{\NP}$}{Vector space of $\NP$-particle states. \symref{sym:calVNP}}  

\beq\mylabel{eqn:HFspace}
\calV = \bigoplus_{N_P = 0}^{\No} \calV_{N_P},
\eeq
which is a vector space over complex numbers $\mathbb{C}$, providing a complete kinematical description of the system. An important property of this vector space, which we will exploit,  is that it is ``graded''\mycite{Spencer1959} in a natural fashion by the sectors of different particle number.
\nomcls{a30}{$\calV$}{Hilbert-Fock space of the system. \symeqref{eqn:HFspace}}
The dynamics of this fermionic system is determined by the Hamiltonian which contains up to $\mb$-body interactions where $0 \le \mb \le \No$, and is formally written as
\beq\mylabel{eqn:Hdef}
\vH = \sum_{K=0}^\mb (\vPsi^\dagger)^K \mnbdy{\mH}{K}(\vPsi)^K 
\eeq
where\nomcls{a40}{$\vH$}{The Hamiltonian operator. \symeqref{eqn:Hdef}}\nomcls{a50}{$\mnbdy{\mH}{K}$}{Matrix for $K$-body interaction Hamiltonian. \symeqref{eqn:Hdef}} 
\bea
 (\vPsi^\dagger)^K \mnbdy{\mH}{K}(\vPsi)^K &\equiv & H^{(K)}_{i_1i_2\ldots i_K;j_1j_2\dots j_K}\left(\psi_{i_1}\psi_{i_2}\ldots \psi_{i_K}\right)^\dagger\psi_{j_1}\psi_{j_2}\ldots\psi_{j_K} \notag \\
 &=& H^{(K)}_{i_1i_2\ldots i_K;j_1j_2\dots j_K} \psi^\dagger_{i_K}\psi^\dagger_{i_{K-1}}\ldots \psi^\dagger_{i_1} \psi_{j_1}\psi_{j_2}\ldots\psi_{j_K}\quad, \notag\\
\mylabel{eq:HKthorder}
\eea\nomcls{a60}{$H^{(K)}_{i_1i_2\ldots i_K;j_1j_2\dots j_K}$}{Matrix elements of $\mnbdy{\mH}{K}$. \symeqref{eq:HKthorder}}with repeated indices $i$s and $j$s summed from $1,\ldots,\No$. Note that here we depart slightly from the usual convention for the many-body Hamiltonian where $\dagger$ operation is done on the complete string of $\psi$s; this notation will eventually prove to be useful in the later manipulations. Note also that  $\left(\psi_{i_1}\psi_{i_2}\ldots \psi_{i_K}\right)^\dagger$ should be distinguished from the definition of the  $1 \times \No$ dimensional array $\Psi^\dagger$ in \eqn{eqn:BigPsiDefn}.   
$\mnbdy{\mH}{K}$ is the matrix which describes the $K$-body interactions, and its components $ H^{(K)}_{i_1i_2\ldots i_K;j_1j_2\dots j_K}$ have two properties. First, $H^{(K)}_{i_1i_2\ldots i_K;j_1j_2\dots j_K}$ is fully antisymmetric under permutations of the $i$ indices among themselves, and also under the permutations of the $j$ indices among themselves.  Expressed in an equation
\beq\mylabel{eqn:HKPerm}
H^{(K)}_{i_{X(1)}i_{X(2)}\ldots i_{X(K)};j_{Y(1)}j_{Y(2)}\ldots j_{Y(K)}} = \mbox{sgn}(X) \; \mbox{sgn}(Y) \, H^{(K)}_{i_1i_2\ldots i_K;j_1j_2\dots j_K}
\eeq
where $X$ and $Y$ are arbitrary permutations of $K$ objects, and $\mbox{sgn}$ denotes the sign of the permutation.
Second, the Hermitian character of the Hamiltonian is reflected in the relation
\beq\mylabel{eqn:HKHerm}
H^{(K)}_{j_1j_2\dots j_K;i_1i_2\ldots i_K} = \left(H^{(K)}_{i_1i_2\ldots i_K;j_1j_2\dots j_K}\right)^*.
\eeq
Each $\mnbdy{\mH}{K}$ is an element of an  $\combi{\No}{K}$-dimensional  vector space $\mnbdy{\calH}{K}$\mylabel{sym:calHK} over the {\em real} numbers $\mathbb{R}$. With no further restrictions other than \eqn{eqn:HKPerm} and \eqn{eqn:HKHerm},
in fact, this vector space is endowed with a structure of a {\em Lie algebra}. This is achieved by constructing an isomorphic vector space $\ci \mnbdy{\calH}{K}$ (multiplying every element of $\mnbdy{\calH}{K}$ by $\ci = \sqrt{-1}$\mylabel{sym:ci}). It is evident that for any two matrices $\ci\mnbdy{\mH_{a}}{K}$ and $\ci\mnbdy{\mH_{b}}{K}$ of $\ci \mnbdy{\calH}{K}$, the commutator $[\ci\mnbdy{\mH_{a}}{K}, \ci\mnbdy{\mH_b}{K}]$ is also a matrix in $\ci  \mnbdy{\calH}{K}$, and in fact,
\nomcls{a70}{$\mnbdy{\calH}{K}$}{Vector space of $\mb$-body interaction Hamiltonians. \symref{sym:calHK}}
\nomcls{a80}{$\ci$}{$\ci=\sqrt{-1}$. \symref{sym:ci}}
\beq\mylabel{eqn:HKIso} 
\ci  \mnbdy{\calH}{K} \sim \lieu\left(\combi{\No}{K} \right),
\eeq \nomcls{a90}{$\lieu()$}{Lie algebra of unitary group. \symeqref{eqn:HKIso}}
i.~e., $\ci  \mnbdy{\calH}{K}$  is isomorphic to the well known Lie algebra $\lieu\left(\combi{\No}{K} \right)$ which generates the Lie group $U\left(\combi{\No}{K} \right)$\mylabel{sym:Uliegrp} of $\combi{\No}{K} \times \combi{\No}{K}$ dimensional unitary matrices.\nomcls{b10}{$U\left( \right)$}{Unitary Lie group. \symref{sym:Uliegrp}} The Hamiltonian \eqn{eqn:Hdef} can be described by a $\mb+1$ tuple
\beq\mylabel{eqn:HSpace}
\mH = (\mnbdy{\mH}{0}, \mnbdy{\mH}{1}, \ldots,\mnbdy{\mH}{\mb}) \in \calH 
\eeq\nomcls{b20}{$\mH = (\mnbdy{\mH}{0},..,\mnbdy{\mH}{\mb})$}{$(\mb+1)$-tuple Hamiltonian ``vector''. \symeqref{eqn:HSpace}}
where $\calH$ is the real vector space
\beq
\calH = \bigtimes_{K=0}^\mb \mnbdy{\calH}{K}
\eeq
The problem of classification of a fermionic system of $\No$-orbitals with the Hilbert-Fock space \eqn{eqn:HFspace} and upto $\mb$-body interactions can now be stated precisely. How many ``distinct''  spaces $\calH$ are possible? The symmetries of the system will determine the distinct structures of these spaces, placing them in different classes.

\begin{figure}
\centerline{\includegraphics[width=\columnwidth]{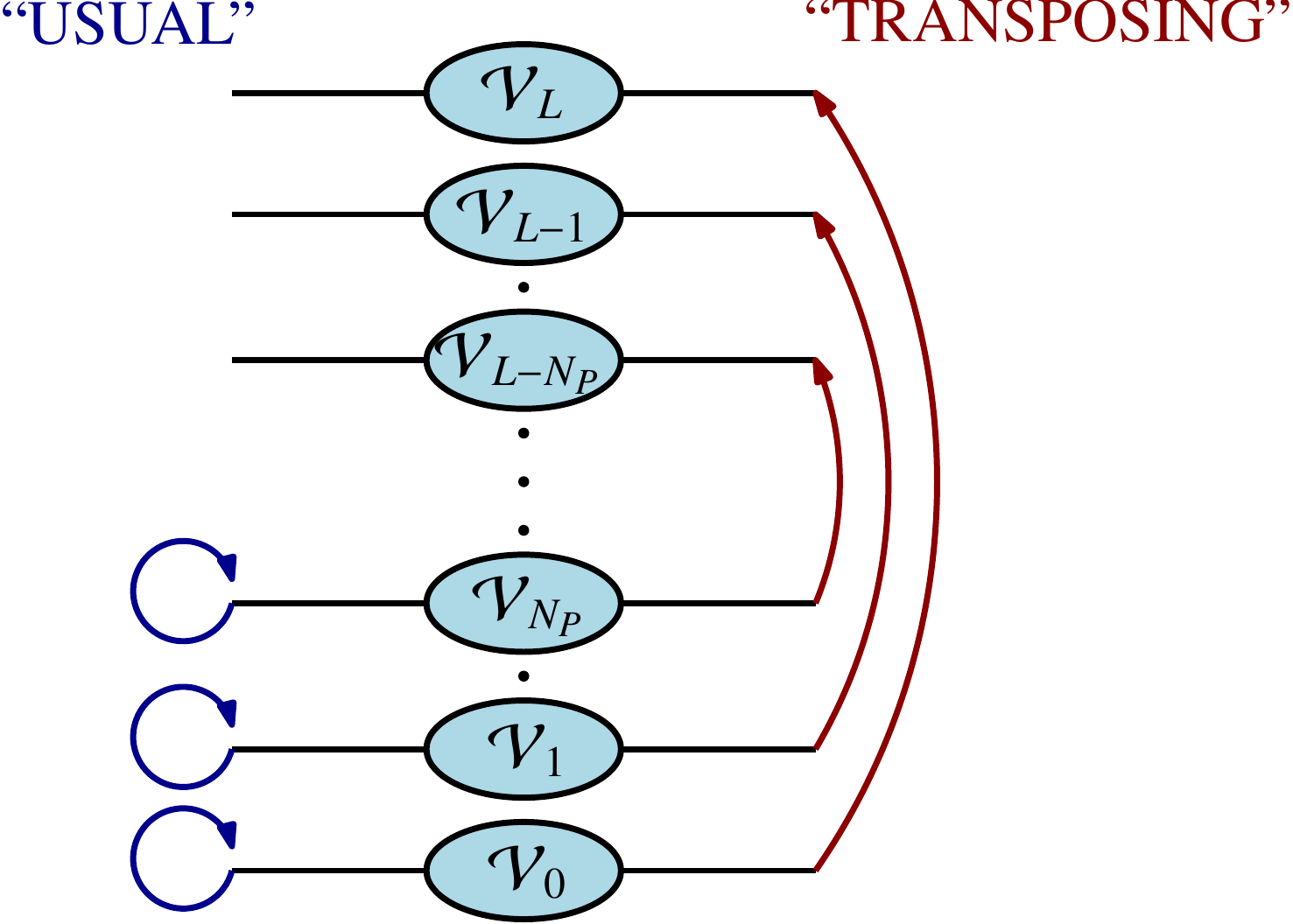}}
\caption{Illustrating ``usual'' and ``transposing'' symmetry operations. A usual symmetry operation maps $\calV_{\NP} \mapsto \calV_{\NP}$, while a transposing symmetry operation maps $\calV_{\NP} \mapsto \calV_{\No-\NP}$.
}
\mylabel{fig:symop}
\end{figure}

\section{Symmetries}\mylabel{sec:Symmetries}
\subsection{Symmetry Operations}\mylabel{sec:SymmOps}

\begin{figure}[!t]
\centerline{\includegraphics[width=\columnwidth]{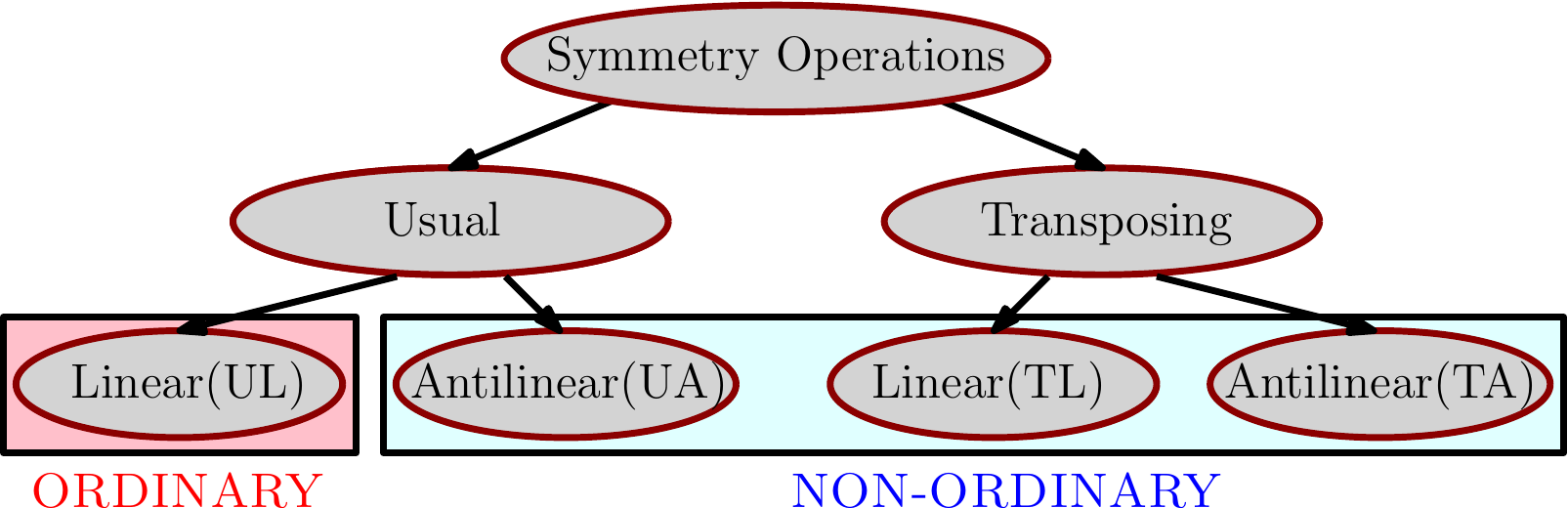}}
\caption{Four types of symmetry operations. }
\mylabel{fig:symflow}
\end{figure}

A symmetry operation as defined by Wigner\mycite{Wigner1959} is a linear {\em or} antilinear (bijective) operator $\vU$\mylabel{sym:vU} acting on the Hilbert space of the system that leaves the magnitude of the inner product invariant. Stated in the context of one particle physics of our $\No$-orbital fermionic system, an operator $\vU$ is a {symmetry operation} if\nomcls{b20}{$\vU$}{Symmetry operator. \symref{sym:vU}}  
\beq\mylabel{eqn:SymmOp}
|\braket{\zeta'}{\varphi'}| = |\braket{\zeta}{\varphi}|, \,\,\,\,\,\, \forall \,\,\,  \ket{\varphi}, \ket{\zeta} \in \calV_1
\eeq 
with $\ket{\varphi'} = \vU \ket{\varphi}$ and $\ket{\zeta'} = \vU \ket{\zeta}$. A symmetry operation must preserve the graded structure of the Hilbert space \eqn{eqn:HFspace}, i.~e., $N_P$ particle states must be mapped only to $N_P$ particle states in a bijective (invertible) fashion. We shall refer to such operations as {\em usual} symmetry operations; note that a usual symmetry operator can be either linear or antilinear. The above discussion can be summarized by the equation
\beq\mylabel{eqn:Uusl}
 \vUUsl(\calV_{N_P}) = \calV_{N_P}, \, \forall \NP
 \eeq 
and the action of $\vUUsl$ is implemented on the operators via
\beq\mylabel{eqn:UPsiUsl}
\vUUsl \vPsi^\dagger \vUUsli = \vPsi^\dagger \mUUsl 
\eeq\nomcls{b30}{$\vUUsl$}{Symmetry operator for usual(\textup{USL}) symmetries. \symeqref{eqn:Uusl}}  
\nomcls{b35}{$\mUUsl$}{Matrix associated with $\vUUsl$. \symeqref{eqn:UPsiUsl}} (in component form $\vUUsl \psi_i^\dagger \vUUsli = \sum_{j = 1}^\No \psi^\dagger_j  (U_{\mbox{\tiny USL}})_{j i}$ ) where $\mUUsl$ is an $\No\times \No$ unitary matrix that encodes the symmetry operation. If such a usual symmetry is linear then $\vUUsl (\ci \vI) \vUUsli = \ci \vI$\mylabel{sym:vI}, and $\vUUsl (\ci \vI) \vUUsli = -\ci \vI$ if it is antilinear where $\vI$ is the identity of operator on $\calV$, with the form \eqn{eqn:UPsiUsl} remaining the same for both linear and antilinear cases. 
\nomcls{b40}{$\vI$}{Identity operator on $\calV$. \symref{sym:vI}}  

The key realization of Altland and Zirnbauer \mycite{Altland1997} is that many particle fermionic systems admit a larger classes of symmetry operations (see \mycite{Zirnbauer2010} for a discussion). The crucial point is that a linear or antilinear operation that maps $\calV_{\NP}$ to $\calV_{\No - \NP}$ that preserves the magnitude of inner product and also {\em preserves the graded} nature of the Hilbert-Fock space \eqn{eqn:HFspace} is also a legitimate symmetry operation. We call such operations {\em transposing} symmetry operations (see \Fig{fig:symop} for a schematic illustration) and they satisfy
\beq\mylabel{eqn:Utrn}
 \vUTrn(\calV_{\NP}) = \calV_{\No-\NP},\, \forall \NP
\eeq \nomcls{b50}{$\vUTrn$}{Symmetry operator for transposing(\textup{TRN}) symmetries. \symeqref{eqn:Utrn}}with $\vUUsl (\ci \vI) \vUUsli = \pm \ci \vI$ for linear ($+$) and antilinear ($-$) operations. Note that such operations are enabled by the fact that $\dim{\calV_{\NP}} = \dim{\calV_{\No - \NP}}$. In fact, operations of the type \eqn{eqn:Uusl} and \eqn{eqn:Utrn} exhaust all possible (anti)linear automorphisms of $\calV$ that preserve its graded structure. From a physical perspective, the transposing symmetry operation maps an $\NP$- ``particle'' state to an $\NP$-``hole'' state. Holes are fermionic excitations obtained by starting from the ``fully filled state'' $\ket{\Omega} = \left(\psi_1 \psi_2 \ldots \psi_\No\right)^\dagger \ket{0}$\mylabel{sym:ketOmega}, and creating hole like excitations (such as $\psi_1 \ket{\Omega}$, a 1-hole state). It is therefore natural to implement the action of a transposing symmetry operation $\vUTrn$ via\nomcls{b60}{$\ket{\Omega}$}{Fully filled state. \symref{sym:ketOmega}}
\beq\mylabel{eqn:UPsiTrn}
\vUTrn \vPsi^\dagger \vUTrni = \vPsi^T \mUTrns 
\eeq\nomcls{b70}{$\mUTrn$}{Matrix associated with $\vUTrn$. \symeqref{eqn:UPsiTrn}}where $^*$ denotes complex conjugation, such that creation operators are mapped to annihilation operators, and the unitary matrix $\mUTrn$ encodes ``relabeling'' of states in this transposing symmetry operation (which can, again, be linear or antilinear).

The main conclusion of the above discussion is that there are four distinct types of symmetry operations as illustrated in \fig{fig:symflow}. They are usual linear (\UL), usual antilinear (\UA), transposing linear (\TL), and transposing antilinear (\TA). We find it useful to introduce additional terminology -- we call \UL~symmetry operations as {\em ordinary} symmetry operations, and all other types of symmetry operations (\UA, \TL~and \TA)\mylabel{sym:ULUATLTA} as {\em non-ordinary} operations.
\nomcls{b80}{\UL,\UA,\TL,\TA}{Usual Linear, Usual Antilinear, Transposing Linear and Transposing Antilinear type of symmetry operations. \symref{sym:ULUATLTA}}

\subsection{Symmetry conditions} \mylabel{sec:SymCond}

A symmetry operation is a symmetry if it leaves the Hamiltonian of the system invariant. For usual symmetries, this is effected by the condition
\beq\mylabel{eqn:UslSym}
\vUUsl \vH \vUUsli = \vH 
\eeq
while for the transposing symmetry operation the condition changes to
\beq\mylabel{eqn:TrnSym}
:\vUTrn \vH \vUTrni:\ = \vH
\eeq\nomcls{b90}{$:~:$}{Normal ordering operation. \symeqref{eqn:TrnSym}}where $:~:$ indicates that the expression $\vUTrn \vH \vUTrni$ has to be normal ordered (all creation operators to the left of annihilation operators) using the anticommutation relations \eqn{eqn:AntiComm}. 
Both of these types of symmetries induces a mapping of $\mH$ in \eqn{eqn:HSpace} to $\mHr$ via
\beq\mylabel{eqn:HMap}
\mH = (\mnbdy{\mH}{0}, \mnbdy{\mH}{1}, \ldots,\mnbdy{\mH}{\mb}) \mapsto \mHr = (\mnbdy{\mHr}{0}, \mnbdy{\mHr}{1}, \ldots,\mnbdy{\mHr}{\mb}).
\eeq\nomcls{c10}{$\mHr$}{Symmetry transformed $\mH$. \symeqref{eqn:HMap}} To obtain $\mnbdy{\mHr}{K}$ for any $K$, we introduce an intermediate quantity $\mnbdy{\mHc}{K}$\mylabel{sym:mHc} which is determined by the type of symmetry operation:
\nomcls{c20}{$\mnbdy{\mHc}{K}$}{Intermediate quantity generated when $\mH$ maps to $\mHr$ under symmetry operation. \symref{sym:mHc}}
\begin{widetext}
\begin{equation}
\nbdy{\breve{H}}{K}{i_1,\ldots,i_K;j_1,\ldots,j_K} = 
\begin{cases}
U_{i_1i_1'} \ldots U_{i_Ki'_K} \nbdy{H}{K}{i'_1,\ldots,i'_K;j'_1,\ldots,j'_K} U^\dagger_{j_1'j_1} \ldots U^\dagger_{j'_K j_K}, &\mbox{\UL} \\[2ex]
U_{i_1i_1'} \ldots U_{i_K i'_K} \left(\nbdy{H}{K}{i'_1,\ldots,i'_K;j'_1,\ldots,j'_K} \right)^* U^\dagger_{j_1'j_1} \ldots U^\dagger_{j'_K j_K}, & \mbox{\UA} \\[2ex]
U^*_{i_1i_1'} \ldots U^*_{i_Ki'_K} \nbdy{H}{K}{i'_1,\ldots,i'_K;j'_1,\ldots,j'_K} U^T_{j_1'j_1} \ldots U^T_{j'_Kj_K}, & \mbox{\TL} \quad .\\[2ex]
U^*_{i_1i_1'} \ldots U^*_{i_Ki'_K} \left(\nbdy{H}{K}{i'_1,\ldots,i'_K;j'_1,\ldots,j'_K} \right)^* U^T_{j_1'j_1} \ldots U^T_{j'_K j_K}, & \mbox{\TA} 
\end{cases}\mylabel{eqn:Htmp}
\end{equation}
\end{widetext}
Here all the primed indices are summed from $1$ to $\No$, and $\mU$s are the unitary matrices that encode the symmetry operations as defined in \eqn{eqn:UPsiUsl} and \eqn{eqn:UPsiTrn}. Note that antilinear symmetry operations lead to a complex conjugation of the matrix elements. The transformation of the Hamiltonian \eqn{eqn:HMap} can now be specified completely. For usual symmetries (\eqn{eqn:UPsiUsl} and \eqn{eqn:UslSym}), both linear and antilinear , we have
\beq\mylabel{eqn:HtUsl}
\mnbdy{\mHr}{K} = \mnbdy{\mHc}{K}.
\eeq
For transposing operation, the result is a bit more involved on account of the normal ordering operation of \eqn{eqn:TrnSym}. We find
\beq\mylabel{eqn:HtTrn}
\mnbdy{\mHr}{K} = \sum_{R=K}^\mb A_{R,K} \left[\tr_{R-K} \mnbdy{\mHc}{R}\right]^T,
\eeq
The trace operation $\tr_{~}$ is defined as
\beq\mylabel{eqn:trPdef}
\tr_{P}\mnbdy{\mHc}{K} \equiv {\breve{H}}^{(K)}_{i_1i_2\ldots i_{(K-P)}k_1k_2\ldots k_P;j_1j_2\ldots j_{(K-P)}k_1k_2\ldots k_P}
\eeq\nomcls{c30}{$\tr_{P}$}{Tracing out operation of $P$ indices. \symdefref{eqn:trPdef}}accomplishing the tracing out (repeated $k$ indices are summed) of $P$ indices in $\mnbdy{\mHc}{K}$, resulting in a $\combi{\No}{K-P} \times \combi{\No}{K-P}$  matrix. The constants $A_{K,R}$ in \eqn{eqn:HtTrn} are computed to be
\beq\mylabel{eqn:ARKdef}
A_{R,K} = 
\begin{cases}
\displaystyle{(-1)^{K}  \frac{1}{(R-K)!} \left(\frac{R!}{K!} \right)^2}, &  0 \le K \le R \\
0, & K > R
\end{cases}
.\nomcls{c40}{$A_{R,K}$}{Combinatorial factor generated when tracing over $R-K$ indices of $\mnbdy{\mHc}{R}$. \symdefref{eqn:ARKdef}}
\eeq
Note that the RHS of \eqn{eqn:HtTrn} involves a matrix transpose, and this is a characteristic of the transposing symmetry operations justifying our terminology. Thus, eqns.~\prn{eqn:HMap}, \prn{eqn:Htmp}, \prn{eqn:HtUsl} and \prn{eqn:HtTrn} completely determine the mapping of $\mH$ to $\mHr$. A symmetry operation (of any type) describes a symmetry if and only if
\beq\mylabel{eqn:SymCond}
\mHr = \mH
\eeq
which is a concise expression of \eqn{eqn:UslSym} and \eqn{eqn:TrnSym}.

\subsection{Symmetry Group}

The set of all symmetry operations $\vU$ of the system forms a group  $G_\calV$\mylabel{sym:GcalV}. This group is a disjoint union of four types of symmetry operations (see \fig{fig:symflow}),
\beq\mylabel{eqn:GcalVsubsets}
G_\calV = {G_\calV^{\ULm}} \cup { G_\calV^{\UAm}} \cup { G_\calV^{\TLm}} \cup { G_\calV^{\TAm}}.
\eeq \nomcls{c50}{$G_\calV$}{Group of all symmetry operations of $\calV$. \symref{sym:GcalV}}\nomcls{c60}{${G_\calV^{\ULm}}, { G_\calV^{\UAm}}, { G_\calV^{\TLm}}, { G_\calV^{\TAm}}$}{Set of all \UL~,\UA~,\TL~,\TA~type symmetry operations. \symeqref{eqn:GcalVsubsets}}The set of ordinary symmetries ${G_\calV^{\ULm}}$ form a normal subgroup of $G_{\calV}$. The non-ordinary symmetries have the following interesting properties. First, for any non-ordinary $\vU$ we have
\beq\mylabel{eqn:NO2}
\vU \vU \in G_\calV^{\ULm}, \,\,\,\,\,\,\, \forall \vU \in { G_\calV^{\UAm}} \cup { G_\calV^{\TLm}} \cup { G_\calV^{\TAm}}.
\eeq
which can be paraphrased ``the square of a non-ordinary symmetry operation is an ordinary symmetry operation''.
Second, the product of two distinct types of non-ordinary symmetry operations is the third type -- for example
\beq
\vU_{\UAm} \vU_{\TLm} = \vU_{\TAm};
\eeq
this is summarized in Fig.~\ref{fig:TypeMult}. The content of that figure can be restated as: the factor group $G_{\calV}/ G^{\textup{UL}}_\calV $ is the Klein 4-group $\Klein$\mylabel{sym:Klein}.\nomcls{c70}{$\Klein$}{Klein 4-group. \symref{sym:Klein} \symdefref{eqn:V4}}

\renewcommand{\arraystretch}{1.5}
\begin{figure}
\begin{tabular}{||c||c|c|c|c||}
\hline
\hline
$G_1 \downarrow \, | \, G_2 \rightarrow$  &  $\GUL$ & $\GUA$ & $\GTL$ & $\GTA$ \\
\hline
\hline
$\GUL$ & $\GUL$ & $\GUA$ & $\GTL$ & $\GTA$ \\
\hline
$\GUA$ & $\GUA$ & $\GUL$ & $\GTA$ & $\GTL$ \\
\hline
$\GTL$ & $\GTL$ & $\GTA$ & $\GUL$ & $\GUA$ \\
\hline
$\GTA$ & $\GTA$ & $\GTL$ & $\GUA$ & $\GUL$ \\
\hline
\hline
\end{tabular}
\caption{Multiplication table of types of symmetries.}\mylabel{fig:TypeMult}
\end{figure}

The set of all the symmetry operations that satisfy either \eqn{eqn:UslSym} or \eqn{eqn:TrnSym} makes up the symmetry group $G$ of the system and is a subgroup of $G_{\calV}$.
 
\section{\Grot~Fermionic Systems}\mylabel{sec:Grot}
We will now focus attention on a special class of $\No$-orbital fermionic systems. Our systems of interest do not possess any nontrivial ordinary symmetries -- we dub such systems as ``{\em \grot} fermionic systems (\GFS)'' to highlight this property. This would suggest that the only ordinary symmetry operation allowed is the trivial identity operation $\vI$. However, since we are working with the Hilbert-Fock vector space (not a projective or ``ray'' space of Wigner, see \mycite{Parthasarathy1969}), the operator 
\beq\mylabel{eqn:Itheta}
\vI_\theta = e^{\ci \theta \vN}
\eeq\nomcls{c80}{$\vI_\theta$}{$e^{\ci \theta \vN}$. \symdefref{eqn:Itheta}}with
\beq\mylabel{eqn:Number}
\vN = \sum_{i=1}^\No \psi_i^\dagger \psi_i
\eeq\nomcls{c90}{$\vN$}{Number operator. \symdefref{eqn:Number}}is always an allowed symmetry operation, and indeed {\em any} Hamiltonian of the form \eqn{eqn:Hdef} will be invariant under this operation.\footnote{This consideration can be easily generalized to superconducting Bogoliubov-de Gennes (BdG) Hamiltonians which are quadratic in fermion operators by expanding the operators \eqn{eqn:BigPsiDefn} to the Nambu representation, and treating the problem as a quadratic fermion problem. In this case, $\vI_{\theta}$ is restricted to $\theta=0, \piup$.} We will later discuss its relationship  to projective representations (see sec.~\ref{sec:GrpCohom}). Thus, for the \GFS, $\GUL = \vI_\theta, \forall \theta \in[0,2\piup]=U(1)$, i.~e., the only ordinary symmetries are the trivial ones. 

We will now demonstrate an important property of a \GFS. A GFS can possess {\em at most} one each of \UA, \TL~ and \TA~ symmetries. In other words, {\em non-ordinary symmetries of a \GFS~are solitary}. To prepare to prove this statement, we adopt some useful  notation. We will denote \UA~symmetry operations by $\TR$, \TL~by $\CC$, and \TA~by $\SL$ --\mylabel{sym:TRCCSL} this choice anticipates later discussion.  Suppose, now, that we have two distinct \UA~symmetries of the \GFS, say $\TR_1$ and $\TR_2$, then $\TR_1 \TR_2$ is also a symmetry of our system. But from figure.~\ref{fig:TypeMult}, we know that $\TR_1 \TR_2$ is an \UL~type symmetry. By definition, in a \GFS~ any unitary symmetry has to be a trivial one, $\vI_\theta$ for some $\theta$. Thus $\TR_1 \TR_2 =\vI_\theta$ or $\TR_2 = \vI_{-\theta} \TR_1^{-1}$ and hence $\TR_2$ is not a distinct \UA~symmetry -- it is simply a product of a trivial symmetry with $\TR_1^{-1}$. The same argument works for $\CC$ and $\SL$ symmetries and the solitarity of non-ordinary symmetries is proved (see also \mycite{Ludwig2015}). In fact, we can conclude many more interesting facts about symmetries of a \GFS.\nomcls{d10}{$\TR$,$\CC$,$\SL$}{Time-reversal, charge-conjugation and sublattice operators. \symref{sym:TRCCSL}}

Consider the symmetry operator $\TR^2$. From the previous paragraph we know $\TR^2 = \vI_\theta$ for some $\theta$. We can determine $\theta$ from $\TR^{-1} \TR^2 = \TR = \TR^2 \TR^{-1}$ which implies $\vI_{-\theta} \TR^{-1} = \TR = \vI_{\theta} \TR^{-1}$. Applying this relation on $\calV_1$ immediately forces $e^{\ci 2\theta} = 1$ (due to the antilinearity of $\TR$) or $e^{\ci\theta} = \pm 1$, and thus we immediately see that the action of $\TR^2$ on ${\cal V}$ is
\beq\mylabel{eqn:T2precise}
\TR^2 = (\pm 1)^{\vN} \vI.
\eeq
 We will mostly write \eqn{eqn:T2precise} simply as
\beq\mylabel{eqn:T2}
\TR^2 = \pm \vI.
\eeq

We turn now to $\CC^2$ which should also be equal to $\vI_\theta$. We get, again, $ \CC^{-1}\vI_\theta = \vI_\theta \CC^{-1}$. Applying this last relation on $\calV_{\NP}$, we get
\beq\mylabel{eqn:CCond}
e^{\ci N_P \theta} = e^{\ci (\No - N_P) \theta}, \,\,\, \forall \, \, N_P = 0, \ldots , \No.
\eeq
resulting in $e^{\ci 2 \theta} = 1$, and
\beq\mylabel{eqn:C2}
\CC^2 = (\pm 1)^{\vN} \vI.
\eeq
Note also that \eqn{eqn:CCond} implies that $\No$ must be even when $\theta = \piup$, i.~e., a \TL~type of symmetry with the $(-)$ signature can only be implemented in a \GFS~with even number of orbitals.

Finally, we discuss $\SL^2$. Noting that $\SL$ is an \TA~ symmetry, we get 
\beq\mylabel{eqn:SCond}
e^{-\ci N_P \theta} = e^{\ci (\No - N_P) \theta}, \,\,\, \forall \, \, N_P = 0, \ldots , \No\quad,
\eeq
and thus $e^{\ci \No \theta} = 1$, implies $\theta= \frac{2\piup \ell}{\No}$ where $\ell$ is one of $0,1,\ldots, \No-1$. 
We will later show that $\theta$ can always be chosen to be zero, leading to 
\beq\mylabel{eqn:S2}
\SL^2 = \vI.
\eeq
We conclude this discussion on the properties of \GFS~symmetries by noting that if a \GFS~has a $\TR$ and $\CC$ type symmetry, then $\SL$ is equal to $\TR \CC$.
If a \GFS~has a $\TR$ symmetry, we say that it possesses {\em time reversal symmetry}, $\CC$ implies the presence of {\em charge conjugation symmetry}, and $\SL$ endows a {\em sublattice symmetry} on the \GFS.

\section{The Tenfold Way}\mylabel{sec:TenFold}

In this section, we show how the ten symmetry classes of fermions arise and obtain the canonical representations of the symmetries in these classes.

\subsection{Symmetry Classes}\mylabel{sec:SymClasses}
Based on the discussion of the previous section, we see that a \GFS~has to be of one of three types. 
\begin{description}
\item[{\bf Type 0}]{Possesses no non-ordinary symmetries.}
\item[{\bf Type 1}]{Possesses one non-ordinary symmetry.}
\item[{\bf Type 3}]{Possesses all three non-ordinary symmetries.}
\end{description}
The resulting symmetry classes and the class hierarchy is shown in  \fig{fig:tree}.
\begin{figure}

\centerline{\includegraphics[width=\columnwidth]{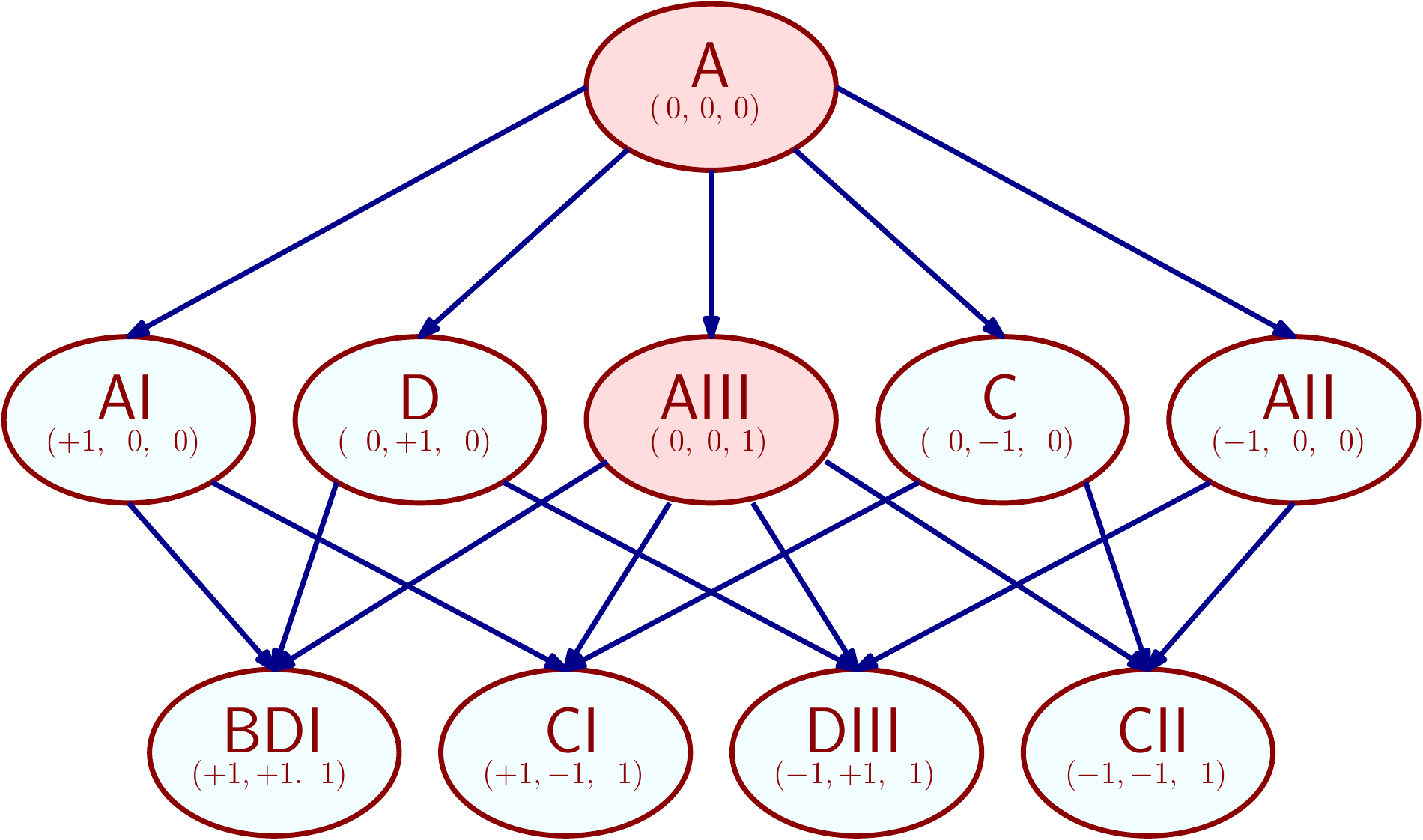}}
\caption{Ten symmetry classes, and their interrelationships. The two classes shaded in red are the complex classes, while all others are real classes. The symmetry signature is denoted by a triple $(\tT,\tC,\tS)$ as discussed in the text.}
\mylabel{fig:tree}
\end{figure}

Whenever time reversal is present it be realized as $\TR^2 = \pm \vI$, and we denote this by $\tT= \pm 1$, similarly $\CC^2= \pm \vI$ is denoted by $\tC= \pm 1$ and presence of $\SL$ is shown by $\tS=1$. Absence of these symmetries is denoted by $\tT=0$, $
\tC=0$, or $\tS=0$ as the case may be. The ``symmetry signature'' of any class is denoted by a triple $(\tT,\tC,\tS)$\mylabel{sym:TCS} (see \fig{fig:tree}). It is now immediately clear that there is only one class of type 0 \GFS, called \Ac~with symmetry signature $(0,0,0)$. There are five classes of type 1 with a single non-ordinary symmetry: \AI $(+1,0,0)$, \AII $(-1,0,0)$, \Dc $(0,+1,0)$, \C $(0,-1,0)$ and \AIII $(0, 0,1)$. The type 3 systems come in four classes: \BDI $(+1,+1,1)$,  \CI $(+1,-1,1)$,   \DIII $(-1,+1,1)$, and \CII $(-1,-1,1)$.\nomcls{d20}{$\tT$,$\tC$,$\tS$}{are equal to $\pm 1$, $0$ or $1$, describing signature of the $\TR$,$\CC$,$\SL$ symmetries. \symref{sym:TCS}}

\subsection{Canonical Representation of Symmetries}\mylabel{sec:Canonical}
An important step towards determining the structure of Hamiltonians 
(\eqn{eqn:HSpace}) in each of these ten symmetry classes is the determination of the canonical representation of the symmetries of each class. This question is addressed in this section.

Each of the non-ordinary symmetry operations, $\TR$, $\CC$ and $\SL$ is represented by a $\No \times \No$ unitary matrix. Time reversal $\TR$ is represented as
\beq\mylabel{eqn:UTdef}
\TR \vPsi^\dagger \TR^{-1} = \vPsi^\dagger \mUT
\eeq\nomcls{d30}{$\mUT$}{Unitary matrices associated with $\TR$ operator. \symdefref{eqn:UTdef}}
using \eqn{eqn:UPsiUsl}. Charge conjugation and sublattice, being transposing operations, are represented using \eqn{eqn:UPsiTrn} respectively as
\beq\mylabel{eqn:UCdef}
\CC \vPsi^\dagger \CC^{-1}  = \vPsi^T \mUCs
\eeq\nomcls{d40}{$\mUC$}{Unitary matrices associated with $\CC$ operator. \symdefref{eqn:UCdef}}and
\beq\mylabel{eqn:USdef}
\SL  \vPsi^\dagger \SL^{-1}  = \vPsi^T \mUSs.
\eeq\nomcls{d50}{$\mUS$}{Unitary matrices associated with $\SL$ operator. \symdefref{eqn:USdef}}

\subsubsection{Type 0}\mylabel{sec:Type0}

\noindent
{\bf Class \Ac: } Class \Ac~has no non-ordinary symmetries and hence nothing to represent. There are, therefore, no restrictions on the $\No$-orbital systems -- any $\No$ orbital system can be in class \Ac.

\subsubsection{Type 1}\mylabel{sec:Type1}

\noindent
{\bf Classes \AI~and \AII: } Time reversal symmetry is the sole non-ordinary symmetry present in these classes with $\tT = \pm 1$. The condition \eqn{eqn:T2} gives (using the antilinearity of $\TR$)
\beq\mylabel{eqn:T2Psi}
\TR^2 \vPsi^\dagger \TR^{-2} = \vPsi^\dagger \mUT \mUTs = \tT\ \vPsi^\dagger
\eeq
leading to 
\beq\mylabel{eqn:UT}
\mUT \mUTs =  \tT \, \mOne
\eeq\nomcls{d60}{$\mOne$}{$\No \times \No$ unit matrix. \symeqref{eqn:UT}}where $\mOne$ is an $\No \times \No$ unit matrix. An immediate consequence of this is that $|\det{\mUT}|^2 = \tT^\No$, leading us to the conclusion that $\tT=+1$ can be realized in any $\No$-orbital \GFS, while $\tT=-1$ requires $\No$ to be an even number.

To construct a canonical $\mUT$ that satisfies \eqn{eqn:UT}, we consider a change of basis of the \GFS~ as described by the matrix $\mR$ defined in \eqn{eqn:BasisChange}. The unitary $\mtUT$ representing $\TR$ in this new basis can be obtained from
\beq\mylabel{eqn:UTtrans}
\mR \mtUT \mR^T = \mUT\ .
\eeq\nomcls{d70}{$\mtUT$}{Basis transformed $\mUT$. \symdefref{eqn:UT}}It is known that any unitary that satisfies \eqn{eqn:UT} with $\tT=+1$ can be written as $\mUT = \mA \mA^T$ where $\mA$ is a unitary matrix. This result from  matrix theory, usually called the Takagi decomposition (see appendix D of \mycite{Dreiner2010}), allows us to conclude that we can always choose ($\mR = \mA$) a basis\footnote{This new basis is {\em not} unique. In fact, existence of such a basis implies that any other basis related by a real orthogonal matrix will also be an equally valid one.}\protect{\mylabel{ftn:NonUnique}}  where $\mtUT = \mOne$ for $\tT = +1$. The outcome of this discussion is that $\tT= + 1$ admits a canonical representation  $\mUT = \mOne$.

In the case of $\tT=-1$, Takagi decomposition provides that any $\mUT$ satisfying \eqn{eqn:UT} can decomposed as $\mA \mJ \mA^T = \mUT$ where 
\beq\mylabel{eqn:J}
\mJ = 
\left(
\begin{array}{cc}
\mZero_{\Mo\Mo} & \mOne_{\Mo\Mo} \\
-\mOne_{\Mo\Mo} & \mZero_{\Mo\Mo}
\end{array}
\right),\ \ \No=2 \Mo,\;\;\;\;\; \mJ \mJ^* = - \mOne,
\eeq\nomcls{d80}{$\mJ$}{Matrix of dimension $\No$ of the form $\left(
\begin{array}{cc}
\mZero_{\Mo\Mo} & \mOne_{\Mo\Mo} \\
-\mOne_{\Mo\Mo} & \mZero_{\Mo\Mo}
\end{array}
\right)$. \symdefref{eqn:J}}and $\mA$ is unitary.
This is consistent with the fact that $\No$ must be necessarily even for $\tT=-1$ as concluded above. The subscripts on the matrices denote their sizes. Taken together with \eqn{eqn:UTtrans} allows us to conclude that $\tT=-1$ case is canonically represented by $\mUT = \mJ$.

We can also obtain a ``one particle'' or ``first quantized'' representation of the time reversal operator. To see this consider a \GFS~with only a one-body Hamiltonian, form \eqn{eqn:HtUsl} and the symmetry condition \eqn{eqn:SymCond}, we obtain
$\mUT \conj{\mnbdy{\mH}{1}} \mUTd = \mnbdy{\mH}{1}$ which can be written as 
\beq
\mT \mnbdy{\mH}{1} \mT^{-1} = \mnbdy{\mH}{1}
\eeq
where $\mT$ is 
\beq\mylabel{eqn:Tmat}
\mT = \mUT \mK
\eeq
where $\mK$ is the complex conjugation ``matrix'', recovering a well known result (see e.g., \cite{Ludwig2015}).\nomcls{d90}{$\mK$}{Complex conjugation operation. \symdefref{eqn:Tmat}}

\noindent
{\bf Classes \Dc~and \C: } These classes respectively with symmetry signatures $(0,+1,0)$ and $(0,-1,0)$ possess the sole non-ordinary symmetry of charge conjugation. As noted just after \eqn{eqn:C2} the latter symmetry can be realized only in \GFS s with even number of orbitals.

As in the previous para, considering the action of $\CC^2$ on the fermion operators, owing to \eqn{eqn:C2}, gives
\beq
\CC^2 \vPsi^\dagger \CC^{-2} = \CC \vPsi^T \mUCs \CC^{-1} = \vPsi^\dagger \mUC \mUCs \,
\eeq
resulting in
\beq\mylabel{eqn:UC}
\mUC \mUCs=  \tC \, \mOne.
\eeq
Note that this relation, despite $\CC$ being a {\em linear} operation, looks very similar to \eqn{eqn:UT} of an usual {\em antilinear} operator. 

Under a change of basis, it can be shown that $\mUC$ transforms in exactly the same manner as $\mUT$, i.~e., via \eqn{eqn:UTtrans}.
Precisely the same considerations using Takagi decomposition of the previous para then allow us to conclude that there is a basis for the \GFS~where $\mUC$ can be represented canonically as $\mUC = \mOne$  for $\tC = +1$ and $\mUC = \mJ$ for $\tC = -1$. Finally, restricting again to a \GFS~described solely by one particle interactions, provides the first quantized version of the $\CC$ operator. Indeed, \eqn{eqn:HtUsl} and \eqn{eqn:SymCond} provide that $-\mUC \transp{\mnbdy{\mH}{1}} \mUCd = \mnbdy{\mH}{1}$ rewritten as
\beq
-\mC \mnbdy{\mH}{1} \mC^{-1} = \mnbdy{\mH}{1}.
\eeq
This is consistent with the first quantized form
\beq\mylabel{eqn:Cmat}
\mC = \mUC \mK.
\eeq
It is easily seen that this feature is generic -- any transposing linear symmetry operation has an {\em antilinear} first quantized representation.

\noindent
{\bf Class \AIII : } The sole non-ordinary symmetry in this class with the symmetry signature $(0,0,1)$ is the sublattice symmetry. 

Investigating the action of $\SL^2$, noting that $\SL$ is an  antilinear operator, we obtain from \eqn{eqn:SCond}
\beq
\SL^2 \vPsi^\dagger \SL^{-2} = \SL \vPsi^T \mUSs \SL^{-1} = \vPsi^\dagger \mUS \mUS  = e^{\ci \frac{2 \piup \ell}{\No}} \vPsi^\dagger
\eeq
implying
\beq\mylabel{eqn:UStmp}
\mUS \mUS = e^{\ci \frac{2 \piup \ell  }{\No}} \mOne.
\eeq
It is evident that it  can be redefined ($e^{-\ci \frac{\piup\ell }{\No}} \mUS \mapsto \mUS$) as 
\beq\mylabel{eq:US}
\mUS \mUS = \mOne.
\eeq
Since $\mUS$ is unitary, we obtain that $\mUSd = \mUS$, and thus, $\mUS$ is Hermitian. 
 This condition implies that all eigenvalues of $\mUS$ are real and of unit magnitude. Quite interestingly, under a change of basis, $\mUS$ transforms as
\beq\mylabel{eqn:UStrans}
\mR \mtUS \mR^\dagger = \mUS
\eeq
This implies that there is a basis in which $\mUS$ has the following canonical form
\beq\mylabel{eqn:US}
\mUS = \mOne_{p,q}
\eeq
where 
\beq\mylabel{eqn:Ipq}
\mOne_{p,q} = \left( 
\begin{array}{cc}
\mOne_{pp} & \mZero_{pq} \\
\mZero_{q p} & - \mOne_{qq}
\end{array} 
\right)
\eeq
where $p+q = \No$. This development makes the meaning of the sublattice symmetry clear. The orbitals of the system are divided into two groups -- ``sublattices $A$ and $B$''. The sublattice symmetry operation, a transposing antilinear operation, maps the ``particle states'' in the $A$ orbitals to ``hole states'' on $A$ orbitals, while $B$-particle states are mapped to {\em negative} $B$-hole states. In the first quantized language, the transposing antilinear operator $\SL$, is, therefore, represented by a {\em linear} matrix, i.~e., 
\beq
\mS = \mUS
\eeq
and in a noninteracting system the symmetry is realized when $-\mS\mnbdy{\mH}{1} \mS^{-1} = \mnbdy{\mH}{1}$.

\subsubsection{Type 3}\mylabel{sec:Type3}

The remainder of the four classes are of type 3, i.~e., they possess all three non-ordinary symmetries. As noted at the end of section~\ref{sec:Grot}, $\SL = \TR \CC$ in these classes implying that
 \beq\mylabel{eqn:UTCS}
\mUS = \mUT \mUCs\ .
\eeq
Our strategy in analyzing these classes is to choose a basis where $\mUS = \mOne_{p,q}$, and to determine the structure of $\mUT$ and $\mUC$. In this basis, we can write 
\beq\mylabel{eqn:UTSbasis}
\begin{split}
\mUT &  = \left( 
\begin{array}{cc}
\mV_{pp} & \mV_{pq} \\
\tT \mV_{pq}^T & \mV_{qq}
\end{array}
\right),\;\;\;\;\mV_{pp}^T = \tT \mV_{pp}, \mV_{qq}^T = \tT \mV_{qq},\;\;\; \\
&\mbox{for} \;\;\;\;\; \tT = \pm 1,
\end{split}
\eeq
where $\mV$s are complex matrices of the dimension indicated by the suffix. This from  automatically satisfies \eqn{eqn:UT}. Once, we fix $\mUT$, $\mUC$ is fixed for every class in type ${\bf 3}$ via \eqn{eqn:UTCS}, as
\beq\mylabel{eqn:UCSbasis}
\begin{split}
\mUC = \mUTt\mUSs &
 = \left( 
\begin{array}{cc}
\tT \mV_{pp} & -\tT \mV_{pq} \\
 \mV_{pq}^T & -\tT \mV_{qq}
\end{array}
\right) 
;\ \tT=\pm 1\ .
\end{split} 
\eeq
The conditions \eqn{eqn:UT} and  \eqn{eqn:UC} constrain  $\mV_{pp}, \mV_{pq}, \mV_{qq}$ of \eqn{eqn:UTSbasis} very strongly. Two possible cases arise.

\begin{itemize}
\item[\protect{$\tT=\tC$}:]{ This case results in the condition $\mUT \mUTs = \mUC \mUCs$ which provides the following constraints
\beq\mylabel{eqn:TeqC}
\begin{split}
\mV_{pp} \mV^*_{pp} = \tT \mOne_{pp}, \;\;\; \mV_{qq} \mV_{qq}^* &= \tT \mOne_{qq}, \\
\mV_{pq} \mV^\dagger_{pq} = \bzero_{p  p}, \;\;\; \mV^T_{pq} \mV^*_{pq} &= \bzero_{q  q}, \\ 
\mV_{pp} \mV^*_{pq} = \bzero_{p  q}, \;\;\; \mV^T_{pq} \mV^*_{pp} &= \bzero_{q  p}, \\
\mV_{pq} \mV_{qq}^* = \bzero_{p  q}, \;\;\; \mV_{qq} \mV_{pq}^\dagger & = \bzero_{q  p},\;\;\;\; \mbox{for~} \tT=\tC.
 \end{split}
\eeq 
}
\item[$\tT=-\tC$:]{ Note that since either $\tT$ or $\tC$ is $-1$, $\No$ is already even $\No=2\Mo$\mylabel{sym:Mo}. The condition $\mUT \mUTs = -\mUC \mUCs$ obtains the constraints\nomcls{d75}{$\Mo$}{$\Mo=\No/2$, when $\No$ is even. \symeqref{eqn:J}}  
\beq
\mylabel{eqn:TneC}
\begin{split}
\mV_{pp} \mV^*_{pp} =  \bzero_{pp}, \;\;\; \mV_{qq} \mV_{qq}^* &=  \bzero_{qq}, \\
\mV_{pq} \mV^\dagger_{pq} =\tT \mOne_{p  p}, \;\;\; \mV^T_{pq} \mV^*_{pq} &=\tT \mOne_{q  q}, \\ 
\mV_{pp} \mV^*_{pq} = \bzero_{p  q}, \;\;\; \mV^T_{pq} \mV^*_{pp} &= \bzero_{q  p}, \\
\mV_{pq} \mV_{qq}^* = \bzero_{p  q}, \;\;\; \mV_{qq} \mV_{pq}^\dagger & = \bzero_{q  p},\;\;\;\; \mbox{for~} \tT=-\tC.
 \end{split}
\eeq 
}\nomcls{e30}{$p,q$}{Labels for indices of sub-blocks of $\mU$ of various symmetry operators. Also indicates the label for states making up the blocks, together they form $\No$ states. \symeqref{eqn:TneC}} 
The second line of \eqn{eqn:TneC}, forces $p=q=\Mo$ and $\mV_{pq}$ to be a $\Mo \times \Mo$ unitary matrix which we will call $\mV_{\Mo\Mo}$. The other conditions provide $\mV_{pp} = \bzero_{p  p}$ and $\mV_{qq} = \bzero_{q  q}$.

\end{itemize}

\noindent
{\bf Class \BDI :} The class has a symmetry signature $(1,1,1)$. Since $\tT=\tC=+1$, we see from \eqn{eqn:TeqC}that $\mV_{pq} = \bzero_{p  q}$, along with $\mV_{pp} \mV^*_{pp} = + \mOne_{pp}$ and $\mV_{qq} \mV^*_{qq} = + \mOne_{qq}$, providing
\beq\mylabel{eqn:TpCpGen}
\mUT   = \left( 
\begin{array}{cc}
\mV_{pp} & \bzero_{p  q} \\
\bzero_{q  p} & \mV_{qq}
\end{array}
\right), \;\;\;\;
\mUC   = \left( 
\begin{array}{cc}
\mV_{pp} & \bzero_{p  q} \\
\bzero_{q  p} & -\mV_{qq} 
\end{array}
\right)
\eeq
Apart from $p+q=\No$, there are no additional constraints on $p$ and $q$. The natural splitting of the orbitals into a $p$-subspace and $q$-subspace now allows us to find canonical forms of $\mUT$ and $\mUC$. Again, Takagi decomposition  allows us to find a unitary matrix $\mr_{pp}$ such that $\mr_{pp} \mr_{pp}^T = \mV_{pp}$, another similar matrix $\mr_{qq}$ that does $\mr_{qq} \mr_{qq}^T = \mV_{qq}$. We can define a basis change matrix
\beq\mylabel{eqn:Rspl}
\mR = \left(
\begin{array}{cc}
\mr_{pp} & \bzero_{p  q} \\
\bzero_{q  p} & \mr_{qq} 
\end{array}
\right)
\eeq
from which we immediately see that
\beq
\mUT = \mR \mOne \mR^T, \;\;\;\; \mUC = \mR (\mOne_{p,q}) \mR^T.
\eeq
Moreover, we see that $\mUS$ transforms under this basis change via \eqn{eqn:UStrans} as
\beq
\mR^\dagger \mOne_{p,q} \mR = \mOne_{p,q} 
\eeq
i.~e., such a basis change does not alter the form of $\mUS$. The conclusion of this discussion is that when $\tT=\tC=+1$, we can always choose a basis in which 
\beq\mylabel{eqn:TpCp}
\mUT = \mOne, \;\;\; \mUC = \mOne_{p,q}, \;\;\; \mUS = \mOne_{p,q}.
\eeq

\noindent
{\bf Class \CII:} This symmetry class has signature $(-1,-1,1)$. From \eqn{eqn:TeqC}, we have $\mV_{pq} = \bzero_{p  q}$,  $\mV_{pp} \mV^*_{pp} = - \mOne_{pp}$ and $\mV_{qq} \mV^*_{qq} = - \mOne_{qq}$, resulting in
\beq\mylabel{eqn:TmCmGen}
\mUT   = \left( 
\begin{array}{cc}
\mV_{pp} & \bzero_{p  q} \\
\bzero_{q  p} & \mV_{qq}
\end{array}
\right), \;\;\;\;
\mUC   = \left( 
\begin{array}{cc}
-\mV_{pp} & \bzero_{p  q} \\
\bzero_{q  p} & \mV_{qq}
\end{array}
\right).
\eeq
Even as $p+q = \No = 2 \Mo$, there are additional constraints on $p$ and $q$. The conditions $\mV_{pp} \mV^*_{pp} = - \mOne_{pp}$ and $\mV_{qq} \mV^*_{qq} = - \mOne_{qq}$, force $p=2r$ and $q=2s$\mylabel{sym:rs}, i.~e., both $p$ and $q$ have to be even numbers, $2(r+s) = 2\Mo$.\nomcls{e40}{$r,s$}{Half of $p,q$ when they are even. \symref{sym:rs}} 

We can find canonical forms for $\mUT$ and $\mUC$ much like \eqn{eqn:TpCp}. For this, noting that $p$ and $q$ are even, we know that there are unitary matrices $\mr_{pp}$ and $\mr_{qq}$ such that 
$\mr_{pp} \mJ_{pp} \mr_{pp}^T = \mV_{pp}$ and $\mr_{qq} \mJ_{qq} \mr_{qq}^T = \mV_{qq}$. Thus one can perform a basis change using a transformation similar to \eqn{eqn:Rspl} to obtain natural forms 
\beq\mylabel{eqn:TmCm}
\mUT = \left(
\begin{array}{cc}
\mJ_{pp} & \bzero_{p  q} \\
\bzero_{q  p} & \mJ_{qq}
\end{array}
\right), \;\;\; \mUC = \left(
\begin{array}{cc}
-\mJ_{pp} & \bzero_{p  q} \\
\bzero_{q  p} & \mJ_{qq}
\end{array}
\right), \;\;\;\; \mUS = \mOne_{p,q},
\eeq
which provide the canonical representation of the three symmetries in class \CII.

\noindent
{\bf Class \CI:} This class with the symmetry signature $(+1,-1,1)$ can be analyzed using \eqn{eqn:TneC} which provides \
\beq\mylabel{eqn:TpCmGen}
\mUT   = \left( 
\begin{array}{cc}
\bzero_{\Mo  \Mo} & \mV_{\Mo\Mo} \\
 \mV_{\Mo\Mo}^T & \bzero_{\Mo  \Mo}
\end{array}
\right), \;\;\;\;\; \mUC   = \left( 
\begin{array}{cc}
\bzero_{\Mo  \Mo} & -\mV_{\Mo\Mo} \\
 \mV_{\Mo\Mo}^T & \bzero_{\Mo  \Mo}
\end{array}
\right)
\eeq
We can obtain canonical forms of $\mUT$ and $\mUC$ by the following manipulations. Consider a basis transformation of the type \eqn{eqn:Rspl} of the from 
\beq\mylabel{eqn:RM}
\bR = \left(
\begin{array}{cc}
\mV_{\Mo\Mo} & \bzero_{\Mo \Mo} \\
\bzero_{\Mo  \Mo} & \bOne_{\Mo \Mo} \\
\end{array}
\right)
\eeq
This basis change effects the following changes (see \eqn{eqn:UTtrans}, \eqn{eqn:UStrans})
\beq\mylabel{eqn:TpCmBC}
\begin{split}
\mUT \mapsto \mR^\dagger \mUT \mR^* & = \left(\begin{array}{cc} 
\bzero_{\Mo \Mo} & \bOne_{\Mo \Mo} \\
\bOne_{\Mo \Mo} & \bzero_{\Mo  \Mo} 
\end{array}
\right) \equiv \mF \\
\mUC \mapsto \mR^\dagger \mUC \mR^* & = -\mJ \\
\mUS \mapsto \mR^\dagger \mOne_{\Mo,\Mo} \mR & =  \mOne_{\Mo,\Mo} \\
\end{split}
\eeq\nomcls{e50}{$\mF$}{Matrix of dimension $\No$ of the form $\left(\begin{array}{cc} 
\bzero_{\Mo \Mo} & \bOne_{\Mo \Mo} \\
\bOne_{\Mo \Mo} & \bzero_{\Mo  \Mo} 
\end{array}
\right)$. \symdefref{eqn:TpCmBC}}where the first line defines the $\No \times \No$ matrix $\mF$.
Thus the canonical forms of class \CII~ are 
\beq\mylabel{eqn:TpCm}
\mUT = \mF, \;\;\; \mUC = -\mJ, \;\;\; \mUS = \mOne_{\Mo,\Mo}
\eeq

\noindent
{\bf Class \DIII:} The symmetry signature $(-1,+1,1)$ provides $\tT=-1=-\tC$.
Form \eqn{eqn:TneC}, we get 
\beq\mylabel{eqn:TmCp}
\mUT   = \left( 
\begin{array}{cc}
\bzero_{\Mo\Mo} & \mV_{\Mo\Mo} \\
 -\mV_{\Mo\Mo}^T & \bzero_{\Mo\Mo}
\end{array}
\right), \;\;\;\;\; \mUC   = \left( 
\begin{array}{cc}
\bzero_{\Mo\Mo} & \mV_{\Mo\Mo} \\
 \mV_{\Mo\Mo}^T & \bzero_{\Mo\Mo}
\end{array}
\right)
\eeq
Using the basis change \eqn{eqn:RM}, we can find canonical structure of the $\mU$ matrices as
\beq\mylabel{eqn:TmCpa}
\mUT = \mJ, \;\;\; \mUC = \mF, \;\;\; \mUS = \mOne_{\Mo,\Mo}.
\eeq

In the classes \CI~ and \DIII, there is no further constraint on $\Mo$ which can be odd or even.
\begin{table*}
\begin{tabular}{||c|| c | c | c ||c|c|c|c||}
\hline\hline
Class & $\tT$ & $\tC$ & $\tS$ & $\No$ & $\mUT$ & $\mUC$ & $\mUS$ \\
\hline\hline
\Ac    & 0   & 0   & $\ 0\ $ & $\No$ & $-$ & $-$ &  $-$ \\ 
\hline
\AI  & $+1$   & 0   & 0  & $\No$ & $\mOne$ & $-$ &  $-$ \\ 

\hline
\AII  & $-1$   & 0   & 0 & $\No=2\Mo$ & $\mJ$ & $-$ &  $-$ \\ 
\hline
\Dc  &  $0$  & $+1$  & 0  & $\No$ & $-$ & $\mOne$ &  $-$ \\
\hline
\C  &  $0$  & $-1$  & 0  & $\No=2\Mo$ & $-$ & $\mJ$ &  $-$ \\
\hline
\AIII  &  0  & 0   & 1  & $\No=p+q$ & $-$ & $-$ &  $\mOne_{p,q}$ \\ 
\hline
\BDI  &  $+1$  & $+1$  & 1  & $\No=p+q$ & $\mOne$ & $\mOne_{p,q}$ & $\mOne_{p,q}$  \\ 
\hline
\CII  &  $-1$  & $-1$  & 1  & $\begin{array}{c}\No=p+q\\ p=2r;\ q=2s\end{array}$ & $\tbtmat{\mJ_{pp}}{\bzero_{pq}}{\bzero_{qp}}{\mJ_{qq}}$ & $\tbtmat{-\mJ_{pp}}{\bzero_{pq}}{\bzero_{qp}}{\mJ_{qq}}$&  $\mOne_{p,q}$ \\
\hline
\CI  &  $+1$  & $-1$  & 1  & $\No=2\Mo$ & $\mF$ & $-\mJ$ &  $\mOne_{\Mo,\Mo}$ \\

\hline
\DIII  &  $-1$  & $+1$  & 1 & $\No=2\Mo$ & $\mJ$ & $\mF$ &  $\mOne_{\Mo,\Mo}$ \\

\hline\hline
\end{tabular}
\caption{Canonical representations of symmetry operators $\mUT$, $\mUC$ and $\mUS$ are shown in the ten symmetry classes. $\mOne$ is the $\No \times \No$ identity matrix, $\mJ$ defined in \eqn{eqn:J}, $\mOne_{p,q}$ defined in \eqn{eqn:Ipq}, 
$\mF$ defined in \eqn{eqn:TpCmBC}. }
\mylabel{tab:tenfold}
\end{table*}

The determination of the canonical structure of the symmetry operation accomplishes a great deal in the determination of the structure of the Hamiltonians in each class. The summary of the findings of this section are given in table~\ref{tab:tenfold}.

\section{A Group Cohomolgy Perspective}\mylabel{sec:GrpCohom}

The results of the previous section can be obtained from a different perspective as we describe in this section. Readers interested in the structure of many body Hamiltonians in each class may proceed directly to sec.~\ref{sec:NonInt}. The main idea of the section is to demonstrate that the ten classes arise from {\em projective} representations\cite{Weinberg1995} of an (abstract) symmetry group. For the \GFS~ discussed in this paper, the abstract group of symmetry operations (see discussion near \eqn{eqn:NO2} 
 is the Klein 4-group $\Klein$. This group consists of four elements $\absI$ (identity), $\absT$ (time reversal), $\absC$ (charge conjugation), and $\absS$ (sublattice) and has the following multiplication table
\beq\mylabel{eqn:V4}
\begin{split}
\begin{tabular}{c|cccc}
  $\Klein$       & \absI & \absT & \absC & \absS \\
\hline
\absI    & \absI & \absT & \absC & \absS \\
\absT    & \absT & \absI & \absS & \absC\\
\absC    & \absC & \absS & \absI & \absT \\
\absS    & \absS & \absC & \absT & \absI 
\end{tabular}\quad\quad.
\end{split}\nomcls{e60}{$\absI,\absT,\absC,\absS$}{Group elements of $\Klein$ denoting identity, time reversal, charge conjugation and sublattice symmetry operations. \symeqref{eqn:V4}}  
\eeq
The abstract symmetry group $\symS$\mylabel{sym:symS} of the \GFS~must be one of the (proper/improper) subgroups of $\Klein$. Thus, $\symS$ is one of $\symI = \{\absI\}, \symZ_2^T = \{\absI,\absT\}, \symZ_2^C = \{\absI,\absC\}, \symZ_2^S = \{\absI,\absS\} $ or $\Klein$.\nomcls{e70}{$\symS$}{Abstract symmetry group of \textup{GFS}. \symref{sym:symS}}\nomcls{e80}{$\symZ_2^T$}{$\symZ_2^T = \{\absI,\absT\}$, sub-group of $\Klein$. \symref{sym:symS}}\nomcls{e90}{$\symZ_2^C$}{$\symZ_2^C = \{\absI,\absC\}$, sub-group of $\Klein$. \symref{sym:symS}}\nomcls{f10}{$\symZ_2^S$}{$\symZ_2^S = \{\absI,\absS\}$, sub-group of $\Klein$. \symref{sym:symS}}
 
We look for projective representations of the group $\symS$ on a graded vector space $\calW = \calV_1 \oplus \bar{\calV}_1$, where $\calV_1$ is an $\No$-dimensional $\mathbb{C}$ vector space. We are restricting here to the one particle sector $\calV_1$ and its ``transposed'' hole space $\bar{\calV_1}$ (which is isomorphic to $\calV_{\No-1}$, see \eqn{eqn:HFspace}). The group elements are represented by $\mathbb{C}$ valued matrices of the form
\beq\mylabel{eqn:ProjMat}
\begin{split}
D(\absI) = \tbtmat{\mOne}{\mZero}{\mZero}{\mOne}, \;\;\;\;\; D(\absT) &= \tbtmat{\mUT}{\mZero}{\mZero}{\mUT} \mK \\
D(\absC) = \tbtmat{\mZero}{\mUC}{\mUC}{\mZero} \mK, \;\;\;\;\;
D(\absS) &= \tbtmat{\mZero}{\mUS}{\mUS}{\mZero}. 
\end{split}
\eeq\nomcls{f20}{$D(.)$}{$\mathbb{C}$ valued matrices representing elements of $\symS$ group. \symdefref{eqn:ProjMat}}
$\mK$ is the usual complex conjugation operator, $\mUT$, $\mUC$, and $\mUS$ are unitary matrices that represent the symmetries appearing in their subscripts. This choice of representations is motivated by discussions of the previous section. 

Denoting the group elements of $\symS$ by $\absg_\ell$\mylabel{sym:gell}, a projective representation reproduces the group multiplication table up to a $U(1)$ phase factor, i.~e.,\nomcls{f30}{$\absg_\ell$}{Group element of $\symS$. \symref{sym:gell}}
\beq\mylabel{eqn:Proj}
D(\absg_1) D(\absg_2) = \omega(\absg_1,\absg_2) D(\absg_1 \absg_2)\ . 
\eeq
With $|\symS|$\mylabel{sym:cardinality} as the number of elements in $\symS$, the $|\symS|^2$ numbers $\omega(\absg_1,\absg_2) \in U(1)$, usually called ``2-cocylces" or ``Schur multipliers"  must satisfy the following co-cycle condition
\beq\mylabel{eqn:CoCycle}
\omega(\absg_1,\absg_2) \omega(\absg_1 \absg_2, \absg_3) = \omega(\absg_1,\absg_2 \absg_3) \omega^{\absg_1}(\absg_2, \absg_3)\ .
\eeq\nomcls{f50}{$\omega(\absg_1,\absg_2)$}{2-cocylces or Schur multipliers. \symref{sym:cardinality}}\nomcls{f60}{$\omega^{\absg}$}{Short form for action of the group element $\absg$ on an element of $U(1)$. \symeqref{eqn:CoCycle}}Here $\omega^{\absg}$ is a short form for action of the group element $\absg$ on an element of $U(1)$, i.~e., for every group element $\absg$, there is an invertible function $\varphi_\absg$ on $U(1)$ which maps $\omega \in U(1)$ to $\varphi_\absg(\omega) \equiv 
\omega^{\absg}$. The function must satisfy 
\beq
\varphi_{\absg_1}(\varphi_{\absg_2}(\omega)) = \varphi_{\absg_1\absg_2}(\omega). 
\eeq
More technically, $\varphi$ is a homomorphism from $\symS$ to $\textup{Aut}{\,U(1)}$, the group of automorphisms (invertible maps that preserve the group structure) of $U(1)$. In our case, the homomorphism $\varphi$ is defined as
\beq\mylabel{eqn:vphi}
\begin{split}
\varphi_{{\tiny \absI}}(\omega) = \omega, \;\;\;\; \varphi_{{\tiny \absT}}(\omega) & = \omega^*  \\
\varphi_{{\tiny \absC}}(\omega) = \omega^*, \;\;\;\; \varphi_{{\tiny \absS}}(\omega) & = \omega  \\
\end{split} 
\eeq
where $(~)^*$ denotes complex conjugation. The check of the homomorphic character of $\varphi$ is routine. This particular choice of $\varphi$ is made keeping in mind the physical aspects of the group elements, consistent with \eqn{eqn:ProjMat}. Now two multiplier system $\omega_1$ and $\omega_2$ (both of which satisfy \eqn{eqn:CoCycle}) can be used to produce a third one $\omega'$, in fact $\omega'(\absg_1,\absg_2) = \omega_1(\absg_1,\absg_2)\omega_2(\absg_1,\absg_2)$. Note that $\omega(\absg_1,\absg_2) = 1$, the unit multiplier, satisfies the cocycle condition \eqn{eqn:CoCycle}. Furthermore, any given $\omega$ has an inverse $\omega^{-1} (\absg_1,\absg_2) =\omega^*(\absg_1,\absg_2)$ such that $\omega \omega^{-1}$ is the unit multiplier. Finally, two multipliers $\omega_1$ and $\omega_2$ are equivalent if there exists  a $U(1)$ valued function on $\symS$, $a(\absg)$, such that 
\beq\mylabel{eqn:Bd}
\frac{\omega_1(\absg_1,\absg_2)}{\omega_2(\absg_1,\absg_2)} = \frac{a(\absg_1)a^{\absg_1}(\absg_2)}{a(\absg_1\absg_2)}. 
\eeq
Suppose we denote the set of all multipliers equivalent to $\omega$ by $[\omega]$. It is easy to see that  the set of equivalence classes of multipliers  $\{[\omega]\}$\mylabel{sym:equivomega} forms an abelian group called the second cohomolgy group (see \cite{Janssen1972} and chapter 7 of \cite{Wickless2004}) $H^2_\varphi(U(1),\symS)$. A key question of importance to us is how many elements (number of equivalence class of multipliers) does $\symS$ admit -- this is answered by computing the group $H^2_\varphi(U(1),\symS)$\mylabel{sym:H2phi}.\nomcls{f70}{$\{[\omega]\}$}{Set of equivalence classes of Shur multipliers. \symref{sym:equivomega}}

We can now make the central point: {\em The number of symmetry classes corresponds to number of distinct equivalence classes of multiplier systems corresponding to the symmetry group $\symS$} which runs over of $\symI, \symZ_2^T, \symZ_2^C, \symZ_2^S$ and $\Klein$. To obtain the structure of each class, we find the multiplier system that labels the class, and find the irreducible representations, ``smallest'' vector space $\calW$, and the matrices of \eqn{eqn:ProjMat}. Again, we organize the classes via the types introduced in sec.~\ref{sec:SymClasses}

\subsection{Type 0}

Here the symmetry group $\symS = \symI$, and $H^2_\varphi(U(1),\symS)$ is trivial containing a single element, and hence a trivial multiplier system.\nomcls{f80}{$H^2_\varphi(U(1),\symS)$}{The second cohomology group. \symref{sym:H2phi}}  

\class{\Ac} The sole irreducible representation corresponding to this trivial multiplier system is 1-dimensional. This labels class \Ac,  and we see that the $\No$-orbtial \GFS~ in class \Ac~ of table~\ref{tab:tenfold} is a reducible representation consisting of $\No$ copies of irreducible one obtained just now.

\subsection{Type 1}

\centerline{\underline{$\symS = \symZ_2^T$}}

The second cohomology group $H^2_\varphi(U(1),\symZ_2^T) = Z_2$ consists two multiplier classes.\mycite{Chen2013}. The multipliers are
\beq \mylabel{eqn:Z2T}
\begin{array}{c|cc}
\omega(\absg_1 \, \downarrow, \absg_2 \, \rightarrow) & \absI & \absT \\
\hline
\absI & 1 & 1 \\
\absT & 1 & \tT
\end{array}, \;\;\;\;\;\;\;\; \tT = \pm 1
\eeq

\class{\AI} The multiplier of $H^2_\varphi(U(1),\symZ_2^T) $ with $\tT=+1$ labels the class \AI. To find the irreducible representation, we need to solve
\beq
\mUT \mUTs = \mOne
\eeq
and the solution with smallest dimension (in the sense of dimension of vector space $\calV_1$ that makes up $\calW$) is 1, and $\mUT = e^{\ci \alpha}$. By the change of basis \eqn{eqn:UTtrans}, $\alpha$ can always chosen to the zero, and $\mUT = 1$, and $D(\absT) = \tbtmat{1}{0}{0}{1} \, \mK$. We see that the $\No$-orbtial \GFS~ of table~\ref{tab:tenfold} is a reducible representation with $\No$-copies of this irrep (irreducible representation).

\class{\AII} The multiplier in \eqn{eqn:Z2T} with $\tT=-1$ corresponds to class \AII. This leads to the equation $\mUT \mUTs = - \mOne$. The solution with the smallest dimension is 2, with
\beq\mylabel{eqn:J2}
\mUT = \mJ_2 = \tbtmat{0}{1}{-1}{0}\ .
\eeq\nomcls{f90}{$\mJ_2$}{$2\times 2$ version of $\mJ$. \symdefref{eqn:J2}}We see that the  \AII~\GFS~  of table~\ref{tab:tenfold} with $\No=2\Mo$ consists $\Mo$ copies of this this irrep corresponding to this multiplier system.

\centerline{\underline{$\symS = \symZ_2^C$}}

The second cohomology group $H^2_\varphi(U(1),\symZ_2^C) = Z_2$ -- the mathematics of this group is same as $\symZ_2^T$, and the multipliers are
\beq \mylabel{eqn:Z2C}
\begin{array}{c|cc}
\omega(\absg_1 \, \downarrow, \absg_2 \, \rightarrow) & \absI & \absC \\
\hline
\absI & 1 & 1 \\
\absC & 1 & \tC
\end{array}, \;\;\;\;\;\;\;\; \tC = \pm 1
\eeq
The key physical difference is that the representation of $\absC$ is via a {\em transposing} antilinear matrix (see \eqn{eqn:ProjMat}). Very similar consideration as in the $\symZ_2^T$ case, leads to two classes.

\class{\Dc} The multiplier with $\tC=+1$ gives this class. The irrep is 1-dimensional with $\mUC = 1$ and $D(\absC) = \tbtmat{0}{1}{1}{0} \, \mK$, $\No$ copies of which are present in the \GFS~ shown in table~\ref{tab:tenfold}.

\class{\C} This class is labeled by the $\tC =-1$ multiplier of \eqn{eqn:Z2C}, with a 2-dimensional irrep such that $D(\absC) = \tbtmat{\mZero}{\mJ_2}{\mJ_2}{\mZero} \mK$. The \C~entry in table~\ref{tab:tenfold} consists of $\Mo$ copies of this irrep.

\centerline{\underline{$\symS = \symZ_2^S$}}

In the case of the sole non-ordinary sublattice symmetry, $H^2_\varphi(U(1),\symZ_2^S)$ is trivial and has only the unit multiplier. 

\class{\AIII} Being an abelian group of order 2,  $\symZ_2^S$ has {\em two} distinct 1-dimensional irreducible representations, which arise from the lowest dimensional solution to the equation $\mUS \mUS = \mOne$. The first irrep has $\mUS = + 1$ and the second has $\mUS = -1$ (note that these two irreps have the same multipliers). We immediately see that the $\No$ orbital \AIII~ system in table~\ref{tab:tenfold} is made of the $p$ copies of the $+$irrep and $q$ copies of the $-$irrep.

\subsection{Type 3}

All the classes in type 3 arise from the multipliers of the symmetry group $\symS = \Klein$. Quite interestingly the second cohomology group of the Klein group is the Klein group itself\footnote{We obtained this result by an explicit calculation.}: $H^2_\varphi(U(1),\Klein) = \Klein$! This results in four multipliers labeled by $\tT$ and $\tC$

\beq\mylabel{eqn:K4}
\begin{array}{c|cccc}
 \omega(\absg_1 \, \downarrow, \absg_2 \, \rightarrow)   & \absI & \absT & \absC & \absS \\
\hline
\absI    & 1 & 1 & 1 & 1 \\
\absT    & 1 & \tT & 1 & \tT \\
\absC    & 1 & \tT\tC & \tC & \tT \\
\absS    & 1 & \tC & \tC & 1 
\end{array} , \;\;\; \tT=\pm1, \tC=\pm1\ .
\eeq

To obtain the irreps, we seek smallest dimensional quantities $\mUT$, $\mUC$, $\mUS$ that simultaneously satisfy
\beq\mylabel{eqn:K4tmp}
\mUT \mUTs = \tT \mOne,\ \mUC \mUCs = \tC \mOne,\ \mUS \mUS = \mOne,\ \mUT \mUCs = \mUS
\eeq
(these are obtained by a straightforward application of \eqn{eqn:ProjMat}).

\noindent
\class{\BDI} The multiplier \eqn{eqn:K4} with $\tT=\tC=+1$ obtains this class. Eqn.~\prn{eqn:K4tmp} gives $\mUT\mUT^* = \mOne,  \mUC \mUC^* =  \mOne, \mUS \mUS = \mOne, \mUT \mUC^* = \mUS$. There are two 1-dimensional irreps, one with $\mUT=1, \mUC=1$ and $\mUS=1$, and the other $\mUT=1, \mUC=-1$ and $\mUS=-1$. We see that class \BDI~in table~\ref{tab:tenfold} contains $p$ copies of the first irrep and $q$ copies of the second one.

\noindent
\class{\CII} This class is characterized by the multiplier in \eqn{eqn:K4} with $\tT=\tC=-1$. Irreps are obtained from the solution of $\mUT\mUT^* = -\mOne,  \mUC \mUC^* =  -\mOne, \mUS \mUS = \mOne, \mUT \mUC^* = \mUS$. These now provide two distinct 2-dimensional representations. First one has $\mUT = \mJ_2, \mUC=-\mJ_2, \mUS=-\mOne$, and the second has $\mUT = \mJ_2, \mUC=\mJ_2, \mUS=\mOne$. The \CII~entry in table~\ref{tab:tenfold} has $r$-copies of the first 2-dimensional irrep, and $s$-copies of the second one.

\noindent
\class{\CI} When $\tT=-\tC=+1$ in the multiplier \eqn{eqn:K4}, we get the \CI~class. The solution of \eqn{eqn:K4tmp} gives a single 2-dimensional representation with $\mUT =\mF_2, \mUC=-\mJ_2$ and $\mUS=\mOne_{1,1}$. $\mF_2 = \tbtmat{0}{1}{1}{0}$\mylabel{sym:mF2} and $\mOne_{1,1}$ is defined in \eqn{eqn:Ipq}. The \GFS~of the \CI~class in table~\ref{tab:tenfold} has $\Mo$ copies of this irrep.\nomcls{g10}{$\mF_2$}{$2\times 2$ version of $\mF$. \symref{sym:mF2}}

\class{\DIII} Class \DIII~is obtained from the multiplier \eqn{eqn:K4}  when $\tT=-\tC=-1$. Eqn.~\prn{eqn:K4tmp}, again, provides a single 2-dimensional representation with $\mUT =\mJ_2, \mUC= \mF_2$ and $\mUS=\mOne_{1,1}$.  \DIII~class in table~\ref{tab:tenfold} is made of $\Mo$ copies of this irrep.

The discussion of this section provides a deeper insight of how the ten symmetry classes arise from projective representations on a graded vector space (this is where the fermionic nature of the states enters). Moreover, all the results of the canonical representations of the symmetry operations derived from elementary considerations can be better understood as copies of irreducible projective representations.

\section{Noninteracting Systems}\mylabel{sec:NonInt}

\begin{table*}
\begin{tabular}{||c||c|c|c|c|c||}
\hline\hline
Class & $\No$ & $\mnbdy{\mH}{1} $& $\dim{\ci\calH^{(1)}}$ & $\ci\calH^{(1)}$   & $\Schrod(t)$ \\
\hline\hline
\Ac $(0,0,0)$ & $\No$ & $\mnbdy{\mH}{1}  = [\mnbdy{\mH}{1}]^\dagger$  & $\No^2$& $\lieu(\No)$ &  $U(\No)$ \\
\hline
\AI $(+1, 0, 0)$  & $\No$ & $\mnbdy{\mH}{1}=[\mnbdy{\mH}{1}]^*$ &$\No(\No+1)/2$ & $\lieu(\No) \setminus \lieo(\No)$  & $U(\No)/O(\No)$  \\
\hline
\AII$(-1,0,0)$ & $\No=2\Mo$ & $\displaystyle{\tbtmat{\mh_{aa}}{\mh_{ab}}{-\mh_{ab}^*}{\mh_{aa}^*}}$  &$\Mo(2\Mo-1)$& $\lieu(2\Mo) \setminus \lieusp(2\Mo)$   & $U(2\Mo)/\USp(2\Mo)$ \\
\hline
\Dc $(0, +1, 0)$  & $\No$ & $\mnbdy{\mH}{1}=-[\mnbdy{\mH}{1}]^*$ &$\No(\No-1)/2$& $\lieo(\No)$   & $O(\No)$  \\
\hline
\C $(0,-1,0)$ & $\No=2\Mo$ & $\displaystyle{\tbtmat{\mh_{aa}}{\mh_{ab}}{\mh_{ab}^*}{-\mh_{aa}^*}}$  &$\Mo(2\Mo+1)$ & $\lieusp(2\Mo)$  & $\USp(2\Mo)$  \\
\hline
\AIII $(0,0,1)$ & $\No=p+q$ & $\displaystyle{\tbtmat{\mZero_{p p}}{\mh_{pq}}{\mh_{pq}^\dagger}{\mZero_{q q}}}$& $2pq$ & $\lieu(p+q) \setminus (\lieu(p)\oplus\lieu(q))$    & $U(p+q)/(U(p)\times U(q))$\\
\hline
\BDI $(+1,+1,1)$ & $\No=p+q$ & $\displaystyle{\tbtmat{\mZero_{p p}}{\mh_{pq}}{\mh_{pq}^T}{\mZero_{q q}}}, \mh_{pq}^* = \mh_{pq}$ &$pq$ & $\lieo(p+q) \setminus (\lieo(p)\oplus\lieo(q))$   & $O(p+q)/(O(p)\times O(q))$\\
\hline
\CII $(-1,-1,1)$ & $\begin{array}{c}\No=p+q,\\ p=2r, q=2s\end{array}$ & $\tbtpart{\bzero_{p p}}{\tbtarray{\mh_{rr}}{\mh_{rs}}{-\mh^*_{rs}}{\mh^*_{rr} }}{ \text{h.c.}}{\bzero_{q q}}$&$4rs$ & $\lieusp(p+q) \setminus (\lieusp(p)\oplus\lieusp(q))$   & $\USp(2(r+s))/(\USp(2r)\times \USp(2s))$\\
\hline
\CI (+1, -1, 1)  & $\No = 2 \Mo$ & $ \displaystyle{\tbtmat{\mZero_{\Mo \Mo}}{\mh_{\Mo\Mo}}{\mh_{\Mo\Mo}^*}{\mZero_{\Mo \Mo}}}, \mh_{\Mo\Mo}^T = \mh_{\Mo\Mo}$& $\Mo(\Mo+1)$& $\lieusp(2\Mo)\setminus \lieu(\Mo)$  & $\USp(2\Mo)/U(\Mo)$\\
\hline
\DIII (-1, +1, 1) & $\No =2 \Mo$ & $ \displaystyle{\tbtmat{\mZero_{\Mo \Mo}}{\mh_{\Mo\Mo}}{-\mh_{\Mo\Mo}^*}{\mZero_{\Mo \Mo}}}, \mh_{\Mo\Mo}^T = -\mh_{\Mo\Mo}$ & $\Mo(\Mo-1)$& $\lieo(2\Mo) \setminus \lieu(\Mo)$  & $O(2\Mo)/U(\Mo)$\\
\hline\hline
\end{tabular}
\caption{Structure of noninteracting Hamiltonians in the ten symmetry classes.}
\mylabel{tab:NonInt}
\end{table*}

We devote this section to obtaining the structure of the Hamiltonians of each class when there are no interactions present. Here $\mH$ of \eqn{eqn:HSpace} has only two entries $\mH = \left(\mnbdy{\mH}{0}, \mnbdy{\mH}{1}\right)$. Another important quantity of interest is Sch\"odinger time evolution operator
\beq\mylabel{eqn:Schord}
\Schrod(t) = e^{-\ci t \Hmat }
\eeq\nomcls{g20}{$\Schrod(t)$}{Sch\"odinger time evolution operator. \symdefref{eqn:Schord}}where $t$ is the time, $\Hmat$ is the Hamiltonian matrix constructed out of $\mH = \left(\mnbdy{\mH}{0}, \mnbdy{\mH}{1}\right)$. For example, the in this noninteracting setting the components of $\Hmat_{ij} = \mnbdy{H}{0} \delta_{ij} + \nbdy{H}{1}{i;j}$. For a fixed time, say $t=1$, the time evolution operator spans out a ``geometric structure'' as $\mH$ runs over all of the space $\calH$. As pointed out by \mycite{Altland1997} (generalizing the pioneering work of Dyson\cite{Dyson1962}), this geometric structure realized is a {\em symmetric space} (see \mycite{Caselle2004} for a review symmetric spaces from a physicist's perspective, \cite{Gilmore1974} for a review of Lie groups and algebras). The ten classes realize the the ten different symmetric spaces classified by Cartan, and in fact the names of the classes borrows Cartan's nomenclature. Our development of the ten fold symmetry classification not only allows us to recover these known results in a simple and direct manner but also provides a very clear structure of the Hamiltonian space that could be particularly useful for model building. Our results are recorded in table~\ref{tab:NonInt}.
 
To proceed with the discussion, we record the symmetry conditions \eqn{eqn:HMap} specifically for the noninteracting systems. Under the action of usual symmetries $\mH$ transforms as 
\beq\mylabel{eqn:SCUslOne}
\mH = \left(\mnbdy{\mH}{0}, \mnbdy{\mH}{1}\right)  \mapsto \mHr = (\mnbdy{\mH}{0}, \mnbdy{\mHc}{1})
\eeq
and for transposing symmetries we obtain
\beq\mylabel{eqn:SCTrnOne}
\mH = \left(\mnbdy{\mH}{0}, \mnbdy{\mH}{1}\right) \mapsto \mHr =  \left(\mnbdy{\mH}{0} + \tr_1{\mnbdy{\mHc}{1}}, -\left[\mnbdy{\mHc}{1} \right]^T \right)
\eeq
(see eqns.~\prn{eqn:Htmp}, \prn{eqn:HtUsl}, and \prn{eqn:HtTrn}).
The structure of $\mH$ is obtained by imposing the symmetry condition \eqn{eqn:SymCond} for all the appropriate symmetries. An immediate consequence is that $\mnbdy{\mH}{0}$, the ``vacuum energy'', is allowed to have any real value $H^{(0)}$ in {\em all} the ten classes. This is a feature that is always true, i.~e., even when interactions are present. As the values of $H^{0}$ spans the reals, the time evolution operator (at a fixed time) acquires a phase factor represented by the $U(1)$ group (see discussion on class \Ac~in the next para).

\class{\Ac} There are no constraints on the Hamiltonian in this class. Any $\No \times \No$ Hermitian matrix is in $\mnbdy{\calH}{1}_{\textup{\Ac}}$. Consequently $\ciHone_{\textup{\Ac}} = \lieu(\No)$, and $\Schrod(t)$ acquires the structure of $U(1)\times U(\No)$. The $U(1)$ factor arises from $\mnbdy{\mH}{0}$ which spans all of real numbers. As discussed in the previous para, this $U(1)$ factor will appear in {\em all} cases (interacting and noninteracting). Henceforth, we suppress this factor (which is equivalent to setting $H^{0} = 0$ (or any other fixed real number). The symmetric space of $\Schrod(t)$ in table~\ref{tab:NonInt} for this case, therefore is shown just as $U(\No)$.

\class{\AI} Eqn.~\prn{eqn:SymCond} appropriate for this class (with $\mUT = \mOne$) provides that real symmetric matrices $\Hone = [\Hone]^*$ are the allowed entries of $\mnbdy{\calH}{1}_{\textup{\AI}}$. This implies that the $\ciHone_ {\textup{\AI}} = \lieu(\No) \setminus \lieo(\No)$\mylabel{sym:lieo}, where $\lieo(\No)$ is Lie algebra of the group of orthogonal matrices. The time evolution operator $\Schrod(t)$ spans the coset space $U(\No)/O(\No)$.\nomcls{g30}{$\lieo()$}{ Lie algebra of the group of orthogonal matrices. \symref{sym:lieo}}\nomcls{g40}{$O()$}{Group of orthogonal matrices. \symref{sym:lieo}}

\class{\AII} Since $\mUT = \mJ$ in this class with $\No=2\Mo$, we can split the orbitals into two types and write the Hamiltonian as
\beq
\Hone = \tbtmat{\mh_{aa}}{\mh_{ab}}{\mh^\dagger_{ab}}{\mh_{bb}}.
\eeq
The symmetry condition gives $\mJ [\Hone]^* \mJ^\dagger = \Hone$ resulting in $\mh_{ab} = - \mh_{ab}^T, \mh_{bb} = \mh_{aa}^*$. Here, $\mh_{aa},\mh_{ab}, \mh_{bb}$ are $\Mo \times \Mo$ matrices. The space $\calH_{\textup{\AII}}$ is made of matrices of the type
\beq\mylabel{eqn:HAIIonep}
\Hone = \tbtmat{\mh_{aa}}{\mh_{ab}}{-\mh_{ab}^*}{\mh_{aa}^*}.
\eeq
It can be seen by explicit calculation that $\ciHone_{\textup{\AII}} = \lieu(2\Mo) \setminus \lieusp(2\Mo)$, leading to the time evolution $\Schrod(t)$ spanning the coset space $U(2\Mo)/\USp(2\Mo)$. Here $\USp(2\Mo)$ is the symplectic group and $\lieusp(2\Mo)$\mylabel{sym:lieusp} is its associated Lie algebra.\nomcls{g50}{$\lieusp()$}{Lie algebra of the group of symplectic matrices. \symref{sym:lieusp}}\nomcls{g60}{$\USp()$}{Symplectic group. \symref{sym:lieusp}}

\noindent
\class{\Dc} This class with the charge conjugation described by $\mUC = \mOne$ is made of Hamiltonians with vanishing trace $\tr{\Hone} = 0$ that satisfy, $\mnbdy{\mH}{1} = - {[\mnbdy{\mH}{1}]}^*$. The crucial distinction between the time reversal class \AI$(\tT=+1)$ is that the transposing nature of the charge conjugation give the negative sign in the just stated symmetry condition. The space $\ciHone_{\textup{D}}$ is made of real antisymmetric matrices which is the Lie algebra $\lieo(\No)$. The time evolution operator spans $O(\No)$.

\class{\C} The charge conjugation operation is described by $\mUC = \mJ$, and the symmetry condition $\mJ [\Hone]^* \mJ^\dagger = -\Hone$, leads to matrices with $\tr{\Hone} =0$,
\beq
\Hone = \tbtmat{\mh_{aa}}{\mh_{ab}}{\mh_{ab}^*}{-\mh_{aa}^*},
\mylabel{eqn:HConep}
\eeq
i.~e., $\ciHone_{\textup{\C}} = \lieusp(2\Mo)$, and $\Schrod(t)$ spans $\USp(2\Mo)$. 

\class{\AIII} The physics of this class is governed by the sublattice symmetry represented by the unitary $\mUS = \mOne_{p,q}$ that naturally partitions the orbitals into two groups (sublattices). It is natural to write the Hamiltonian as
\beq\mylabel{eqn:AIIIStruc}
\Hone = \tbtmat{\mh_{pp}}{\mh_{pq}}{\mh^\dagger_{pq}}{\mh_{qq}}.
\eeq
The symmetry condition leads immediately to $\mh_{pp} = \bzero_{p  p}$, $\mh_{qq} = \bzero_{q  q}$. Thus, $\calH_{\textup{\AIII}}$ is made of matrices of the form
\beq\mylabel{eqn:AIIIOnep}
\Hone =  \tbtmat{\bzero_{p  p}}{\mh_{pq}}{\mh^\dagger_{pq}}{\bzero_{q  q}}.
\eeq
Indeed, we have $\ciHone_{\textup{\AIII}} = \lieu(p+q) \setminus (\lieu(p) \oplus \lieu(q))$, resulting in $\Schrod(t)$ being in the coset space $U(p+q)/(U(p)\times U(q))$.

All of the type 3 classes can be viewed as descendants of class $\AIII$, and thus all of these classes -- often referred to as chiral classes -- have Hamiltonians that imbibe the structure in \eqn{eqn:AIIIOnep}.

\class{\BDI} The symmetry condition \eqn{eqn:SCUslOne} with $\mUT = \mOne$ gives $\mh_{pq} = \mh_{pq}^*$, and thus elements of $\calH_{\textup{\BDI}}$ are of the form
\beq\mylabel{eqn:BDIOnep}
\Hone =  \tbtmat{\bzero_{p  p}}{\mh_{pq}}{\mh^T_{pq}}{\bzero_{q  q}},\,\,\, \mh_{pq} = \mh_{pq}^*
\eeq
We obtain $\ciHone_{\textup{\BDI}} =  \lieo(p+q) \setminus (\lieo(p) \oplus \lieo(q))$, with $\Schrod(t)$ spanning the the coset space $O(p+q)/(O(p)\times O(q))$.

\class{\CII} The symmetry condition \eqn{eqn:SCUslOne} with $\mUT$ as shown in table~\ref{tab:tenfold} and $\Hone$ of the form \eqn{eqn:AIIIOnep} provides
\beq
\mJ_{pp} \mh^*_{pq} = \mh_{pq} \mJ_{qq} .
\mylabel{eqn:CIIOnep}
\eeq
The key point in this class is that the $p$ orbitals themselves arise as $2r$ orbitals and $q$ as $2s$ orbitals. Thus, $\mh_{pq}$ can be written as
\beq
\mh_{pq} = \tbtmat{\mh_{rr}}{\mh_{rs}}{\mh_{sr}}{\mh_{ss}}.
\eeq
resulting in $\mh_{sr} = -\mh_{rs}^*$ and $\mh_{ss} = \mh_{rr}^*$
\beq
\mh_{pq} = \tbtmat{\mh_{rr}}{\mh_{rs}}{-\mh^*_{rs}}{\mh^*_{rr}}.
\eeq
We see that $\ciHone_{\textup{\CII}} = \lieusp(2(r+s)) \setminus (\lieusp(2r)) \oplus \lieusp(2s)) $ and the symmetric space generated by the time evolution operator is $\USp(2(r+s))/(\USp(2r)\times \USp(2s))$.

\class{\CI} Here $\No=2\Mo$, and thus \eqn{eqn:AIIIOnep} provides 
\beq\mylabel{eqn:CI}
\Hone = \tbtmat{\bzero_{\Mo  \Mo}}{\mh_{\Mo\Mo}}{\mh^\dagger_{\Mo\Mo}}{\bzero_{\Mo  \Mo}}.
\eeq
The symmetry condition \eqn{eqn:SCUslOne} with $\mUT=\mF$ gives $\mh_{\Mo\Mo}^T = \mh_{\Mo\Mo}$ with
\beq\mylabel{eqn:CIOnep}
\Hone = \tbtmat{\mZero_{\Mo  \Mo}}{\mh_{\Mo \Mo}}{\mh^*_{\Mo  \Mo}}{\mZero_{\Mo  \Mo}}.
\eeq
It is now clear that the $\ciHone_{\textup{\CI}} = \lieusp(2\Mo)\setminus \lieusp(\Mo)$, and the symmetric space corresponding to the time evolution is the coset space $\USp(2\Mo)/U(\Mo)$.

\class{\DIII} In this $2\Mo$ dimensional \GFS~with, the symmetry condition \eqn{eqn:SCUslOne} leads to $\mh_{\Mo\Mo}^T = -\mh_{\Mo\Mo}$, and
\beq\mylabel{eqn:DIIIOnep}
\Hone = \tbtmat{\mZero_{\Mo  \Mo}}{\mh_{\Mo \Mo}}{-\mh^*_{\Mo  \Mo}}{\mZero_{\Mo  \Mo}}
\eeq
Here $\ciHone_{\textup{\DIII}}$ is isomorphic to $\lieo(2\Mo) \setminus \lieu(\Mo)$. The symmetric space spanned by the time evolution operator is $O(2\Mo)/U(\Mo)$.

The classes $\Ac$ and $\AIII$ are complex classes, while the remainder of the classes involve a ``reality condition'' of the form $\mH = \mH^*$ and are the real classes.

\section{Framework for Systems with Interactions} \mylabel{sec:IntGen}

In this section we establish ideas that allow for the determination of the structure of Hamiltonians with upto $N$-body interactions as in \eqn{eqn:Hdef}. For usual symmetries, the conditions that determine the spaces $\mnbdy{\ciH}{K}$ are straightforward. From, \eqn{eqn:HtUsl} we obtain
\beq\mylabel{eqn:HKUslStruc}
\mnbdy{\mH}{K} = \mnbdy{\mHc}{K}\ .
\eeq  
These are a set of homogeneous equations in the matrix elements $H^{(K)}_{i_1\ldots\i_K;j_1\ldots,j_K}$ that make $\mnbdy{\ciH}{K}$ of the appropriate class a vector subspace of $\lieu\left(\combi{\No}{K}\right)$ following the discussion near \eqn{eqn:HKIso}. Eqn.~(\ref{eqn:HKUslStruc}) provides conditions to completely determine the structure of the admissible Hamiltonians.

Transposing symmetries have a more involved story. To make progress, we rewrite \eqn{eqn:SymCond} using \eqn{eqn:HtTrn} to obtain
\beq\mylabel{eqn:HMtmp}
\begin{split}
\mnbdy{\mH}{K} &= \sum_{R=K}^\mb A_{R,K} \left[\tr_{R-K} \mnbdy{\mHc}{R}\right]^T \\
&=  (-1)^K\left[\mnbdy{\mHc}{K}\right]^T  + (-1)^{K}  \sum_{R>K}^\mb  \frac{1}{(R-K)!} \left(\frac{R!}{K!} \right)^2\left[\tr_{R-K} \mnbdy{\mHc}{R}\right]^T 
\end{split}
\eeq 
We now define $\mHPLe{K}$ and $\mHMIe{K}$ as
\beq\mylabel{eqn:HkpHkm}
\begin{split}
\mHPLe{K} & \equiv \mnbdy{\mH}{K}  + (-1)^K \left[\mnbdy{\mHc}{K}\right]^T   \\
\mHMIe{K} & \equiv  \mnbdy{\mH}{K}  - (-1)^K \left[\mnbdy{\mHc}{K}\right]^T
\end{split}
\eeq \nomcls{g70}{$\mHPLe{K}$,$\mHMIe{K}$}{Decompositions of $\mnbdy{\mH}{K}$. \symdefref{eqn:HkpHkm}}with $\mnbdy{\mH}{K} = \half \left(\mHPLe{K} + \mHMIe{K} \right)$ and $(-1)^K \left[\mnbdy{\mHc}{K}\right]^T   = \half \left(\mHPLe{K} - \mHMIe{K} \right)$. With these definitions \eqn{eqn:HtTrn} becomes 
\beq\mylabel{eqn:HMSol}
\mHMIe{K} = \frac{(-1)^{K}}{2}  \sum_{R>K}^\mb  \frac{(-1)^R}{(R-K)!} \left(\frac{R!}{K!} \right)^2\left[\tr_{R-K} \mHPLe{R} - \tr_{R-K} \mHMIe{R}\right]\ .
\eeq
Two points emerge from these considerations: (i) the transposing symmetry condition \eqn{eqn:HMtmp} puts no constraint on $\mHPLe{K}$ for any $K$, and (ii) for every $K$, $\mHMIe{K}$ is solely determined  by $R$-body interactions terms where $R>K$. This implies that the set of independent parameters of $\mH$ is entirely determined by $\mHPLe{K}$ for $K$ from $0$ to $N$. Further note that every $\mHPLe{K}$ belongs to a vector subspace of $\lieu\left(\combi{L}{K}\right)$ defined by $\mHMIe{K}= \bzero$, which we call $\ciH^{(K)}_+$. Since, $\mnbdy{\ciH}{K}$ is made of objects of the type $\half \left(\mHPLe{K} + \mHMIe{K} \right)$ where $\mHMIe{K}$ is a ``constant'' determined by \eqn{eqn:HMSol}, $\mnbdy{\ciH}{K}=\frac{\ci}{2} \left(\mHPLe{K} + \mHMIe{K} \right)$, is no longer a vector subspace of $\lieu\left(\combi{L}{K}\right)$. In fact, $\mnbdy{\ciH}{K}$ is an affine subspace of $\lieu\left(\combi{L}{K}\right)$ whose dimension (as a manifold) is same as the subspace $\ciH^{(K)}_+$.

These observations show that $\calH$ in any class can be completely determined by starting from the $N$-body (highest multi-body interaction) term in $\mH$ which satisfies $\mHMIe{N} = \bzero$, and recursively using \eqn{eqn:HMSol} to determine $\mHMIe{K}$ for $K < N$. Again, for each $K$ the equation 
$\mHMIe{K}= \bzero$ gives the subspace that defines $\ciH^{(K)}_+$ which then makes up the affine space $\mnbdy{\ciH}{K}$. The problem of finding the structure of $\calH$ is then reduced solely to finding the subspace $\ciH^{(K)}_+$ for each $K$. In the subsequent sections we shall demonstrate the determination of this subspace $\mHMIe{N}= \bzero$ for the highest multi-body  ($N$-body) interaction in our system. Clearly, the same results will apply mutatis mutandis to all $K < N$. Finally, we note that $\mnbdy{\mH}{R}$ s satisfy the identity 
\beq
\frac{1}{2}  \sum_{R>0}^\mb  (-1)^R R! \left[\tr_{R} \mHPLe{R} - \tr_{R} \mHMIe{R}\right]=0.
\eeq

  In the next section we show how this is achieved for $N=2$, and generalize this to higher $N$ in the subsequent sections. The main results of this exercise are summarized in tables \ref{tab:NevenH} and \ref{tab:NoddH}.

\begin{widetext}
\begin{turnpage}
\begin{table*}
\begin{tabular}{||c|c|c|c|c|c|c|}
\hline\hline
Class &  $\No$ & $P$ & $Q$ & $\mHmb{\mb}$ & $\dim{\ci \calH^{(\mb)}}$ & $\ci\calH^{\mb}_+$  \\

\hline\hline

\sysign{\Ac}{0}{0}{0} & $\No$ & $\combi{\No}{\mb}$ & $-$ &  $\mHm = \dagg{\mHm}$ & $P^2$  & $\lieu(P)$  \\
\hline

\sysign{\AI}{+1}{0}{0} & $\No$ & $\combi{\No}{\mb}$ & $-$ & $\mHm = \conj{\mHm}$ & $P(P+1)/2$  & $\lieu(P)\setminus\lieo(P)$ \\
\hline

\sysign{\AII}{-1}{0}{0} & $\No=2\Mo$ & $\frac{1}{2}\left\{\combi{\No}{\mb} + \combi{\Mo}{\mb/2}\right\}$ &  $\frac{1}{2}\left\{\combi{\No}{\mb} - \combi{\Mo}{\mb/2}\right\}$ &  $\tbtmat{\mhm_{PP}}{\mhm_{PQ}}{\dagg{\mhm_{PQ}}}{\mhm_{QQ}}$ \footnotesize{$\begin{array}{l} \mhm_{PP} = \conj{\mhm_{PP}} \\ \mhm_{QQ} = \conj{\mhm_{QQ}} \\  \mhm_{PQ} = -\conj{\mhm_{PQ}} \end{array}$}  
 & $\frac{P(P+1)}{2} + \frac{Q(Q+1)}{2} + PQ$ & $\begin{array}{c}\lieu(P+Q)\\\setminus\lieo(P+Q)\end{array}$  \\
\hline

\sysign{\Dc}{0}{+1}{0}  & $\No$ & $\combi{\No}{\mb}$ & $-$ & $\mHm =\conj{\mHm}$ & $P(P+1)/2$  & $\lieu(P)\setminus\lieo(P)$  \\

\hline

\sysign{\C}{0}{-1}{0}  & $\No=2\Mo$ & $\frac{1}{2}\left\{\combi{\No}{\mb} + \combi{\Mo}{\mb/2}\right\}$ &  $\frac{1}{2}\left\{\combi{\No}{\mb} - \combi{\Mo}{\mb/2}\right\}$ &  $\tbtmat{\mhm_{PP}}{\mhm_{PQ}}{\dagg{\mhm_{PQ}}}{\mhm_{QQ}}$ \footnotesize{$\begin{array}{l} \mhm_{PP} = \conj{\mhm_{PP}} \\ \mhm_{QQ} = \conj{\mhm_{QQ}} \\  \mhm_{PQ} = -\conj{\mhm_{PQ}} \end{array}$}  
 & $\frac{P(P+1)}{2} + \frac{Q(Q+1)}{2} + PQ$ & $\begin{array}{c}\lieu(P+Q)\\\setminus\lieo(P+Q)\end{array}$  \\

\hline

\sysign{\AIII}{0}{0}{1} & $\No=p+q$ & $\sum\limits_{a=1,3,\dots}^{\mb-1}\combi{p}{a}\combi{q}{\mb-a}$ &  $\sum\limits_{a=0,2,\dots}^{\mb}\combi{p}{a}\combi{q}{\mb-a}$  &  $\tbtmat{\mhm_{PP}}{\bzero_{PQ}}{\bzero_{QP}}{\mhm_{QQ}}$ \footnotesize{$\begin{array}{l} \mhm_{PP} = \dagg{\mhm_{PP}} \\ \mhm_{QQ} = \dagg{\mhm_{QQ}} \end{array}$}  
 & $P^2 + Q^2$ & $\lieu(P)\oplus\lieu(Q)$  \\

\hline

\sysign{\BDI}{+1}{+1}{1} & $\No=p+q$ & $\sum\limits_{a=1,3,\dots}^{\mb-1}\combi{p}{a}\combi{q}{\mb-a}$ &  $\sum\limits_{a=0,2,\dots}^{\mb}\combi{p}{a}\combi{q}{\mb-a}$  &  $\tbtmat{\mhm_{PP}}{\bzero_{PQ}}{\bzero_{QP}}{\mhm_{QQ}}$ \footnotesize{$\begin{array}{l} \mhm_{PP} = \conj{\mhm_{PP}} \\ \mhm_{QQ} = \conj{\mhm_{QQ}} \end{array}$}  
 & $\frac{P(P+1)}{2} + \frac{Q(Q+1)}{2}$ & $\begin{array}{c}(\lieu(P)\setminus\lieo(P))\\\oplus\\(\lieu(Q)\setminus\lieo(Q))\end{array}$  \\

\hline

\sysign{\CII}{-1}{-1}{1}  & $\begin{array}{c} \No=p+q \\ p=2r \;\ q=2s \end{array}$

& $\begin{array}{c} P = \sum\limits_{a=1,3,\dots}^{\mb-1}\combi{p}{a}\combi{q}{\mb-a} \\ A(B) = P/2  \end{array}$

& $\begin{array}{c} Q = \sum\limits_{a=0,2,\dots}^{\mb}\combi{p}{a}\combi{q}{\mb-a}\\ C(D) = \frac{Q}{2} \pm \sum\limits_{a=0,2,\dots}^{m} \frac{1}{2} \combi{r}{a/2}\combi{s}{(m-a)/2} \end{array}$

& $\tbtpart{ \tbtarray{\mhm_{AA}}{\mhm_{AB}}{\dagg{\mhm_{AB}}}{\mhm_{BB}} }{\bzero_{PQ}}{\bzero_{QP}}{ \tbtarray{\mhm_{CC}}{\mhm_{CD}}{\dagg{\mhm_{CD}}}{\mhm_{DD}} }$

\footnotesize{$\begin{array}{l} \mhm_{AA(CC)} = \conj{\mhm_{AA(CC)}} \\ \mhm_{BB(DD)} = \conj{\mhm_{BB(DD)}} \\  \mhm_{AB(CD)} = -\conj{\mhm_{AB(CD)}} \end{array}$}
 
 & $\begin{array}{c} \frac{A(A+1)}{2} + \frac{B(B+1)}{2} + AB \\ + \frac{C(C+1)}{2} + \frac{D(D+1)}{2} + CD\end{array}$ & $\begin{array}{c}(\lieu(A+B)\\\setminus\lieo(A+B))\\\oplus\\(\lieu(C+D)\\\setminus\lieo(C+D))\end{array}$  \\

\hline

\sysign{\CI}{+1}{-1}{1}  & $\No=2 \Mo$  & 

$\begin{array}{c} P = \sum\limits_{a=1,3,\dots}^{\mb-1}\combi{\Mo}{a}\combi{\Mo}{\mb-a} \\ A(B) = \begin{cases} P/2 & ; \mb/2  \ \text{even} \\ \frac{P}{2} \mp \combi{\Mo}{\mb/2}  & ; \mb/2 \ \text{odd}  \end{cases} \end{array}$

& $\begin{array}{c} Q = \sum\limits_{a=0,2,\dots}^{\mb}\combi{\Mo}{a}\combi{\Mo}{\mb-a}\\ C(D) =  \begin{cases} \frac{Q}{2} \pm \combi{\Mo}{\mb/2} & ; \mb/2 \ \text{even} \\ Q/2 & ; \mb/2 \ \text{odd} \end{cases} \end{array}$

& $\tbtpart{ \tbtarray{\mhm_{AA}}{\mhm_{AB}}{\dagg{\mhm_{AB}}}{\mhm_{BB}} }{\bzero_{PQ}}{\bzero_{QP}}{ \tbtarray{\mhm_{CC}}{\mhm_{CD}}{\dagg{\mhm_{CD}}}{\mhm_{DD}} }$

\footnotesize{$\begin{array}{l} \mhm_{AA(CC)} = \conj{\mhm_{AA(CC)}} \\ \mhm_{BB(DD)} = \conj{\mhm_{BB(DD)}} \\  \mhm_{AB(CD)} = -\conj{\mhm_{AB(CD)}} \end{array}$}
 
 & $\begin{array}{c} \frac{A(A+1)}{2} + \frac{B(B+1)}{2} + AB \\ + \frac{C(C+1)}{2} + \frac{D(D+1)}{2} + CD\end{array}$ & -do- \\
\hline

\sysign{\DIII}{-1}{+1}{1} & $\No=2 \Mo$  & 

$\begin{array}{c} P = \sum\limits_{a=1,3,\dots}^{\mb-1}\combi{\Mo}{a}\combi{\Mo}{\mb-a} \\ A(B) = \begin{cases} P/2 & ; \mb/2  \ \text{even} \\ \frac{P}{2} \pm \combi{\Mo}{\mb/2}  & ; \mb/2 \ \text{odd}  \end{cases} \end{array}$

& $\begin{array}{c} Q = \sum\limits_{a=0,2,\dots}^{\mb}\combi{\Mo}{a}\combi{\Mo}{\mb-a}\\ C(D) =  \begin{cases} \frac{Q}{2} \pm \combi{\Mo}{\mb/2} & ; \mb/2 \ \text{even} \\ Q/2 & ; \mb/2 \ \text{odd} \end{cases} \end{array}$ 


& $\tbtpart{ \tbtarray{\mhm_{AA}}{\mhm_{AB}}{\dagg{\mhm_{AB}}}{\mhm_{BB}} }{\bzero_{PQ}}{\bzero_{QP}}{ \tbtarray{\mhm_{CC}}{\mhm_{CD}}{\dagg{\mhm_{CD}}}{\mhm_{DD}} }$

\footnotesize{$\begin{array}{l} \mhm_{AA(CC)} = \conj{\mhm_{AA(CC)}} \\ \mhm_{BB(DD)} = \conj{\mhm_{BB(DD)}} \\  \mhm_{AB(CD)} = -\conj{\mhm_{AB(CD)}} \end{array}$}
 
 & $\begin{array}{c} \frac{A(A+1)}{2} + \frac{B(B+1)}{2} + AB \\ + \frac{C(C+1)}{2} + \frac{D(D+1)}{2} + CD\end{array}$ & -do- \\

\hline\hline
\end{tabular}
\caption{Structure of $N$-body interaction hamiltonian $\mHmb{\mb}$ ($\mb$ even) in each symmetry class. The space of $K$-body ($K$ even) interaction Hamiltonians $\mnbdy{\ciH}{K}$ in a class is an affine subspace $\mnbdy{\ciH}{K} = \mnbdy{\ciH}{K}_+ + \mHMIe{K}$ where $\mnbdy{\ciH}{K}_+$ is to be read from the last column of this table and $\mHMIe{K}$ is given in \eqn{eqn:HMSol}.} 
\mylabel{tab:NevenH}
\end{table*}

\end{turnpage}
\begin{turnpage}

\begin{table}
\begin{tabular}{||c|c|c|c|c|c|c|}
\hline\hline
Class & $\No$ & $P$ & $Q$ & $\mHmb{\mb}$ & $\dim{\ci \calH^{(\mb)}}$ & $\ci\calH^{(\mb)}_+$   \\

\hline\hline

\sysign{\Ac}{0}{0}{0}  & $\No$ & $\combi{\No}{\mb}$ & $-$ &  $\mHm = \dagg{\mHm}$  & $P^2$ & $\lieu(P)$ \\

\hline

\sysign{\AI}{+1}{0}{0} & $\No$ & $\combi{\No}{\mb}$ & $-$ & $\mHm = {[\mHm]}^{*}$ & $P(P+1)/2$  & $\lieu(P)\setminus\lieo(P)$\\

\hline

\sysign{\AII}{-1}{0}{0} & $\No=2\Mo$ & $\frac{1}{2}\combi{2\Mo}{\mb}$ & $\frac{1}{2}\combi{2\Mo}{\mb}$  &  $\tbtmat{\mhm_{PP}}{\mhm_{PQ}}{- \conj{\mhm_{PQ}}}{\conj{\mhm_{PP}}}$   
 & $P^2 + 2\times \frac{P(P-1)}{2} $ & $\lieu(2P)\setminus\lieusp(2P)$  \\
\hline

\sysign{\Dc}{0}{+1}{0}  & $\No$ & $\combi{\No}{\mb}$ & $-$ & $\mHm = -\conj{\mHm}$ & $P(P-1)/2$  & $\lieo(P)$  \\

\hline

\sysign{\C}{0}{-1}{0} & $\No=2\Mo$ & $\frac{1}{2}\combi{2\Mo}{\mb}$ & $\frac{1}{2}\combi{2\Mo}{\mb}$  &  $\tbtmat{\mhm_{PP}}{\mhm_{PQ}}{ \conj{\mhm_{PQ}}}{-\conj{\mhm_{PP}}}$   
 & $P^2 + 2\times \frac{P(P+1)}{2} $ & $\lieusp(2P)$  \\

\hline

\sysign{\AIII}{0}{0}{1} & $\No=p+q$ & $\sum\limits_{a=1,3,\dots}^{\mb}\combi{p}{a}\combi{q}{\mb-a}$ &  $\sum\limits_{a=0,2,\dots}^{\mb-1}\combi{p}{a}\combi{q}{\mb-a}$  &  $\tbtmat{\bzero_{PP}}{\mhm_{PQ}}{\dagg{\mhm_{PQ}}}{\bzero_{QQ}}$  
 & $2PQ$ & $\lieu(P+Q)\setminus(\lieu(P)\oplus \lieu(Q))$ \\

\hline

\sysign{\BDI}{+1}{+1}{1} & $\No=p+q$ & $\sum\limits_{a=1,3,\dots}^{\mb}\combi{p}{a}\combi{q}{\mb-a}$ &  $\sum\limits_{a=0,2,\dots}^{\mb-1}\combi{p}{a}\combi{q}{\mb-a}$  &  $\tbtmat{\bzero_{PP}}{\mhm_{PQ}}{\transp{\mhm_{PQ}}}{\bzero_{QQ}}$  
 & $PQ$ & $\lieo(P+Q)\setminus(\lieo(P)\oplus \lieo(Q))$  \\

\hline

\sysign{\CII}{-1}{-1}{1} & $\begin{array}{c} \No=p+q \\ p=2r \;\ q=2s \end{array}$

& $\begin{array}{c} P = \sum\limits_{a=1,3,\dots}^{\mb}\combi{p}{a}\combi{q}{\mb-a} \\ A(B) = P/2  \end{array}$

& $\begin{array}{c} Q = \sum\limits_{a=0,2,\dots}^{\mb-1}\combi{p}{a}\combi{q}{\mb-a}\\ C(D) = \frac{Q}{2} \end{array}$

& $\tbtpart{\bzero_{PP}}{\tbtarray{\mhm_{AC}}{\mhm_{AD}}{-\conj{\mhm_{AD}}}{\conj{\mhm_{AC}}}}{ \text{h.c.}}{\bzero_{QQ}}$

 & $PQ$ & $\lieusp(P+Q)\setminus(\lieusp(P)\oplus \lieusp(Q))$ \\

\hline

\sysign{\CI}{+1}{-1}{1} & $\No=2\Mo$ & $\frac{1}{2}\combi{2\Mo}{\mb}$ & $\frac{1}{2}\combi{2\Mo}{\mb}$  &  $\tbtmat{\bzero_{PP}}{\mhm_{PQ}}{ \conj{\mhm_{PQ}}}{\bzero_{QQ}}$   
 & $P(P+1) $ & $\lieusp(2P)\setminus\lieu(P)$  \\
 
\hline

\sysign{\DIII}{-1}{+1}{1} & $\No=2\Mo$ & $\frac{1}{2}\combi{2\Mo}{\mb}$ & $\frac{1}{2}\combi{2\Mo}{\mb}$  &  $\tbtmat{\bzero_{PP}}{\mhm_{PQ}}{-\conj{\mhm_{PQ}}}{\bzero_{QQ}}$   
 & $P(P-1) $ & $\lieo(2P)\setminus\lieu(P)$ \\

\hline\hline
\end{tabular}
\caption{The ten symmetry classes of fermions and the structure of many-body Hamiltonians $\mHmb{\mb}$ when $\mb$ is odd. The space of $K$-body ($K$ odd) interaction Hamiltonians $\mnbdy{\ciH}{K}$ in a class is an affine subspace $\mnbdy{\ciH}{K} = \mnbdy{\ciH}{K}_+ + \mHMIe{K}$ where $\mnbdy{\ciH}{K}_+$ is to be read from the last column of this table and $\mHMIe{K}$ is given in \eqn{eqn:HMSol}.} 
\mylabel{tab:NoddH}
\end{table}
\end{turnpage}

\end{widetext}

\section{Structure of Two-body Hamiltonians}\mylabel{sec:Twobody}

In this section we find the structure of two-body Hamiltonians in each class. This will serve as a precursor to our study of a generic even $\mb$-body Hamiltonian. We define a \emph{strictly} two-body Hamiltonian as
\beq \mylabel{eqn:Ham2}
\vH = \vPsi^\dagger \vPsi^\dagger \mHmb{2} \vPsi \vPsi,
\eeq
where $\vPsi \vPsi$ is a notational short-hand for a column vector comprised of $\combi{\No}{2}$ distinct product terms of two fermionic annihilation operators, written as
\beq 
\vPsi \vPsi \equiv \left( \begin{array}{c}
_{\No-1}\begin{cases}
\psi_1 \psi_2 \\
\vdots \\
\psi_1 \psi_\No \end{cases}
\\ 
_{\No-2}\begin{cases}
\psi_2 \psi_3 \\
\vdots \\
\psi_3 \psi_\No  \\
\end{cases} \\
\vdots \\
_1\begin{cases} \psi_{\No-1} \psi_\No \\ \end{cases}
\end{array}
\right).
\mylabel{eq:twoPBigPsi}
\eeq Henceforth, we will call these product terms as states.
The definition of $\vPsi\vPsi$ also defines $\vPsi^\dagger \vPsi^\dagger \equiv $
\beq
\left( \begin{array}{cccc} 
 (\psi_1\psi_2)^\dagger \ldots (\psi_1\psi_\No)^\dagger & (\psi_2\psi_3)^\dagger \ldots (\psi_2\psi_\No)^\dagger & \ldots & (\psi_{\No-1} \psi_\No)^\dagger 
\end{array} \right).
\mylabel{eq:twoPBigPsiDag}
\eeq
The preceding definitions are in the same spirit as our definition for the states in \eqn{eqn:BigPsiDefn}.
%
%
%

We can also write $\mHmb{2}$ as (see \eqn{eq:HKthorder})
\beq
 (\vPsi^\dagger)^2 \mnbdy{\mH}{2}(\vPsi)^2 =  \sum_{i_1,i_2,j_1,j_2} (\psi_{i_1} \psi_{i_2})^\dagger H^{(2)}_{i_1,i_2;j_1,j_2} \psi_{j_1} \psi_{j_2}
\eeq
Note here that although $\mHmb{2}$ is a four indexed object, fermionic anticommutation demands that it is a matrix of dimension $\combi{\No}{2} \times \combi{\No}{2}$.
As done in the case of one-body Hamiltonians we now construct the $\combi{\No}{2} \times \combi{\No}{2}$ canonical representations of the symmetry operations($\mUTmb{2},\mUCmb{2},\mUSmb{2}$) from their one-body counterparts(see table~\ref{tab:tenfold}) and apply them to determine the structure of $\mnbdy{\mH}{2}$. 

\class{\Ac}
This class has no symmetries, therefore any hermitian matrix of dimension $\combi{\No}{2} \times \combi{\No}{2}$ belongs to this class.

\class{\AI}For the $\AI$ class we have $\mUT = \mOne$. Hence for this case, 
\bea
\TR \vPsi^\dagger \TR^{-1} &=& \vPsi^\dagger \mUT = \vPsi^\dagger \notag \\
\TR \vPsi \TR^{-1} &=& \mUTd \vPsi = \vPsi
\eea
and for the two-body states (see \eqn{eq:twoPBigPsi} and \eqn{eq:twoPBigPsiDag}) we have 
\bea
\TR \vPsi^\dagger \vPsi^\dagger \TR^{-1} &=&  \vPsi^\dagger \vPsi^\dagger \notag \\
\TR \vPsi \vPsi \TR^{-1} &=& \vPsi \vPsi,
\eea
making $\mUTmb{2}=\mOne$. Time reversal symmetry implies the following
\beq
\mHmb{2}={\left[\mHmb{2}\right]}^*
\eeq
which puts $\mHmb{2}$ in the class of real symmetric $\combi{\No}{2} \times \combi{\No}{2}$ matrices. The dimension of this class is $\frac{1}{2}\combi{\No}{2}\left\{\combi{\No}{2}+1\right\}$.

\class{\AII}
For class $\AII$ we have canonical $\mUT = \mJ$ and 
\bea
\TR \vPsi^\dagger \TR^{-1} &=& \vPsi^\dagger \mUT = \vPsi^\dagger \mJ\notag \\
\TR \vPsi \TR^{-1} &=& \mUTd \vPsi = -\mJ \vPsi.
\eea
Also we had seen that for this case to be realizable $\No=2\Mo$. We can divide the $2\Mo$ states as $\Mo$ states of flavor $\ta$ and  $\Mo$ of flavor $\tb$\mylabel{sym:tatb}. Explicitly these transform in the following way\nomcls{g71}{$\ta$, $\tb$}{Labels the $p$, $q$ type of states respectively. \symref{sym:tatb}} 
\bea
\TR  \left(
\psi^\dagger_{1\ta} \ldots \psi^\dagger_{\Mo\ta}~~\psi^\dagger_{1\tb} \ldots \psi^\dagger_{\Mo\tb}
\right)\TR^{-1}&\notag\\=&  
\left( 
-\psi^\dagger_{1\tb} \ldots -\psi^\dagger_{\Mo\tb}~~\psi^\dagger_{1\ta} \ldots \psi^\dagger_{\Mo\ta}\right)\notag
\eea
\bea
\TR
\left( \begin{array}{c}
\psi_{1\ta} \\
\vdots \\
\psi_{\Mo\ta} \\
\psi_{1\tb} \\
\vdots \\
\psi_{\Mo\tb} 
\end{array}
\right) \TR^{-1} &=& 
\left( \begin{array}{c}
-\psi_{1\tb} \\
\vdots \\
-\psi_{\Mo\tb} \\
\psi_{1\ta} \\
\vdots \\
\psi_{\Mo\ta} 
\end{array}
\right)
\eea
implying, 
\bea
\TR \psi^\dagger_{i\ta} \TR^{-1} &=& -\psi^\dagger_{i\tb} \notag \\
\TR \psi^\dagger_{i\tb} \TR^{-1} &=& \psi^\dagger_{i\ta} \notag \\
\TR \psi_{i\ta} \TR^{-1} &=& -\psi_{i\tb} \notag \\
\TR \psi_{i\tb} \TR^{-1} &=& \psi_{i\ta}.
\mylabel{eq:AIIcondi}
\eea
Since $\vPsi\vPsi$ is formed by the product of these one-body operators, its elements can be divided into the following three kinds 
\beq 
\vPsi \vPsi = \left( \begin{array}{c}
_{\combi{\Mo}{2}}\begin{cases}
\psi_{i\ta} \psi_{j\ta} 
\end{cases}
\\ 
_{\Mo^2}\begin{cases}
\psi_{i\ta} \psi_{j\tb} 
\end{cases}
\\
_{\combi{\Mo}{2}}\begin{cases}
\psi_{i\tb} \psi_{j\tb} 
\end{cases}
\end{array}
\right).
\eeq
The number of states in each kind add upto $\combi{\No}{2}$, i.e., $2\combi{\Mo}{2}+\Mo^2=\Mo(\Mo-1)+\Mo^2=\Mo(2\Mo-1)=\frac{\No(\No-1)}{2}=\combi{\No}{2}.$ We find the effect of the $\TR$ operators on these states to be
\beq 
\TR \vPsi \vPsi \TR^{-1} = \TR \left( \begin{array}{c}
\psi_{i\ta} \psi_{j\ta} \\ 
\psi_{i\ta} \psi_{j\tb} \\
\psi_{i\tb} \psi_{j\tb} \\
\end{array}
\right) \TR^{-1}
= \left( \begin{array}{c}
\psi_{i\tb} \psi_{j\tb} \\ 
-\psi_{i\tb} \psi_{j\ta} \\
\psi_{i\ta} \psi_{j\ta} \\
\end{array}
\right) 
= \left( \begin{array}{c}
\psi_{i\tb} \psi_{j\tb} \\ 
\psi_{j\ta} \psi_{i\tb} \\
\psi_{i\ta} \psi_{j\ta} \\
\end{array}
\right).
\eeq
This suggests that we can re-organize the two-body basis 
into ``symmetric"($\Sys$) and ``antisymmetric"($\Ans$) states as follows\footnote{For the clarity of the equations, we will not explicitly put the factors of $\frac{1}{\sqrt{2}}$ in front of symmetric and antisymmetric states. It should be assumed that all the states are properly normalized.}
\beq \mylabel{eqn:symantisym}
\left( \begin{array}{c}
_{\combi{\Mo}{2}}\begin{cases}
\psi_{i\ta} \psi_{j\ta} 
\end{cases}
\\ 
_{\Mo^2}\begin{cases}
\psi_{i\ta} \psi_{j\tb}
\end{cases}
\\
_{\combi{\Mo}{2}}\begin{cases}
\psi_{i\tb} \psi_{j\tb}
\end{cases}
\end{array}
\right)
\longrightarrow
\left( \begin{array}{c}
_\Sys\begin{cases}
_{\combi{\Mo}{2}}\begin{cases}
\psi_{i\ta} \psi_{j\ta} + \psi_{i\tb} \psi_{j\tb}
\end{cases}
\\ 
_{\Mo}\begin{cases}
\psi_{i\ta} \psi_{i\tb}
\end{cases}
\\ 
_{\combi{\Mo}{2}}\begin{cases}
\psi_{i\ta} \psi_{j\tb} + \psi_{j\ta}\psi_{i\tb} \:\:_{(i\neq j)}
\end{cases}
\end{cases}
\\
_\Ans\begin{cases}
_{\combi{\Mo}{2}}\begin{cases}
\psi_{i\ta} \psi_{j\tb} - \psi_{j\ta}\psi_{i\tb} \:\:_{(i\neq j)}
\end{cases}
\\
_{\combi{\Mo}{2}}\begin{cases}
\psi_{i\ta} \psi_{j\ta}  -\psi_{i\tb} \psi_{j\tb}
\end{cases}
\end{cases}
\end{array}
\right).
\eeq\nomcls{g80}{$\Sys$, $\Ans$}{``symmetric'' and ``antisymmetric'' states. \symdefref{eqn:symantisym}} Conveniently, in this way of organizing our Hilbert space, we have
\bea
\TR \Sys \TR^{-1} &=& \Sys \notag \\
\TR \Ans \TR^{-1} &=& -\Ans.
\mylabel{eqn:SymmAnitsymmTransf}
\eea 
Also note that $\Mo^2=2\combi{\Mo}{2}+\Mo$. 

It will be useful to mention here that the transformation properties of these states under symmetry gives us the definitions of ``symmetric'' and ``antisymmetric'' states. This terminology will be used in later sections when the $\mb$-body Hamiltonians will be discussed. Note that the word ``symmetric" may not necessarily imply a ``+" sign in the linear combination of its constituent states and vice-versa.

Armed with this, we look for the structure of $\mHmb{2}$
\bea
\TR \vPsi^\dagger \vPsi^\dagger \mHmb{2} \vPsi \vPsi \TR^{-1} &=& \vPsi^\dagger \vPsi^\dagger \mHmb{2} \vPsi \vPsi \notag \\
\mUTmb{2} \conj{\mHmb{2}} \mUTdmb{2} &=& \mHmb{2},
\mylabel{eqn:twopsymm}
\eea
where $\mUTmb{2}$ in this new basis is obtained from \eqn{eqn:SymmAnitsymmTransf} and is given by
\beq
\dagg{\mUTmb{2}} = \mOne_{P,Q},
\eeq
where $P=2\combi{\Mo}{2}+\Mo$ and $Q=2\combi{\Mo}{2}$. Assuming,
\beq
\mHt = \left( 
\begin{array}{cc}
\mht_{PP} & \mht_{PQ} \\
\dagg{\mht_{PQ}} & \mht_{QQ}
\end{array}
\right)
\mylabel{eqn:TwoPHstruc}
\eeq
and implementing the condition in \eqn{eqn:twopsymm}, we get the following constraints, 
\bea
\mht_{PP} = \conj{\mht_{PP}}, \quad 
\mht_{QQ} = \conj{\mht_{QQ}}, \quad  
\mht_{PQ} = -\conj{\mht_{PQ}}.
\mylabel{eqn:TwoPHcond}
\eea
This leads to total $\frac{P(P+1)}{2} + \frac{Q(Q+1)}{2} + PQ$ independent parameters in $\mHt$, which is equal to $\dim{\ci \calH^{(2)}_{\AII}}$.

\class{\Dc}
We now look at transposing symmetries of the \TL~ kind. The states transform according to \eqn{eqn:UPsiTrn} and since class \Dc~ has canonical $\mUC=\mOne$, we have 
\bea
\vU \Psi^\dagger \vU^{-1} &=& \vPsi^T \notag \\
\vU \Psi \vU^{-1} &=&  (\vPsi^\dagger)^T
\eea 
which implies
\bea
\vU \psi^\dagger_i \vU^{-1} &=& \psi_i \notag \\
\vU \psi_i \vU^{-1} &=&  \psi^\dagger_i .
\mylabel{eq:transD}
\eea 
Thus even for the two-body Hamiltonian the symmetry condition(\eqn{eqn:TrnSym}) is 
\beq
 \mHt = \conj{\mHt},
\eeq
making $\mHt$ a real symmetric matrix with $\frac{1}{2}\combi{\No}{2}\left\{\combi{\No}{2}+1\right\}$ parameters.

\class{\C}
Class \C~ comes with canonical $\mUC= \mJ$ and requires $\No=2\Mo$. The states transform as
\bea
\vU \Psi^\dagger \vU^{-1} &=& \vPsi^T \mJ \notag \\
\vU \Psi \vU^{-1} &=&  -\mJ (\vPsi^\dagger)^T.
\eea 
Similar to class \AII~ we label the states with $\ta,\tb$ flavors and find there transformations to be 
\bea
\CC \psi^\dagger_{i\ta} \CC^{-1} &=& -\psi_{i\tb}\notag \\
\CC \psi^\dagger_{i\tb} \CC^{-1} &=& \psi_{i\ta} \notag\\
\CC \psi_{i\ta} \CC^{-1} &=& -\psi^\dagger_{i\tb} \notag\\
\CC \psi_{i\tb} \CC^{-1} &=& \psi^\dagger_{i\ta} .
\mylabel{eq:condC}
\eea
However, unlike class $\AII$, one must remember that here we are dealing with a transposing symmetry. The two-body basis in this case can be rearranged in the following manner 
\beq 
\left( \begin{array}{c}
_{\combi{\Mo}{2}}\begin{cases}
\psi_{i\ta} \psi_{j\ta} 
\end{cases}
\\ 
_{\Mo^2}\begin{cases}
\psi_{i\ta} \psi_{j\tb}
\end{cases}
\\
_{\combi{\Mo}{2}}\begin{cases}
\psi_{i\tb} \psi_{j\tb}
\end{cases}
\end{array}
\right)
\longrightarrow
\left( \begin{array}{c}
_\Ans\begin{cases}
_{\combi{\Mo}{2}}\begin{cases}
\psi_{i\ta} \psi_{j\ta} + \psi_{i\tb} \psi_{j\tb}
\end{cases}
\\ 
_{\Mo}\begin{cases}
\psi_{i\ta} \psi_{i\tb}
\end{cases}
\\ 
_{\combi{\Mo}{2}}\begin{cases}
\psi_{i\ta} \psi_{j\tb} + \psi_{j\ta}\psi_{i\tb} \:\:_{(i\neq j)}
\end{cases}
\end{cases}\\
_\Sys\begin{cases}
_{\combi{\Mo}{2}}\begin{cases}
\psi_{i\ta} \psi_{j\tb} - \psi_{j\ta}\psi_{i\tb} \:\:_{(i\neq j)}
\end{cases}
\\
_{\combi{\Mo}{2}}\begin{cases}
\psi_{i\ta} \psi_{j\ta}  -\psi_{i\tb} \psi_{j\tb}
\end{cases}
\end{cases}
\end{array}
\right).
\eeq
The action of $\CC$ symmetry on these states are
\bea
\CC \Sys \CC^{-1} &=& \Sys^\dagger \notag \\
\CC \Ans \CC^{-1} &=& -\Ans^\dagger.
\eea 
From this we get $2\combi{\Mo}{2}$ number of $\Sys$ states and  $\Mo^2$ number of $\Ans$ states. Imposing $\CC$ symmetry on the Hamiltonian gives
\bea
:\CC \vPsi^\dagger \vPsi^\dagger \mHmb{2} \vPsi \vPsi \CC^{-1}: &=& \vPsi^\dagger \vPsi^\dagger \mHmb{2} \vPsi \vPsi \notag \\
\mUCdmb{2} \conj{\mHmb{2}} \mUCmb{2} &=& \mHmb{2}
\mylabel{eqn:twopCsymm}
\eea
where
\beq
\mUCmb{2} = -\mOne_{P,Q}
\eeq
and $P=2\combi{\Mo}{2} + \Mo$ and $Q=2\combi{\Mo}{2}$. Denoting the internal structure of $\mHt$ with \eqn{eqn:TwoPHstruc}, the constraints turn out to be same as \eqn{eqn:TwoPHcond}. This again leads to $\frac{P(P+1)}{2} + \frac{Q(Q+1)}{2} + PQ$ independent parameters.

%
%

\class{\AIII}
For transposing symmetries of the \TA~ kind, we know 
\beq
:(\vU \vH \vU^{-1}): \;\; =  \vH.
\eeq 
Class $\AIII$ has canonical $\mUS=\mOne_{p,q}$ and
\bea
\vU \Psi^\dagger \vU^{-1} &=& \vPsi^T \mOne_{p,q} \notag \\
\vU \Psi \vU^{-1} &=&  \mOne_{p,q} (\vPsi^\dagger)^T.
\eea 
Here, $\No=p+q$. We label $p$-type states with $\ta$ and the $q$-type states with $\beta$ and they transform as  
\bea
\vU \psi^\dagger_{i\ta} \vU^{-1} &=& \psi_{i\ta} \notag \\
\vU \psi^\dagger_{i\tb} \vU^{-1} &=& -\psi_{i\tb} \notag \\
\vU \psi_{i\ta} \vU^{-1} &=&  \psi^\dagger_{i\ta} \notag \\
\vU \psi_{i\tb} \vU^{-1} &=&  -\psi^\dagger_{i\tb}. 
\mylabel{eq:condClassAIII}
\eea 
The two-body states look like 
\beq 
\vPsi \vPsi = \left( \begin{array}{c}
_{\combi{p}{2}}\begin{cases}
\psi_{i\ta} \psi_{j\ta} 
\end{cases}
\\ 
_{pq}\begin{cases}
\psi_{i\ta} \psi_{j\tb} 
\end{cases}
\\
_{\combi{q}{2}}\begin{cases}
\psi_{i\tb} \psi_{j\tb} 
\end{cases}
\end{array}
\right).
\eeq
Also note that $\combi{p}{2}+pq+\combi{q}{2}=\combi{\No}{2}$ and
\beq 
\SL \vPsi \vPsi \SL^{-1} = \SL \left( \begin{array}{c}
\psi_{i\ta} \psi_{j\ta} \\ 
\psi_{i\ta} \psi_{j\tb} \\
\psi_{i\tb} \psi_{j\tb} \\
\end{array}
\right) \SL^{-1}
= \left( \begin{array}{c}
-(\psi_{i\ta}\psi_{j\ta})^\dagger \\ 
 (\psi_{i\ta}\psi_{j\tb})^\dagger \\
-(\psi_{i\tb}\psi_{j\tb})^\dagger \\
\end{array}
\right)\ , 
\eeq
which shows that terms with odd and even $\ta$s will transform differently under the $\SL$ symmetry. We reorganize the states as 
\beq
\left( \begin{array}{c}
\psi_{i\ta} \psi_{j\ta} \\ 
\psi_{i\ta} \psi_{j\tb} \\
\psi_{i\tb} \psi_{j\tb} \\
\end{array}
\right)
\rightarrow
\left( \begin{array}{c}
_P\begin{cases}
\psi_{i\ta} \psi_{j\tb} \\
\end{cases} \\
_Q\begin{cases}
\psi_{i\ta} \psi_{j\ta} \\ 
\psi_{i\tb} \psi_{j\tb} \\
\end{cases}
\end{array}
\right),
\eeq
where $P=pq$ and $Q=\combi{p}{2}+\combi{q}{2}$. Symmetry then demands 
\bea
:\SL \vPsi^\dagger \vPsi^\dagger \mHmb{2} \vPsi \vPsi \SL^{-1}: &=& \vPsi^\dagger \vPsi^\dagger \mHmb{2} \vPsi \vPsi \notag \\
\mUSdmb{2} \mHmb{2} \mUSmb{2} &=& \mHmb{2}
\mylabel{eqn:twopSsymm}
\eea
where 
\beq
\mUSmb{2} = \mOne_{P,Q}.
\eeq
Using the structure of $\mHt$ as given in \eqn{eqn:TwoPHstruc} and imposing \eqn{eqn:twopSsymm}, the constraints come out as
\bea
\mht_{PQ} = \bzero,  \quad 
\mht_{PP} = \dagg{\mht_{PP}},  \quad 
\mht_{QQ} = \dagg{\mht_{QQ}}. 
\mylabel{eq:AIIIcondtwo}
\eea
The off-diagonal blocks of $\mHt$ are forced to be \emph{zero}, which is in contrast with structure of the one-body Hamiltonian in class \AIII (see table~\ref{tab:NonInt}).  The number of independent parameters are $P^2+Q^2$.

Before delving into the classes \BDI,~\CII,~\CI~ and \DIII, we bring to the attention of the reader that since all of them respect the sublattice symmetry($\tS=1$), the Hamiltonian for all four classes will be of the form
\beq
\mHt = \left( 
\begin{array}{cc}
\mht_{PP} & \mZero \\
\mZero & \mht_{QQ}
\end{array}
\right).
\mylabel{eq:SLH2struc}
\eeq
The presence of $\tT=\pm 1$ and $\tC=\pm 1$ symmetries will only put constraints on the non-zero sub-blocks of $\mHt$.

\class{\BDI}
This class has $\mUS=\mOne_{p,q}$, $\mUC=\mOne_{p,q}, \mUT=\mOne$. Since time reversal symmetry operation has a canonical $\mUT=\mOne$, we now have additional conditions on the sub-blocks(see \eqn{eq:SLH2struc}), 
\bea
\mht_{PP} = \conj{{\mht_{PP}}}, \quad 
\mht_{QQ} = \conj{{\mht_{QQ}}}.
\eea
The number of independent parameters are $\frac{P(P+1)}{2} + \frac{Q(Q+1)}{2}$.

\class{\CII}
Here,
\beq
\mUT = \left(
\begin{array}{cc}
\mJ_{pp} & \bzero_{p  q} \\
\bzero_{q  p} & \mJ_{qq}
\end{array}
\right), \;\;\; \mUC = \left(
\begin{array}{cc}
-\mJ_{pp} & \bzero_{p  q} \\
\bzero_{q  p} & \mJ_{qq}
\end{array}
\right), \;\;\;\; \mUS = \mOne_{p,q}\ .
\eeq
Given sublattice symmetry, we have the structure of the Hamiltonian as
shown in \eqn{eq:SLH2struc}. We also have $p=2r$ and $q=2s$. We label the $p$ states as $\pa$ and $\pb$ both of which run from $1 \ldots r$. Similarly the $q$ states can be labeled with $\qa$ and $\qb$, each running from $1 \ldots s$. Transformations of these new flavors under symmetry are 
\bea
\TR \psi_{i\pa} \TR^{-1} &=& -\psi_{i\pb} \notag \\
\TR \psi_{i\pb} \TR^{-1} &=& \psi_{i\pa} \notag \\
\TR \psi_{i\qa} \TR^{-1} &=& -\psi_{i\qb} \notag \\
\TR \psi_{i\qb} \TR^{-1} &=& \psi_{i\qa}.
\mylabel{eq:CondClassCII}
\eea
The $P=pq=4rs$ states can be rearranged as  
\beq 
\left( \begin{array}{c}
_{rs}\begin{cases}
\psi_{i\pa} \psi_{j\qa} 
\end{cases}
\\ 
_{rs}\begin{cases}
\psi_{i\pa} \psi_{j\qb}
\end{cases}
\\
_{rs}\begin{cases}
\psi_{i\pb} \psi_{j\qa}
\end{cases}
\\
_{rs}\begin{cases}
\psi_{i\pb} \psi_{j\qb}
\end{cases}
\end{array}
\right)
\longrightarrow
\left( \begin{array}{c}
_\Sys\begin{cases}
_{rs}\begin{cases}
\psi_{i\pa} \psi_{j\qa} + \psi_{i\pb} \psi_{j\qb}
\end{cases}
\\ 
_{rs}\begin{cases}
\psi_{i\pa} \psi_{j\qb} - \psi_{i\pb}\psi_{j\qa}
\end{cases}
\end{cases}
\\
_\Ans\begin{cases}
_{rs}\begin{cases}
\psi_{i\pa} \psi_{j\qa} - \psi_{i\pb}\psi_{j\qb} 
\end{cases}
\\
_{rs}\begin{cases}
\psi_{i\pa} \psi_{j\qb}+\psi_{i\pb} \psi_{j\qa}
\end{cases}
\end{cases}
\end{array}
\right).
\eeq
Remembering that $Q=\combi{p}{2}+\combi{q}{2}$, both the $\combi{p}{2}$ and $\combi{q}{2}$ states can also be organized into symmetric and antisymmetric states. For example rearrangement of the $\combi{p}{2}$ states leads to
\beq 
\left( \begin{array}{c}
_{\combi{r}{2}}\begin{cases}
\psi_{i\pa} \psi_{j\pa} 
\end{cases}
\\ 
_{r^2}\begin{cases}
\psi_{i\pa} \psi_{j\pb}
\end{cases}
\\
_{\combi{r}{2}}\begin{cases}
\psi_{i\pb} \psi_{j\pb}
\end{cases}
\end{array}
\right)
\longrightarrow
\left( \begin{array}{c}
_\Sys\begin{cases}
_{\combi{r}{2}}\begin{cases}
\psi_{i\pa} \psi_{j\pa} + \psi_{i\pb} \psi_{j\pb}
\end{cases}
\\ 
_{r}\begin{cases}
\psi_{i\pa} \psi_{i\pb}
\end{cases}
\\ 
_{\combi{r}{2}}\begin{cases}
\psi_{i\pa} \psi_{j\pb} + \psi_{j\pa}\psi_{i\pb} \:\:_{(i\neq j)}
\end{cases}
\end{cases}
\\
_\Ans\begin{cases}
_{\combi{r}{2}}\begin{cases}
\psi_{i\pa} \psi_{j\pb} - \psi_{j\pa}\psi_{i\pb} \:\:_{(i\neq j)}
\end{cases}
\\
_{\combi{r}{2}}\begin{cases}
\psi_{i\pa} \psi_{j\pa}  -\psi_{i\pb} \psi_{j\pb}
\end{cases}
\end{cases}
\end{array}
\right),
\eeq
which transform as 
\bea
\TR \Sys \TR^{-1} &=& \Sys \notag \\
\TR \Ans \TR^{-1} &=& -\Ans.
\eea 
It is interesting to note that this rearrangement of the states do not mix the $P$ and $Q$ blocks. 

The two-body $\mUTmb{2}$ can now be written as
\beq
\mUTmb{2} = \left(
\begin{array}{cc}
\mOne_{A,B} & \bzero \\
\bzero & \mOne_{C,D}
\end{array}
\right)
\eeq
where  $A=2rs$ and $B=2rs$, $C=r^2+s^2$ and $D=2\combi{r}{2}+2\combi{s}{2}$. Imposing symmetry demands 
\bea
\TR \vPsi^\dagger \vPsi^\dagger \mHmb{2} \vPsi \vPsi \TR^{-1} &=& \vPsi^\dagger \vPsi^\dagger \mHmb{2} \vPsi \vPsi \notag \\
\mUTmb{2} \conj{\mHmb{2}} \mUTdmb{2} &=& \mHmb{2},
\eea
and $\mht_{PP}$, $\mht_{QQ}$ remain decoupled. Writing
\beq
\mht_{PP} = \left( 
\begin{array}{cc}
\mht_{AA} & \mht_{AB} \\
\dagg{\mht_{AB}} & \mht_{BB}
\end{array}
\right)
\eeq
and implementing the above conditions, we get the following constraints 
\bea
\mht_{AA} = \conj{\mht_{AA}},~~~& 
\mht_{BB} = \conj{\mht_{BB}},~~~& 
\mht_{AB} = -\conj{\mht_{AB}}.
\eea
This leads to total $\frac{A(A+1)}{2} + \frac{B(B+1)}{2} + AB$ independent parameters. Similarly setting 
\beq
\mht_{QQ} = \left( 
\begin{array}{cc}
\mht_{CC} & \mht_{CD} \\
\dagg{\mht_{CD}} & \mht_{DD}
\end{array}
\right)
\eeq
and implementing the symmetry conditions, we get the following constraints 
\bea
\mht_{CC} = \conj{\mht_{CC}},~~~& 
\mht_{DD} = \conj{\mht_{DD}},~~~& 
\mht_{CD} = -\conj{\mht_{CD}},
\eea
leading to $\frac{C(C+1)}{2} + \frac{D(D+1)}{2} + CD$ independent parameters for this part of the Hamiltonian.

\class{\CI}
This class 
arises with
\beq
\mUT = \mF, \;\;\; \mUC = -\mJ, \;\;\; \mUS = \mOne_{p,q}.
\eeq
Sublattice symmetry already implies a structure of Hamiltonian as given in \eqn{eq:SLH2struc}. Also we have $P=\Mo^2$ and $Q=2\combi{\Mo}{2}$. States can be re-organized as 
%
\beq 
\left( \begin{array}{c}
_{\Mo^2}\begin{cases}
\psi_{i\ta} \psi_{j\tb}
\end{cases}
\\
_{\combi{\Mo}{2}}\begin{cases}
\psi_{i\ta} \psi_{j\ta} 
\end{cases}
\\
_{\combi{\Mo}{2}}\begin{cases}
\psi_{i\tb} \psi_{j\tb}
\end{cases}

\end{array}
\right)
\longrightarrow
\left( \begin{array}{c}
_\Sys\begin{cases}
_{\combi{\Mo}{2}}\begin{cases}
\psi_{i\ta} \psi_{j\tb} - \psi_{j\ta}\psi_{i\tb} \:\:_{(i\neq j)}
\end{cases}
\end{cases} \\
_\Ans\begin{cases}
_{\combi{\Mo}{2}}\begin{cases}
\psi_{i\ta} \psi_{j\tb} + \psi_{j\ta}\psi_{i\tb} \:\:_{(i\neq j)}
\end{cases}
\\
_{\Mo}\begin{cases}
\psi_{i\ta} \psi_{i\tb}
\end{cases}
\end{cases} \\
_\Sys\begin{cases}
_{\combi{\Mo}{2}}\begin{cases}
\psi_{i\ta} \psi_{j\ta} + \psi_{i\tb} \psi_{j\tb}
\end{cases}
\end{cases}
\\
_\Ans\begin{cases}
_{\combi{\Mo}{2}}\begin{cases}
\psi_{i\ta} \psi_{j\ta}  -\psi_{i\tb} \psi_{j\tb}
\end{cases}
\end{cases}
\end{array}
\right).
\eeq
Again both the sectors get decoupled and are symmetric and antisymmetric under time reversal. The remaining analysis follows closely the one for class \CII~ with the appropriate replacements of $A=\combi{\Mo}{2}$, $B=\combi{\Mo}{2}+\Mo$, $C=\combi{\Mo}{2}$  and $D=\combi{\Mo}{2}$.

\class{\DIII}
This one possesses 
\beq
\mUT = \mJ, \;\;\; \mUC = \mF, \;\;\; \mUS = \mOne_{p,q}\ ,
\eeq
and $\No=2\Mo$. States can again be reorganized as symmetric and antisymmetric  under time reversal as follows
\beq 
\left( \begin{array}{c}
_{\Mo^2}\begin{cases}
\psi_{i\ta} \psi_{j\tb}
\end{cases}\\
_{\combi{\Mo}{2}}\begin{cases}
\psi_{i\ta} \psi_{j\ta} 
\end{cases}
\\
_{\combi{\Mo}{2}}\begin{cases}
\psi_{i\tb} \psi_{j\tb}
\end{cases}
\end{array}
\right)
\longrightarrow
\left( \begin{array}{c}
_\Sys\begin{cases}
_{\combi{\Mo}{2}}\begin{cases}
\psi_{i\ta} \psi_{j\tb} + \psi_{j\ta}\psi_{i\tb} \:\:_{(i\neq j)}
\end{cases}
\\
_{\Mo}\begin{cases}
\psi_{i\ta} \psi_{i\tb}
\end{cases}
\end{cases} \\
_\Ans\begin{cases}
_{\combi{\Mo}{2}}\begin{cases}
\psi_{i\ta} \psi_{j\tb} - \psi_{j\ta}\psi_{i\tb} \:\:_{(i\neq j)}
\end{cases}
\end{cases} \\
_\Sys\begin{cases}
_{\combi{\Mo}{2}}\begin{cases}
\psi_{i\ta} \psi_{j\ta} + \psi_{i\tb} \psi_{j\tb}
\end{cases}
\end{cases}
\\
_\Ans\begin{cases}
_{\combi{\Mo}{2}}\begin{cases}
\psi_{i\ta} \psi_{j\ta}  -\psi_{i\tb} \psi_{j\tb}
\end{cases}
\end{cases}
\end{array}
\right).
\eeq
Yet again both the sectors get decoupled and are symmetric and antisymmetric under the time reversal. Rest of the analysis proceeds along the same lines as class \CII~ and \CI~with $A=\combi{\Mo}{2}+\Mo$, $B=\combi{\Mo}{2}$, $C=\combi{\Mo}{2}$  and $D=\combi{\Mo}{2}$.

\section{$\mb$-body Interacting Hamiltonians}\mylabel{sec:Nbody}
We now determine the structure of a $\mb$-body Hamiltonian of the generic form (see \eqn{eq:HKthorder}),
\beq
\vH^{(\mb)} =\sum\limits_{\substack{i_1,i_2 \ldots,i_\mb \\ j_1,j_2 \ldots j_\mb}} (\psi_{i_1} \psi_{i_2} \ldots \psi_{i_\mb})^\dagger H^{(\mb)}_{i_1,i_2 \ldots,i_\mb;j_1,j_2 \ldots j_\mb} \psi_{j_1} \psi_{j_2} \ldots \psi_{j_\mb}. 
\mylabel{eq:moddH}
\eeq As discussed in section \sect{sec:IntGen}, this is the key ingredient to determine the structure of $\mH$ in \eqn{eqn:HSpace}. A basis of $\combi{\No}{\mb}$ states is required to describe $\vH^{(\mb)}$. We find that structure determination of $\vH^{(\mb)}$ depends on whether $\mb$ is odd or even. In the next subsection we focus on the cases when $\mb$ is odd followed by the subsection for $\mb$ even. 
In both cases, the strategy is to use table~\ref{tab:tenfold} to find a canonical representation of a symmetry operation $\mU^{(\mb)}$, which then aids in finding the final structure of $\mHmb{\mb}$.

\subsection{Structure of $\mHmb{\mb}$ for $\mb$ odd}
\class{\Ac}
This class has no symmetries and the only condition imposed on the Hamiltonian is 
\beq
\mHmb{\mb}=\dagg{\mHmb{\mb}}.
\eeq
The Hamiltonian has $\left\{\combi{\No}{\mb} \right\}^2$ independent parameters.

\class{\AI}
The implementation of the time reversal demands
\beq
\mHmb{\mb}=\conj{\mHmb{\mb}}.
\eeq
Hence the number of independent parameters are $\frac{1}{2}\combi{\No}{\mb}\left[\combi{\No}{\mb} +1\right]$.

\class{\AII}
Given the conditions, as in \eqref{eq:AIIcondi}, we know $\No=2\Mo$. There are $\Mo$  $\ta$ and $\Mo$ $\tb$ states. A generic many-body state has the form
\beq
\ta_a \tb_{\mb-a} \equiv \underbrace{\psi_{i\ta_1}\ldots \psi_{i\ta_a}}_{a}\underbrace{\psi_{j\tb_1}\ldots \psi_{j\tb_{\mb-a}}}_{\mb-a}.
\mylabel{eq:genstate}
\eeq
This state transforms in the following way, 
\bea
\TR \ta_a \tb_{\mb-a}\TR^{-1} &=& \tb_a \ta_{\mb-a}(-1)^a= \ta_{\mb-a} \tb_a (-1)^{a(\mb-a+1)} \notag \\ &=& 
(-1)^a\ta_{\mb-a}\tb_a.
\eea
The last equality uses the fact that $\mb$ is odd.
Using this we form two kinds of states, where one is made of $even$ number of $\ta$s($\equiv E_\ta$)\mylabel{sym:EaOa} and the other with $odd$ ($\equiv O_\ta$), having\nomcls{g90}{$E_\ta$,$E_\pa$,$E_\qa$}{A state made of even number of $\ta, \pa, \qa$ states respectively. \symref{sym:EaOa}}\nomcls{h10}{$O_\ta$,$O_\pa$,$O_\qa$}{A state made of odd number of $\ta, \pa, \qa$ states respectively. \symref{sym:EaOa}}\nomcls{h20}{$P,Q,A,B,C,D$}{Dimensions of various matrix blocks making up the parent many-body Hamiltonian matrix.}  
\bea
\dim O_\ta &=& \sum_{a=1,3\ldots}^{\mb} \combi{\Mo}{a} \combi{\Mo}{\mb-a} = \frac{1}{2}{\combi{2\Mo}{\mb}} \notag \\
\dim E_\ta &=& \sum_{a=0,2\ldots}^{\mb-1} \combi{\Mo}{a} \combi{\Mo}{\mb-a} = \frac{1}{2}{\combi{2\Mo}{\mb}}. 
\eea
These new states transform conveniently as
\bea
\TR O_\ta \TR^{-1} &=& -E_\ta \notag \\
\TR E_\ta \TR^{-1} &=& O_\ta 
\eea
which overall gives the following transformation
\beq
\TR \Psi \ldots \Psi  \TR^{-1} = - \mnbdy{\mJ}{\mb} \Psi \ldots \Psi
\eeq
where $\mnbdy{\mJ}{\mb}$\mylabel{sym:mJmb} is the $\mb$-body version of $\mJ$(defined in \eqref{eqn:J}), but now with dimensions $\combi{\No}{\mb}$.
This tells us that the Hamiltonian transforms as the one-body case discussed near \eqn{eqn:HAIIonep}. However, the dimension now is given by, $\dim{\calH^{(\mb)}_{\AII}} =  P^2 + 2 \times \frac{P(P-1)}{2}$, where $P=\frac{1}{2}{\combi{2\Mo}{\mb}}$.\nomcls{h30}{$\mnbdy{\mJ}{\mb}$}{$\mb$-body version of $\mJ$ defined in \eqn{eqn:J}. \symref{sym:mJmb}}

\class{\Dc}
In this class the transformation of $\psi$s are shown in \eqn{eq:transD}. The Hamiltonian satisfies
\beq
: \CC \vH^{(\mb)} \CC^{-1}:\ = \vH^{(\mb)}.
\eeq
This implies, $-[\mHmb{\mb}]^{*}=\mHmb{\mb}$, with the number of independent parameters being $\frac{P(P-1)}{2}$, where $P=\combi{\No}{\mb}$.

\class{\C}
Using the transformations for the fermionic operators given in \eqn{eq:condC}, we find out that $O_\ta$ and $E_\ta$ transform as 
\bea
\CC O_\ta \CC^{-1} = (-1)^{\frac{\mb+1}{2}} (E_\ta)^\dagger \notag \\
\CC E_\ta \CC^{-1} = (-1)^{\frac{\mb-1}{2}} (O_\ta)^\dagger.
\eea
Therefore as discussed near \eqn{eqn:HConep}, we again have $\mUCmb{\mb}$ $\propto \mnbdy{\mJ}{\mb}$. Hence the Hamiltonian satisfies the same condition as the single-body case, with the dimension $\dim{\calH_\C^{(\mb)}} = P^2 + 2 \times \frac{P(P+1)}{2} = P (2P+1)$  where $P=\frac{1}{2}{\combi{2\Mo}{\mb}}$.

\class{\AIII}
Given the conditions for Class \AIII~ (\eqn{eq:condClassAIII}) where $p$ and $q$ states are labeled by $\ta$ and $\tb$, the states transform according to
\bea
\SL O_\ta \SL^{-1} = (-1)^{\frac{\mb-1}{2}} (O_\ta)^{\dagger} \notag \\
\SL E_\ta \SL^{-1} = (-1)^{\frac{\mb+1}{2}} (E_\ta)^{\dagger}\ ,
\eea
having
\bea
\dim O_\ta &=& \sum_{a=1,3\ldots}^{\mb} \combi{p}{a}\combi{q}{\mb-a} = P \notag \\
\dim E_\ta &=& \sum_{a=0,2\ldots}^{\mb-1} \combi{p}{a}\combi{q}{\mb-a} = Q.
\eea
Therefore $\mUSmb{\mb} \propto \mOne_{P,Q}$ as we had seen in the one-body case (see near \eqn{eqn:AIIIStruc}). This implies that just like the one-body case, the diagonal blocks of $\mHmb{\mb}$ will be constrained to be $zero$ (see \eqn{eqn:AIIIOnep}) forcing the structure of the Hamiltonian to be 
\beq
\mHmb{\mb}=
\tbtmat{\bzero_{PP}}{\mhm_{PQ}}{\dagg{\mhm_{PQ}}}{\bzero_{QQ}}.
\mylabel{eqn:HAIIIodd}
\eeq
This gives the number of independent parameters as $\dim{\calH^{(\mb)}_{\AIII}} = 2PQ$.

We again bring to attention of the reader that from here onwards and until the end of this subsection all the classes have sublattice symmetry and therefore their respective Hamiltonians will always posses the above form.

\class{\BDI}
The time reversal symmetry implementation on the class \AIII, demands $\mHmb{\mb}$ to be real (see \eqn{eqn:HAIIIodd}) and therefore, the $\dim{\calH^{(\mb)}_{\BDI}} = PQ$.

\class{\CII}
The transformation of states is given by \eqn{eq:CondClassCII}, and in this class we have $p=2r$, $q=2s$. We now show how these affect the $P$ and $Q$ states. The general structure of any state is of the form 
\beq
\underbrace{ \underbrace{\psi_{i\pa_1}\ldots \psi_{i\pa_c}}_{c} \underbrace{\psi_{i\pb_1} \ldots \psi_{i\pb_{a-c}}}_{a-c}}_{a}\underbrace{ \underbrace{\psi_{j\qa_1}\ldots \psi_{j\qa_d}}_d \underbrace{\psi_{j\qb_1} \ldots \psi_{j\qb_{\mb-a-d}}}_{\mb-a-d}}_{\mb-a}.
\eeq
Given sublattice symmetry 
 and the fact that $\pa \leftrightarrow \pb$,  $\qa \leftrightarrow \qb$ under time reversal, the transformed states still live within their respective blocks. Therefore they do not change the constraints put on the Hamiltonian due to sublattice symmetry and hence preserve the structure of the Hamiltonian appearing in \eqn{eqn:HAIIIodd}.

To see this more clearly, we write the above state schematically and find that it transforms under $\TR$ as
\bea
\pa_c\pb_{a-c}\qa_d\qb_{\mb-a-d} \xrightarrow{\TR} \begin{cases} (-1)^{c} \pa_{a-c} \pb_a \qa_{\mb-a-d}\qb_d &   ; a  \ \mathrm{odd} \\ (-1)^{d}\pa_{a-c} \pb_a \qa_{\mb-a-d}\qb_d &   ;a\ \mathrm{even} \end{cases}.\notag\\
\eea
Therefore the odd-even structure of $a$ is still preserved. On further substructuring of states into odd(even)-number of $\pa$ $\equiv O_{\pa}(E_{\pa})$
and odd(even)-number of $\qa$ $\equiv O_{\qa}(E_{\qa})$ within each $P$ and $Q$ block, we find that $\TR$ symmetry acts on them as
\beq 
\left( \begin{array}{c}
_{P}\begin{cases}
O_\ta 
\end{cases}
\\
_{Q}\begin{cases}
E_\ta
\end{cases}
\end{array}
\right)
\longrightarrow
\left( \begin{array}{c}
_1\begin{cases}
_{\frac{P}{2}}\begin{cases}
E_\pa O_\qa  \\
E_\pa E_\qa 
\end{cases}
\end{cases}
\\
_2\begin{cases}
_{\frac{P}{2}}\begin{cases}
O_\pa O_\qa  \\
O_\pa E_\qa 
\end{cases}
\end{cases}
\\ 
_3\begin{cases}
_{\frac{Q}{2}}\begin{cases}
O_\pa E_\qa  \\
E_\pa E_\qa 
\end{cases}
\end{cases} 
\\ 
_4\begin{cases}
_{\frac{Q}{2}}\begin{cases}
O_\pa O_\qa  \\
E_\pa O_\qa 
\end{cases}
\end{cases} 
\end{array}
\right)
{\xrightarrow{\TR}}
\left( \begin{array}{c}
_2\begin{cases}
O_\pa O_\qa  \\
O_\pa E_\qa 
\end{cases}
\\
_{-1}\begin{cases}
-E_\pa O_\qa  \\
-E_\pa  E_\qa 
\end{cases}
\\ 
_4\begin{cases}
O_\pa O_\qa  \\
E_\pa O_\qa 
\end{cases} 
\\ 
_{-3}\begin{cases}
-O_\pa E_\qa \\
-E_\pa E_\qa
\end{cases} 
\end{array}
\right).
\eeq
Hence $\mUTmb{\mb}$ takes the same form as in the one-body case (see \eqn{eqn:TmCm}) which is
\beq
\mUTmb{\mb} = \left(
\begin{array}{cc}
\mJ_{PP} & \bzero_{P  Q} \\
\bzero_{Q  P} & \mJ_{QQ}
\end{array}
\right).
\eeq
The structure of Hamiltonian then follows from the one-body case discussed near \eqn{eqn:CIIOnep} with the number of independent parameters given by $\dim\calH^{(\mb)}_{\CII} = PQ$.

\class{\CI}
The presence of sublattice symmetry enforces that the Hamiltonian has a off-diagonal structure (see \eqn{eqn:HAIIIodd}) with dimension $P=Q=\frac{1}{2}\combi{2\Mo}{\mb}$. As we have seen, $\mUT$ in this case is $\mF$ (see \eqn{eqn:TpCm}), and converts $\ta \leftrightarrow \tb$ and vice-versa, therefore the transformation is
\bea
\TR O_\ta \TR^{-1} &=& E_\ta \notag \\
\TR E_\ta \TR^{-1} &=& O_\ta .
\eea
Hence $\mUTmb{\mb}$ is $\mF^{(\mb)}$(a generalized version of $\mF$ defined in \eqref{eqn:TpCmBC} but with dimension $\combi{\No}{\mb}$)\mylabel{sym:mFmb} and the constraint on $\mHmb{\mb}$ is same as that shown in \eqn{eqn:CIOnep} but with $M$ replaced with $P$. The $\dim$ of $\calH^{(\mb)}_{\CI}$ is $P(P+1)$.\nomcls{h40}{$\mnbdy{\mF}{\mb}$}{$\mb$-body version of $\mF$ defined in \eqn{eqn:TpCmBC}. \symref{sym:mFmb}}

\class{\DIII}
For this class, $\mUT=\mJ$ (see \eqn{eqn:TmCp}) and states transform as
\bea
\TR \psi_{i\ta} \TR^{-1} = -\psi_{i\tb} \notag \\
\TR \psi_{i\tb} \TR^{-1} = \psi_{i\ta}.
\eea
The odd and even $\ta$ states transform in the following way
\bea
\TR O_\ta \TR^{-1} &=& -E_\ta \notag \\
\TR E_\ta \TR^{-1} &=& O_\ta .
\eea
Therefore $\mUTmb{\mb} = \mJ^{(\mb)}$ with matrix dimension $P$. The constraints on the Hamiltonian is same as that shown in \eqn{eqn:DIIIOnep} but now with $\dim \calH_{\DIII}^{(\mb)} = P(P-1)$, where $P =\frac{1}{2}\combi{2\Mo}{\mb}$. The generic structure of $\mHmb{\mb}$ for $\mb$-odd is summarized and tabulated in table~\ref{tab:NoddH}.

\subsection{Structure of $\mHmb{\mb}$ for $\mb$ even}

\class{\Ac}
This class has no symmetries and the only condition that applies is $\mHmb{\mb}=\dagg{\mHmb{\mb}}$.  
The Hamiltonian therefore has $\left\{\combi{\No}{\mb} \right\}^2$ independent parameters.

\class{\AI}
As is in the case of $\mb$-odd, this class only enforces the condition  $\mHmb{\mb}={[\mHmb{\mb}]}^*$ and the number of independent parameters are  $\frac{1}{2}\combi{\No}{\mb}\left\{\combi{\No}{\mb} +1\right\}$.

\class{\AII}
Given the conditions in \eqn{eq:AIIcondi}, we know $\No=2\Mo$. There are $\Mo$ $\ta$ and $\Mo$ $\tb$ states. Note that given $\mb$ is even, we can reorganize states into symmetric and antisymmetric states (like in two-body case, see \eqn{eqn:symantisym}). The transformation on a generic state (see \eqn{eq:genstate}) is given by
\beq
\TR \ta_a \tb_{\mb-a} \TR^{-1} = \ta_{\mb-a} \tb_{a}.
\eeq
Therefore symmetric and antisymmetric states can be made by linearly combining, 
$\ta_a \tb_{\mb-a}$ and  $\ta_{\mb-a} \tb_{a}$ with a $\pm$ sign. Note that $\Ans$ and $\Sys$ don't contain the same number of states. The state $\ta_{\frac{\mb}{2}}\tb_{\frac{\mb}{2}}$ (with same orbital labels) goes back to itself  without any sign change under transformation and therefore is a symmetric state. Then dimensions of symmetric($\Sys$) and antisymmetric($\Ans$) states are
\bea
\dim \Sys &=& \frac{1}{2}\left\{\combi{\No}{\mb} + \combi{\Mo}{\mb/2} \right\} = P\notag\\
\dim \Ans &=& \frac{1}{2}\left\{\combi{\No}{\mb} - \combi{\Mo}{\mb/2} \right\} = Q
\eea
and $\mUTmb{\mb} = \mOne_{P,Q}$. The structure of the Hamiltonian in this basis is given by
\beq
\mHmb{\mb} = \tbtmat{\mhm_{PP}}{\mhm_{PQ}}{\dagg{\mhm_{PQ}}}{\mhm_{QQ}}
\mylabel{eqn:HNevenstruc}
\eeq
with symmetry conditions being
\bea
\TR \vPsi^\dagger \vPsi^\dagger \mHmb{\mb} \vPsi \vPsi \TR^{-1} &=& \vPsi^\dagger \vPsi^\dagger \mHmb{\mb} \vPsi \vPsi \notag \\
\mUTmb{\mb} \conj{\mHmb{\mb}} \mUTdmb{\mb} &=& \mHmb{\mb}.
\mylabel{eqn:mbpsymm}
\eea
This imposes the following constraints,
\bea
 \mhm_{PP} = \conj{\mhm_{PP}},~~~&
 \mhm_{QQ} = \conj{\mhm_{QQ}},~~~&
 \mhm_{PQ} = -\conj{\mhm_{PQ}}.
\mylabel{eqn:Hevencond}
\eea
The total number of independent parameters are  $\frac{P(P+1)}{2} + \frac{Q(Q+1)}{2} + PQ$. This general formula reduces to the specific two-body case discussed above by substituting $\mb=2$. 
 
\class{\Dc}
For this class the transformation of $\psi$s are determined by \eqn{eq:transD}. The constraint on the Hamiltonian is then $\conj{\mHmb{\mb}}=\mHmb{\mb}$, which is same as that for Class \AI.

\class{\C}
The symmetric ($\Sys$) and antisymmetric states ($\Ans$) are again linear combinations of $\ta_a \tb_{\mb-a}$ and  $\ta_{\mb-a} \tb_{a}$ with a $\pm$ sign. However depending on the value of $\mb$, the $+$ linear combination may transform under the $\CC$ symmetry with a $-$ sign and therefore be an antisymmetric state by definition. In general a state will transform as 
\beq
\ta_a \tb_{\mb-a} \xrightarrow{\CC} (-1)^{\mb/2} (\ta_{\mb-a} \tb_a)^{\dagger}.
\eeq
With the definition
\bea
P &=&  \frac{1}{2}\left\{\combi{\No}{\mb} + \combi{\Mo}{\mb/2} \right\} \notag  \\
Q &=&  \frac{1}{2}\left\{\combi{\No}{\mb} - \combi{\Mo}{\mb/2} \right\}\ ,
\eea
the dimensions of symmetric and antisymmetric states now are 
\bea
\dim \Sys &=& \begin{cases} P \;\ ; \mb/2 \ \text{even} \\ Q \;\ ;  \ \text{odd} \end{cases}\notag \\
\dim \Ans &=& \begin{cases} Q \;\ ; \mb/2 \ \text{even} \\ P \;\ ;  \ \text{odd} \end{cases}\ .
\eea
Therefore $\mUCmb{\mb} = (-1)^{\mb/2}\mOne_{P,Q}$. The structure of the Hamiltonian  is again constrained in the same way as we had seen in the \eqn{eqn:HNevenstruc} and \eqn{eqn:Hevencond}.
 
\class{\AIII}
Given the conditions for Class \AIII~(\eqn{eq:condClassAIII}) where $p$ and $q$ states are labeled by $\ta$ and $\tb$, we look at the transformation of $E_{\ta}$ and $O_{\ta}$ states introduced near \eqn{eq:genstate},
\bea
\SL O_\ta \SL^{-1} &=& (-1)^{\frac{\mb}{2}-1} O^{\dagger}_\ta\notag \\
\SL E_\ta \SL^{-1} &=& (-1)^{\frac{\mb}{2}}E^{\dagger}_\ta.
\eea
The dimensions are given by, 
\bea \mylabel{eqn:PQdefAIII}
\begin{split}
\dim O_\ta = \sum_{a=1, 3,\dots}^{\mb-1} \combi{p}{a} \combi{q}{\mb-a} = P \notag \\
\dim E_\ta = \sum_{a=0, 2,\dots}^{\mb} \combi{p}{a} \combi{q}{\mb-a} = Q
\end{split}\quad ,
\eea
and $\mUSmb{\mb} = (-1)^{\frac{\mb}{2}-1}\mOne_{P,Q}$. This imposes the condition that off-diagonal blocks in \eqn{eqn:HNevenstruc} are now $zero$ and the independent parameters, which are $P^{2} + Q^{2}$ in number, belong to the diagonal blocks.

The following classes discussed in this subsection all have sublattice symmetry and the form of the Hamiltonian in each class will satisfy
\beq
\mHmb{\mb} = \tbtmat{\mhm_{PP}}{\bzero_{PQ}}{\bzero_{QP}}{\mhm_{QQ}}
\mylabel{eqn:HNevenstrucSL}
\eeq
\class{\BDI}
Apart from the implementation of the sublattice symmetry as in the previous section, 
time reversal operation for this class is just $\mUTmb{\mb} = \mOne$ and therefore just demands $\mhm_{PP}$ and $\mhm_{QQ}$ to be real. The number of independent parameters are $\frac{P(P+1)}{2} + \frac{Q(Q+1)}{2}$.

\class{\CII}
This class requires that $p=2r$ and $q=2s$. Imposition of sublattice symmetry breaks the basis into blocks $O_\ta$ and $E_\ta$ of dimensions given in \eqn{eqn:PQdefAIII}. Now the $\ta$ states are made of two varieties $\pa$ and $\pb$. While the $\tb$ states are made of $\qa$ and $\qb$. The time-reversal symmetry converts $\pa \leftrightarrow \pb$ and $\qa \leftrightarrow \qb$ states. Therefore one can make symmetric and antisymmetric combinations. The transformations under $\TR$ are given in \eqn{eq:CondClassCII}. 
Now let us discuss the $O_\ta$ states. Dimension of $O_\ta$ states is $P$ and is comprised of following kinds of states
\beq 
\left( \begin{array}{c}
_1\begin{cases}
E_\pa E_\qa \\
O_\pa O_\qa
\end{cases}
\\
_2\begin{cases}
O_\pa E_\qa \\
E_\pa O_\qa
\end{cases}
\end{array}
\right)
{\xrightarrow{\TR}}
\left( \begin{array}{c}
_1\begin{cases}
O_\pa O_\qa \\
E_\pa E_\qa
\end{cases}
\\
_{2}\begin{cases}
-E_\pa O_\qa \\
-O_\pa E_\qa
\end{cases}
\end{array}
\right).
\eeq
Hence the number of symmetric states($\equiv A$) and number of antisymmetric states($\equiv B$) equals $P/2$ within the $P$ block (see table~\ref{tab:NevenH}).

A typical $E_\ta$ state transforms under $\TR$ as
\beq
\pa_c \pb_{a-c} \qa_{d} \qb_{\mb-a-d} \xrightarrow{\TR} \pa_{a-c} \pb_c \qa_{\mb-a-d} \qb_d,
\eeq
using which symmetric($\Sys$) and antisymmetric($\Ans$) combinations can again be formed.  For each $a$, the  total number of states ($\Sys+\Ans$) are,  
\beq
\combi{p}{a}\combi{q}{\mb-a} = \sum_{c,d} \combi{r}{c} \combi{r}{a-c}\combi{s}{d}\combi{s}{\mb-a-d}.
\eeq
The number of symmetric and antisymmetric states are equal except when $c=\frac{a}{2}$ and $d=\frac{\mb-a}{2}$ which gives individual counts of symmetric and antisymmetric states as
\bea
\Sys_a &=& \frac{1}{2}\left\{\combi{p}{a}\combi{q}{\mb-a} + \combi{r}{\frac{a}{2}} \combi{s}{\frac{\mb-a}{2}} \right\} \notag \\
\Ans_a &=& \frac{1}{2}\left\{\combi{p}{a}\combi{q}{\mb-a} - \combi{r}{\frac{a}{2}}\combi{s}{\frac{\mb-a}{2}} \right\}. 
\eea
This brings the total symmetric($C$) and antisymmetric($D$) states in the $Q$ block to be
\bea
C = \sum_{a=0,2,\dots}^{\mb} \Sys_a  \qquad
D = \sum_{a=0,2,\dots}^{\mb} \Ans_a.
\eea
The structure of the Hamiltonian for the $O_\ta$ states is similar to the $\AII$ case with $(P,Q)$ of $\AII$ replaced with $(A,B)$ respectively. Likewise, replacing $(P,Q)$ with $(C,D)$ gives us the Hamiltonian structure for $E_\ta$ states(see table~\ref{tab:NevenH}). The number of independent parameters are 
\bea
\frac{A(A+1)}{2} + \frac{B(B+1)}{2} + AB \notag \\ +
\frac{C(C+1)}{2} + \frac{D(D+1)}{2} + CD\ .
\mylabel{eqn:indepparaevenN}
\eea

\class{\CI} 
In this class, $\No=2\Mo$. Imposition of sublattice symmetry again breaks the basis states into two blocks $P$ and $Q$ with dimensions as seen in \eqn{eqn:PQdefAIII}. It can be seen that $\ta_a \tb_{\mb-a} \rightarrow (-1)^a \ta_{\mb-a} \tb_a$. A state having $a$ as odd(even) belongs to the $P(Q)$ block. Therefore symmetric and antisymmetric states can again be constructed. Given $\frac{\mb}{2}$ is an even integer, then the number of symmetric and antisymmetric states which can be formed using states of the $P(Q)$ block will be equal(unequal). This makes the $P,Q$ Hamiltonian blocks take the structure of the Hamiltonian for the case $\CII$ with
\bea
A = \frac{P}{2} \qquad C = \frac{1}{2} \left\{ Q + \combi{\Mo}{\mb/2} \right\} \notag \\
B =  \frac{P}{2} \qquad D = \frac{1}{2}\left\{Q - \combi{\Mo}{\mb/2}\right\}.
\eea
Where as, for $\mb/2$ odd the number of symmetric and antisymmetric states become unequal(equal) leading to the same Hamiltonian structure(also see table~\ref{tab:NevenH}) as before but with
\bea
A = \frac{1}{2} \left\{ P - \combi{\Mo}{\mb/2} \right\} \qquad  C = \frac{Q}{2} \notag \\
B = \frac{1}{2} \left\{ P + \combi{\Mo}{\mb/2} \right\} \qquad D =  \frac{Q}{2}. 
\eea
The expression for total number of independent parameters is same as Class \CII.

\class{\DIII}
The reasoning for $\DIII$ class is similar to $\CI$ with the crucial distinction that $\ta_a \tb_{\mb-a} \rightarrow \ta_{\mb-a} \tb_a$.  This distinction readjusts the number of symmetric and antisymmetric states in $P,Q$ blocks. The similarities between the two classes gives us the same cases depending on the oddness and evenness of $\mb/2$, however now with minor changes in the formulae for dimensions of the $A-D$ sub-blocks of $P,Q$, which now become
\beq
 A = \begin{cases} P/2 &  \\ \frac{P}{2} + \combi{\Mo}{\mb/2}  & \end{cases} \qquad B = \begin{cases} P/2 &  ; \mb/2  \ \text{even} \\ \frac{P}{2} - \combi{\Mo}{\mb/2}  &   ; \  \text{odd}  \end{cases} \notag
\eeq 
\beq 
 C =  \begin{cases} \frac{Q}{2} + \combi{\Mo}{\mb/2} \\ Q/2 &  \end{cases} \quad D =  \begin{cases} \frac{Q}{2} - \combi{\Mo}{\mb/2} & ; \mb/2 \ \text{even} \\ Q/2 & ; \ \text{odd} \end{cases} .
\eeq 
The structure of Hamiltonian remains same as discussed for class \CI~with the expression for total number of independent parameters given by \eqn{eqn:indepparaevenN}.
The generic structure of $\mHmb{\mb}$ for $\mb$-even is summarized and tabulated in table~\ref{tab:NevenH}.

\section{Concluding Remarks}\mylabel{sec:Conc}

In this paper we have revisited the tenfold scheme of classification of fermions with the aim of studying interacting systems. We have endeavored to provide a simple and direct approach that makes clear the underlying physical content even while not being tied to the single particle picture. The canonical representation of symmetries in each of the ten classes (see table~\ref{tab:tenfold}) not only allows to obtain the structures of Hamiltonians in each class, but also provides some crucial physical insights into the nature of the symmetry operations. For example, our discussion of the symmetry operations in type 3 classes reveals a physical view of the interplay between the sublattice symmetry with other symmetries. For example, in the chiral classes with $\tT=\tC$ (\BDI, \CII) both time reversal and charge conjugation operation leaves the sublattice flavor unchanged, while for $\tT=-\tC$ (\DIII, \CI) the time reversal and charge conjugation operation flips the sublattice flavor. Furthermore, the results of our group cohomological considerations(\sect{sec:GrpCohom}) brings in crucial insights into the results of table~\ref{tab:tenfold}. From the perspective of interacting fermions, tables~\ref{tab:NevenH} and \ref{tab:NoddH} contain the key results on the structures of the $N$-body Hamiltonians in each class. We believe that the results will be useful to construct models for interacting systems in various classes, studies of which could be used to develop deeper understanding and phenomenology that can aid reaching the ultimate goal -- the topological classification of interacting fermionic systems. Finally, not the least, our analysis provides a natural way to reveal the geometric structure of the time evolution operator (see \eqn{eqn:Schord}) in systems with $N$-body interactions. It is clear that Cartan's symmetric spaces make their appearance again.

We conclude the paper by pointing out another very active area of condensed matter physics where our results would be of value --  many body localization and thermalization\cite{Nandkishore2015}.  Our results can be used, again, to create models in any symmetry class with arbitrary disorder in the kinetic energy (quadratic Hamiltonian) or any $N$-body interaction. In fact, tables~\ref{tab:NevenH} and \ref{tab:NoddH} can be used to generate random matrix ensembles that can create models with different physical content which can be used to investigate outstanding issues in that area.

\section*{Acknowledgments}

The authors thank A.~Altland, G.~Baskaran, S.~Rao and D.~Sen for remarks and comments. AA and AH thank CSIR, India, for support and VBS acknowledges a research grant from SERB, DST, India.

\bibliography{bibtenfold}

\clearpage
\newpage

\appendix
\renewcommand{\nomname}{}
\begin{widetext}
\section{Appendix: List of symbols}\mylabel{sec:ListOfSymbols}
\printnomcl
\end{widetext}

\end{document}